# High Purity Germanium Based Radiation Detectors with Segmented Amorphous Semiconductor Electrical Contacts: Fabrication Procedures

**Mark Amman**

Lawrence Berkeley National Laboratory, Berkeley, CA 94720 USA (retired)

## Article History

## Article Identifiers

## Key Words

## Abstract

Radiation detectors constructed from large volume high purity Ge (HPGe) single crystals are widely used for gamma-ray spectroscopy. The detectors for this application can be simple in that they need only have two electrical contacts for voltage application and signal readout. Such HPGe based detectors have been commercially produced for many decades using standard semiconductor fabrication processes. For the applications of gamma-ray imaging and particle tracking, however, interaction position measurement within the detector as well as the measurement of the deposited energy is required. This necessitates a more complex detector often with the electrical contacts divided into a large number of segments that can be individually instrumented for signal readout. The reliable and cost-effective implementation of contact segmentation with the standard commercial processes is a challenge. An alternative fabrication technology based on thin film amorphous semiconductor layers was developed at Lawrence Berkeley National Laboratory (LBNL) and can be used to produce finely segmented, fully passivated HPGe detectors. Over the last almost two decades, a large number of segmented contact HPGe detectors have been produced at LBNL using this technology. This paper provides a set of procedures that has been used at LBNL for the manufacture of the segmented amorphous semiconductor contact HPGe detectors.

# Table of contents

# 1. Introduction

An introduction to the physics and fabrication of high purity Ge (HPGe) based radiation detectors has been given in an earlier paper [Amman 2018]. This previous paper also described the issues associated with the standard fabrication technologies typically used to produce these detectors and the benefits of the alternative technology that makes use of thin films of amorphous semiconductors. None of that material will be repeated in the current paper, and it is recommended that the reader review the aforementioned paper as well as the papers describing the development of the amorphous semiconductor electrical contact and surface passivation technologies [Hansen 1977] [Hansen 1980] [Luke 1992] [Luke 1994a] [Luke 1994b] [Amman 2000a] [Amman 2000b] [Hull 2005] [Amman 2007] [Looker 2015a] [Looker 2015b]. The purpose of this current paper is to provide a set of procedures developed and used at Lawrence Berkeley National Laboratory (LBNL) for the manufacture of segmented electrical contact HPGe detectors.

A typical detector produced with the LBNL procedures is of the type schematically shown in Figure 1.1, and a photograph of two example detectors is given in Figure 1.2. The detector geometry is planar with two large area electrical contact faces that are approximately square in shape. In addition to this active volume of the detector's HPGe crystal are two extensions (handles) protruding from the bottom side of the crystal. The geometry of the crystal and electrical contacts is such that, during operation as a detector, the depletion region within the crystal never extends significantly into the handles. Since the handles remain undepleted, surface damage to the handles will not introduce leakage current. Consequently, the handles simplify detector fabrication and mounting during testing by providing an area of the crystal that can be handled without negatively affecting the detector performance. This geometry is similar to the top hat geometry that has been extensively used in the past both to provide the handle (brim of the top hat) for processing and to introduce geometric control of any surface channels along the sides of the detector [Llacer 1966]. A typical detector of this type produced at LBNL would be cut from a cylindrical HPGe boule slice with the approximate dimensions of 100 mm diameter and 15 mm thickness. The active area would be approximately 80 mm by 80 mm.

The cross-sectional drawings of Figure 1.1 illustrate the basic structure of the detector. This structure is simple and is created by first coating the properly prepared HPGe crystal with a high resistivity thin film of an amorphous semiconductor such as amorphous Ge (a-Ge) or amorphous Si (a-Si). This is then followed by depositing a patterned layer of metal (electrodes), typically Al, on top of the amorphous film. The amorphous semiconductor film dictates the charge injection blocking behavior of the electrical contacts, whereas the low resistivity metal defines the physical areas of the contacts. The amorphous semiconductor coating also serves as a passivation coating on the detector surfaces that are not covered with the metal layer. The properties of the amorphous semiconductor film affect the performance of the resultant detectors, and these properties substantially depend on and are controllable through the deposition process parameters used to create the film. The optimization of the film depends on its desired function and has been covered in other papers [Amman 2018 and references therein].

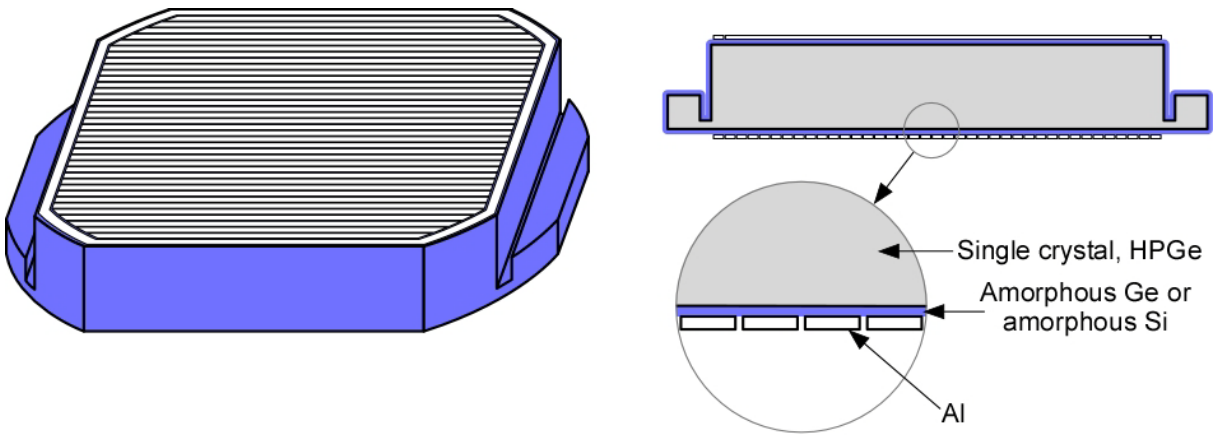

**Figure 1.1** Schematic drawings of an orthogonal strip HPGe based radiation detector. The drawing on the left is a perspective view of the detector while those on the right are cross-sectional views.

The signal readout electrode design of the detectors shown in Figures 1.1 and 1.2 is an orthogonal strip configuration. In this configuration, a set of strips is created on each of the two detector faces, and the two sets of strips are oriented perpendicular to each other. The segmentation of the electrical contacts into strips is done so that the position of each radiation interaction event can be determined. The interaction position in such a detector is determined in the two dimensions parallel to the electrical contact faces by the spatial intersection of the strip collecting the electrons from the event on one detector face with the hole collecting strip on the other detector face. The location can be further localized in the dimension perpendicular to the contacts based on the difference in the charge arrival times at the two strips [Momayezi 1999] [Amman 2000a] [Amman 2000b]. In addition to the strip electrodes is a guard ring surrounding each set of strips. The purpose of the guard rings is to take up the leakage current flowing along the detector surface between the two contact faces [Goulding 1961] and to reduce the negative impact of any side surface channels on detector performance [Kingston 1956] [Llacer 1964] [Dinger 1975] [Malm 1976] [Hansen 1980] [Hull 1995].

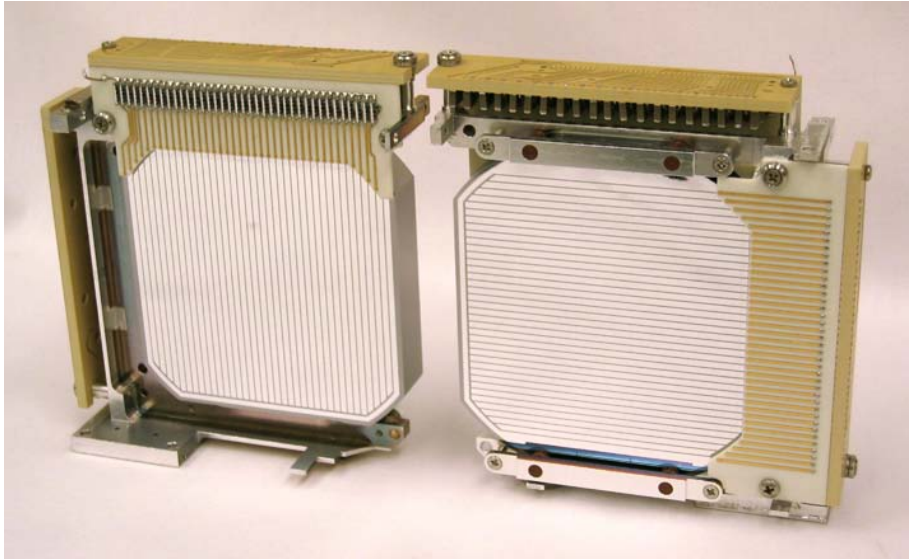

**Figure 1.2** Photograph of two orthogonal strip HPGe based radiation detectors produced for the Compton Spectrometer and Imager (COSI) instrument [Kierans 2016] [Chiu 2017]. One detector is viewed looking at its top face while the other is of its bottom. Each detector has an area of approximately 80 mm by 80 mm and is approximately 15 mm thick. Each face of each detector has 37 strips with a center to center strip spacing of 2 mm and a gap of 0.25 mm between each strip. Surrounding the set of strips on each detector face is a guard ring. Each detector is held in an Al holder, and attached to this holder are circuit boards used to route the detector signals to flexcircuits (not shown). Connections between the detector strips and the circuit board traces were made with ultrasonic wedge wire bonding. The sets of circuit boards near the top of the photograph are associated with the high voltage side of the detectors. These boards house the high voltage coupling capacitor and bias resistor associated with each signal readout channel.

The detector fabrication procedures detailed in the following sections are for detectors of the type shown in Figures 1.1 and 1.2. This planar orthogonal strip detector configuration is only one example of a HPGe detector that can be fabricated using the amorphous semiconductor technologies. Other examples have been presented previously [Amman 2018 and references therein]. The fabrication procedures given below can be adapted for many of these other detector configurations. The content of the remainder of this paper consists of sections covering fabrication preparation, detector fabrication, and detector evaluation and troubleshooting.

# 2. Fabrication preparation

Prior to the actual fabrication of a detector, a significant amount of work should be completed to ensure smooth, cost-effective detector production. A partial list of prefabrication tasks includes process equipment design, procurement, construction, and testing; process validation; detector and detector assembly design; processing sequence design; processing tools design, fabrication, and testing; HPGe material procurement and evaluation; processing materials procurement and preparation; processing sequence testing and refinement; detector testing equipment design, procurement, construction and testing; and detector assembly components procurement, construction, and preparation. A few of these tasks are discussed in this section.

## 2.1 Detector and detector assembly design and component preparation

The detector, the mechanical structure holding the detector, and the electrical and mechanical components attached to this detector holder constitute the detector assembly. Detector assembly examples are shown in Figures 1.2 and 2.1. These are from the Compton Spectrometer and Imager (COSI) instrument [Kierans 2016] [Chiu 2017] and the Gamma-Ray Imager/Polarimeter for Solar flares (GRIPS) instrument [Shih 2012]. Both of these large area orthogonal strip detector assembly examples were designed by the Space Sciences Laboratory at the University of California, Berkeley in collaboration with LBNL for space science applications. In general, the designs of the detector and detector assembly are done with not only the requirements of the application in mind but also with consideration for the aspects that will facilitate detector fabrication and operation. The detector shape used for both of these examples is that shown in Figure 1.1. From an instrument performance perspective, minimizing the dead material near the detector and the detector-to-detector spacing are both desirable. Exclusively using these goals to guide the detector design would preclude the use of the detector handles and a guard ring. However, as discussed previously in the introduction section, these design attributes are included since they improve the detector production yield and potentially the detector performance.

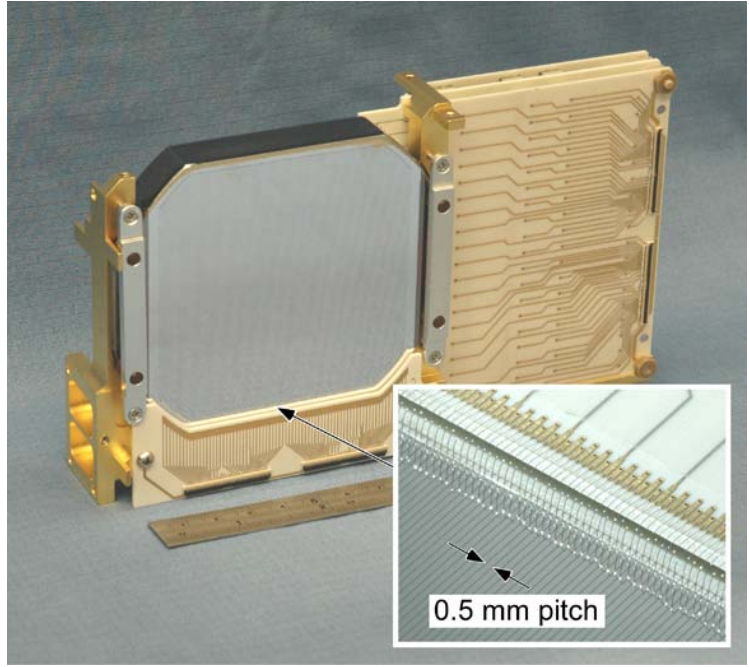

**Figure 2.1** Photograph of a HPGe orthogonal strip detector produced for the Gamma-Ray Imager/Polarimeter for Solar flares (GRIPS) instrument [Shih 2012]. The detector has an area of approximately 80 mm by 80 mm and is approximately 15 mm thick. Each face of the detector has 149 strips with a center to center strip spacing of 0.5 mm and a gap of 0.06 mm between each strip. Surrounding the set of strips on each detector face is a guard ring. The detector is held in an Al holder, and attached to this holder are circuit boards used to route the detector signals to flexcircuits (not shown). Connections between the detector strips and the circuit board traces were made with ultrasonic wedge wire bonding. The larger of the two circuit board assemblies is associated with the high voltage side of the detector. This circuit board assembly is actually a stack of three circuit boards, and this stack houses the high voltage coupling capacitor and bias resistor for each readout channel.

The COSI and GRIPS detector assembly examples of Figures 1.2 and 2.1 are similar in design. They consist of a detector of the Figure 1.1 shape fixed into an Al holder. Bars made of Al clamp the detector at its handles to the holder. Aluminum is chosen for these parts because of its low radiation absorption, good mechanical strength, low cost, and excellent machinability. The detector clamping structure is designed to accommodate differential thermal contraction yet provide lateral support of the detector. Electrical isolation between the detector and holder is implemented through BN sheets between the detector handles and the holder, and plastic (PTFE or Vespel polyimide) buttons between the clamping bars and the handles. The BN provides the desired electrical isolation as well as the necessary good thermal transport for detector cooling. Indium foil is also added at the interfaces between the BN, detector holder, and detector handles to improve the thermal connection. As well as fulfilling structural requirements, the detector holder is designed to provide protection of the detector's sensitive surfaces and allow for low risk, simple integration with the detector. Once the detector has been integrated into the holder, the detector is at less risk of being damaged since all handling procedures will then involve contact with the holder rather than the detector. Depending on the detector fabrication sequence, this can be done in the middle of the sequence or once the detector has been completed. A beneficial goal when devising a fabrication process is to design it so that many of the steps are compatible with having the HPGe crystal contained within a holder, thereby minimizing the risk of direct contact damage to the crystal.

After the detector has been integrated into the holder, the detector assembly is completed by attaching the circuit boards needed to route signals from the detector to flexcircuit connectors and then making wire bond connections between the detector electrodes and the circuit boards. Details on the wire bonding are given later in the detector fabrication section. The circuit boards of the example detector assemblies were constructed from hydrocarbon ceramic laminate materials. Specifically, Rogers RO4003 [Rogers 2020] was used because of its many desirable properties including low dielectric loss, high resistivity, low outgassing, low cost, and compatibility with standard epoxy/glass (FR4) manufacturing processes. Standard epoxy/glass boards are typically not appropriate due to the possible introduction of measurable dielectric noise. The circuit trace metallization on the boards must also be carefully chosen particularly in the areas that will receive wire bonds. In these areas, the base Cu layer must have a surface finish appropriate for ultrasonic wedge bonding with Al wire. A multilayer metallization finish capped with soft bondable Au is appropriate, and a specific example is the electroless Ni/electroless Pd/immersion Au (ENEPIG) surface finish [IPC-4556 2013]. The use of ENEPIG on our boards has allowed us to achieve a robust wire bonding process capable of consistently producing strong bonds.

As can be seen in Figures 1.2 and 2.1, the designs of the circuit boards differ between the two sides of the detector assembly. This is because of the high voltage bias applied to the detector. The electrical contacts on the face of the detector adjacent to the handles are held near ground potential during detector operation and can, therefore, be directly connected to the preamplifiers of the signal readout electronics as shown on the left side of Figure 2.2. The low voltage detector board implementing these connections is simple and contains the electrical traces directly routing signals from the detector to a flexcircuit connector. In contrast to this is the high voltage side of the detector assembly. The contacts on the detector face opposite the low voltage face are held at high voltage. The connection between these contacts and the readout electronics requires a coupling capacitor and a bias resistor for each readout channel as shown in Figure 2.2. These components are located near the detector on the high voltage detector board rather than exterior to the detector assembly in order to minimize microphonics, reduce the chance of high voltage breakdown, and eliminate the need for a large number of high voltage vacuum feedthroughs that would be necessary if the components were outside of the cryostat. These components must be carefully chosen. The coupling capacitors must be low noise, be rated for the high detector voltage (up to 2 kV for the example detectors), have large enough capacitance to effectively pass the signal pulses (both 3 and 8 nF were used), be compact in size (typically needed for applications requiring compact instruments or close-packed detectors), and have high stability with temperature. High reliability testing of the capacitors at the manufacturer (burn-in at elevated voltage and temperature) is also of great benefit since the failure of a capacitor in the completed assembly is costly. The high voltage detector boards for COSI and GRIPS used multilayer ceramic capacitors that are based on an ultrastable C0G/NP0 dielectric. These were procured from Novacap, which is part of Knowles Precision Devices [Knowles 2020]. The bias resistors used in the high voltage detector boards must be low noise, have a large enough value to both effectively prevent shorting the signal through the high voltage supply and achieve a sufficiently low associated Johnson noise, be compact in size, and have high stability with temperature. Thick film, high voltage chip resistors of 1 GΩ value from Vishay (model CRHV) [Vishay 2020] as well as other similar products were successfully used in our detector boards.

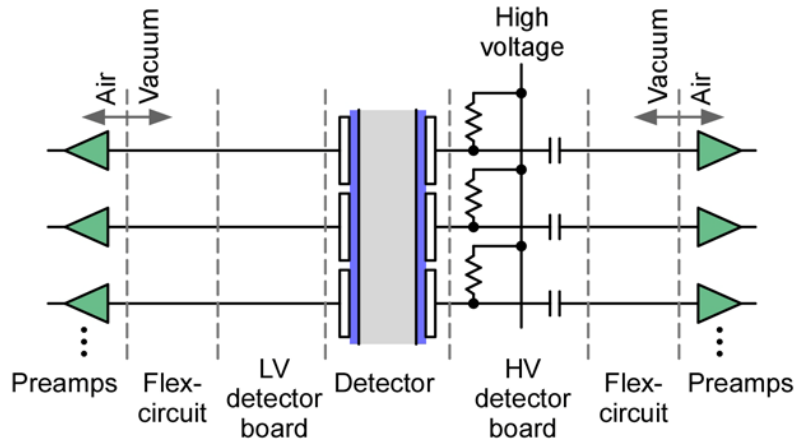

**Figure 2.2** Simplified schematic electrical diagram showing the interconnection of the detector, detector circuit boards, flexcircuits, and preamplifiers. LV stands for low voltage and HV for high voltage.

Another detector board design concern is the spacing and electrical insulation between components and circuit traces held at high voltage and the nearby grounded conductors. Sufficient high voltage standoff is required to prevent electrical breakdown. If such breakdown events occur during detector operation, they are difficult to distinguish from detector breakdown failure and therefore complicate instrument debugging. Furthermore, high voltage breakdown in the detector board or any other part of the high voltage system can cause damage to and failure of the strip detector. For conductors connected by surfaces such as that along the circuit board, breakdown is often along these surfaces and therefore varies greatly with the surface conditions. This necessitates choosing large, conservative spacings, using great care when preparing and handling the surfaces, or coating the surfaces and conductors with an appropriate insulator. These strategies have been applied to our boards. Large spacings were in part achieved in the COSI and GRIPS boards by placing the components and traces held at high voltage on a circuit board separate from those at low voltage/ground. The coupling capacitors then bridge and electrically connect the two boards. The GRIPS high voltage detector board stack shown in Figure 2.1 also made use of a coating of Parylene that was applied after the stack was fully assembled and cleaned.

A final important design consideration to be mentioned is the accommodation of thermal differential contraction between the detector boards and the Al detector holder. When the detector assembly is cooled, the dimensional contraction of the Al holder is greater than that of the detector boards constructed from the RO4003 material. If the boards were firmly attached and pinned to the holder at the board edges, the center of the board would bow up away from the holder and potentially would break the wire bond connections between the boards and the detector. To prevent this, our assemblies firmly attach each board near one edge of the board. At the opposing board edge, a slotted hole in the board allows the board to slide against the holder. Appropriate pressure and friction are maintained against the board through a stack of flat plastic, flat metal, Belleville metal, and shoulder washers held under compression with an attachment screw.

To end this section on the detector assembly design and component preparation, two step-by-step procedures are provided. These procedures were written specifically for the COSI and GRIPS detector assemblies. The first procedure concerns the preparation of the detector holder. The holder as well as all other assembly components should be cleaned and handled as appropriate for use in a high vacuum system. The second procedure covers the high voltage detector board preparation and testing. High voltage capacitor breakdown and excessive leakage current are possible failure modes even for a well-designed board and have been observed in some of our detector boards in the past. It is therefore critically important to thoroughly test the completed detector board prior to integration with the detector. Otherwise, as noted previously, detector testing becomes much more difficult since electrical failure could be the result of either a detector or a high voltage board problem, and it is difficult to distinguish between the two.

**Detector holder preparation procedure**

**(a) Clean parts:** Ultrasonically clean all parts in an appropriate detergent (for example, International Products' Micro-90 for stainless steel and plastic parts and International Products' Surface Cleanse 930 for Al parts). Thoroughly rinse the parts in $H_2O$ and then dry them.

**(b) Test fit parts:** Make sure all the parts fit together as expected.

**(c) Chemical polish Al parts:** Chemical polish etch all of the Al parts of the holder assembly. Though this step is not absolutely necessary, it ensures that the parts are thoroughly degreased and leaves a shiny and smooth Al surface. At LBNL, an acid bath consisting of a mixture of concentrated acids is maintained for this purpose. The mixture consists of the following volume percentages of deionized $H_2O$ and concentrated acids: 11% $H_2O$, 80% $H_3PO_4$, 6% glacial acetic acid $CH_3COOH$, and 3% $HNO_3$. To polish an Al part, both the part and acid bath are first heated to 100°C. The part is then rapidly swirled around in the acid bath for about 30 s. A blackish coating will typically appear on the part during this polish etch. After the appropriate time has elapsed, the part is rinsed and submerged in $H_2O$. While the part is still wet, the blackish surface coating is rubbed off using paper wipes, cotton swabs, and brushes. The part is then placed in an ultrasonic $H_2O$ bath so that any inaccessible surfaces can be cleared of the blackish coating through ultrasonic cleaning. After the ultrasonic cleaning, the part is wiped off and then dried in an oven.

**(d) Au plate (optional):** Have an outside plating shop Au plate the detector holder. This will ensure a good electrical grounding connection to the cryostat. The GRIPS program used the following specification: Bright Au, MIL-G-45204D, Type 11, Class 1, 0.00005" thick.

**High voltage detector board preparation and testing procedure**

**(a) Clean boards:** If the detector board contains only soldered components, wash/soak/brush it in a commercial flux removal solvent followed by an alcohol rinse. Blow it dry with $N_2$ or air. If the high voltage capacitors are attached to the board with silver epoxy, clean the board with alcohol and then blow it dry. If the board has been Parylene coated (as was done for the GRIPS boards), clean only the wire bonding pad areas as needed to ensure good bonds.

**(b) Verify resistor values:** Using a Fluke 87 digital multimeter or equivalent set to measure nS, probe each resistor, and confirm a conductance of 1 nS (+- 10%).

**(c) Verify capacitor values:** Attach the board to a detector holder, load it into the appropriate test cryostat, and connect the flexcircuit assembly to the board. An example setup for the testing of a COSI detector board is shown in Figure 2.3(a). Interconnect all pins at the output of the flexcircuit so that one side of each coupling capacitor is interconnected through the flexcircuit. Using a Fluke 87 digital multimeter or equivalent set to measure capacitance, probe the capacitance between the interconnected flexcircuit end and the opposing end of each capacitor on the board. This measurement is schematically shown in Figure 2.4(a). The capacitance should be 8 nF (+- 20%) for the COSI boards and 3 nF (+- 20%) for the GRIPS boards. This also confirms a good connection between the flexcircuit and each of the board traces. The GRIPS board capacitors are tested without the use of a cryostat. For a GRIPS board, the channels are interconnected using a combination of flexcircuits and interconnected low voltage board (see Figure 2.3(b)).

**(d) Check capacitor high voltage leakage:** With the board connected as described in (c), connect high voltage to the board and a picoammeter to the interconnected flexcircuit end. Such a measurement setup is schematically shown in Figure 2.4(b). Slowly increase the high voltage to 1000 V and then to 2000 V. Eventually, the leakage through the capacitors should drop below about 1 nA at 1000 V and 2 nA at 2000 V. Due to the slow relaxation processes in the capacitors, it can require about 30 min for the current to reach the appropriate level at 2000 V. This test should be made under vacuum with the board cooled. If the low level of leakage is obtained, the board is good, and the test is finished. If the leakage current does not reach the desired low level, the bad capacitors can be identified through high voltage probing as described in (e).

**(e) Identify bad capacitors (non-Parylene coated COSI boards only):** Remove the board from the cryostat and the detector holder. Use the circuit shown in Figure 2.4(c) to measure the leakage current through each capacitor at a voltage of about 100 V. Any bad capacitor will likely have a non-negligible leakage. Note that a visual inspection or probing with a digital multimeter (resistance or capacitance measurement) is not always sufficient to identify problem capacitors. Replace all capacitors identified as bad, and then repeat the cleaning and testing procedures.

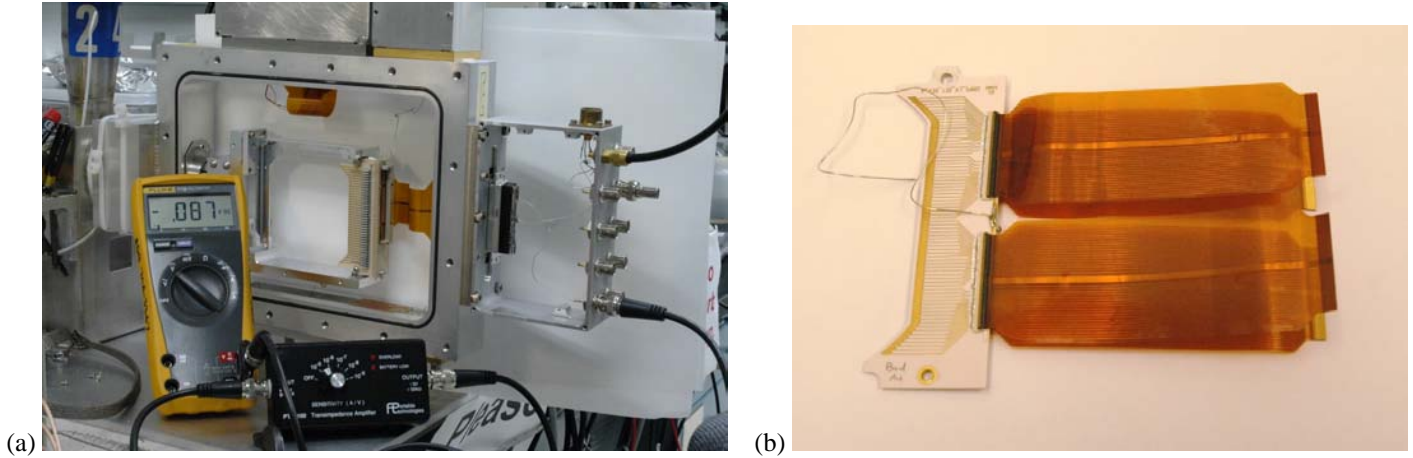

**Figure 2.3 (a)** Measurement setup for the COSI high voltage detector board testing. **(b)** Flexcircuits and interconnected low voltage board used for the GRIPS high voltage detector board testing.

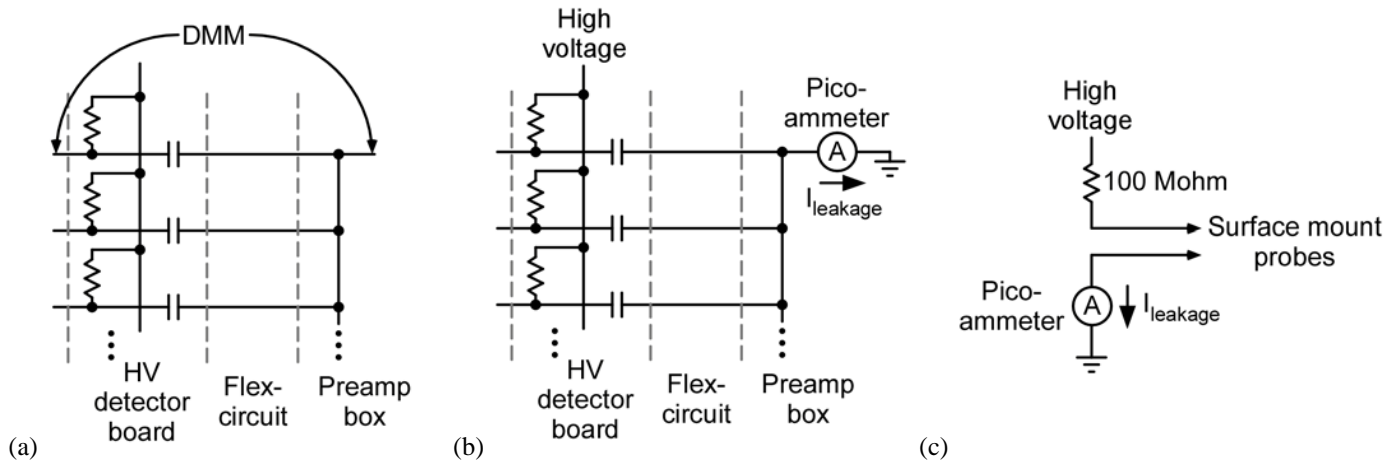

**Figure 2.4** Schematic electrical diagrams of the high voltage detector board testing circuits. **(a)** Capacitor value check using a capacitance meter or digital multimeter (DMM) set to capacitance measurement. **(b)** Through capacitor leakage current check. **(c)** High voltage leakage current probe circuit used to identify individual bad capacitors on a board that has failed the leakage current check of (b).

## 2.2 Detector fabrication sequence and tool design

The design of the detector assembly, detector fabrication sequence, and detector processing tools should all be done concurrently. The detector fabrication sequence consists of the mechanical processing, chemical processing, thin film deposition, thin film lithography, wire bonding, handling, assembly, and evaluation steps required to produce a completed detector assembly. A carefully designed sequence should implement the practices expected to maximize the detector yield and optimize the detector performance. Minimizing the risk of damage to the HPGe crystal and surface coating layers during fabrication is critical. Custom designed tools are integral to this strategy of minimizing risk. The photographs contained in the detector fabrication section exhibit many of these tools that are needed to safely manipulate and support the detector including the etch clamps, sputter fixtures, transfer fixture, flipping fixture, and photoresist spinner detector holder. Once the design of the fabrication sequence has been completed and the tools fabricated, the entire fabrication sequence should be evaluated through dummy fabrication runs made with an inexpensive piece of material of the same starting shape as the HPGe crystal. Such runs can highlight potential weaknesses in the detector assembly, detector fabrication sequence, and detector processing tools designs. These runs should then be the basis for iterative improvements in the designs.

## 2.3 HPGe material procurement

Obtaining the appropriate starting HPGe material is likely the most important step when manufacturing detectors for applications requiring high efficiency and excellent spectral performance. It is a difficult endeavor to overcome inadequacies in the detector material through correction methods involving modifications of the detector design, fabrication, or signal processing. Nearly all of the HPGe crystals for the large volume orthogonal strip detectors produced at LBNL were purchased from Ortec [Ortec 2020]. The set of specifications used for the procurements is given in Figure 2.5. The dimensional specifications are straightforward and have values and tolerances dictated by the geometric needs of the intended instrument. In contrast, the material specifications are less precisely stated and typically rely on the knowledge of a manufacturer that is highly experienced in the growth of HPGe crystals for high resolution spectroscopy detectors. This is reflected by the ill-defined *Quality* specification of *Detector grade*. Implicit in this *Detector grade* designation is the requirement that the HPGe has extremely good electron and hole charge transport. High mobility and minimal charge trapping of both carrier types are needed. Related to this is the need for crystalline perfection and the associated etch pit density (EPD) specification. The EPD provides a measure of the crystalline dislocation density. A high density of dislocations will lead to an excessive amount of charge trapping and hence there is an upper limit placed on the EPD. Though not explicitly called out in the specifications of Figure 2.5, a lower limit on the EPD is also necessary. The HPGe crystals used for radiation detectors are grown by the Czochralski method with a hydrogen atmosphere. Crystals grown without dislocations using this method contain a significant hole trap that is associated with divacancy hydrogen complexes. Due to this charge trapping, the material is not suitable for typical radiation detector applications. For this reason, the HPGe material needs to have a small and uniform density of dislocations of about $10^2$-$10^3$ cm$^{-2}$ in order for sufficient dislocations to be available for the absorption of the vacancies created during the high temperature crystal growth [Haller 1981] [Hansen 1983] [Haller 2006].

Another critical specification is the net ionized impurity concentration. The voltage at which a planar detector becomes fully depleted of free charge carriers $V_{fd}$ is directly proportional to this concentration and is given by $V_{fd} = qNt^2/2\varepsilon$, where $q$ is the magnitude of the electron charge, $N$ is the net ionized impurity concentration, $t$ is the detector thickness, and $\varepsilon$ is the dielectric constant of Ge [Bertolini 1968] [Knoll 1989]. Since there is inevitably a maximum operational voltage constraint on the detector, this relationship between the full depletion voltage and the impurity concentration in part dictates an upper limit restriction on the impurity concentration. For the entire detector volume to be active, the detector must be operated fully depleted. Furthermore, to minimize any charge trapping effects and the charge collection times, the detector should be sufficiently over depleted to achieve saturation velocity movement of the charge carriers [Ottaviani 1975] [Jacoboni 1981]. For the large volume orthogonal strip detectors produced at LBNL, a maximum detector voltage of 2000 V was chosen based on the voltage limit of the dc coupling capacitors associated with the high voltage side of the detector and the desire to avoid detector yield losses due to breakdown failures at higher operational voltages. Assuming an over depletion voltage of no more than 1000 V, the maximum full depletion voltage would be 1000 V. With a detector thickness of 15 mm, this leads to an upper limit on the impurity concentration of about 8 x 10$^9$ cm$^{-3}$. However, experience has shown that the methods employed by the HPGe suppliers to estimate a crystal's impurity concentration can be inaccurate and lead to the delivery of material that is sometimes too impure. Consequently, an even tighter impurity concentration specification with an upper limit of 5 x 10$^9$ cm$^{-3}$ was chosen.

With respect to the HPGe impurities, the net impurity type is also of concern. Both p- and n-type crystals have been successfully used at LBNL to make the large volume orthogonal strip detectors. However, limited data appear to show that p-type detectors produced using the processes described in this paper are less likely to have deleterious side surface channel issues as compared to n-type detectors. Furthermore, based on the typical net impurity concentration profile along the length of a HPGe boule [Haller 1981] [Hansen 1983] [Haller 2006], it is expected that more p-type than n-type material of a suitable net impurity concentration should be available from each boule. For these reasons, the specifications state a preference for p-type material.

Once a HPGe crystal has been procured, quality screening checks are made of it prior to its conversion to a strip detector. The primary purpose of these checks is to identify any fatal flaw in the material since early identification can save significant effort. Two useful checks are (1) a post polish etch surface inspection to identify crystallographic defects and (2) a simple full area electrode planar or guard ring detector electrical characterization to determine the impurity concentration and ability to operate at the desired high voltage. Both of these screening tests are described later in the detector fabrication and detector evaluation sections.

| Specification | Value |
|---|---|
| Dimensional | |
| Shape | Right cylinder |

| Diameter | 100 mm -0 mm / + 1 mm |
|---|---|
| Thickness | 15 mm - 0 mm / + 1 mm |
| | |
| Material | |
| Quality | Detector grade |
| Net impurity concentration | $< 5 \times 10^9$ cm$^{-3}$ |
| Net impurity type | P |
| Etch pit density | $< 8000$ cm$^{-2}$ |
| Additional notes | No gross lineages |

**Figure 2.5** Specifications used in the procurement of the HPGe crystals for the large volume orthogonal strip detectors of the type shown in Figures 1.1, 1.2, and 2.1.

# 3. Detector fabrication

In this paper, detector fabrication refers to the process of converting a commercially procured HPGe crystal into a radiation detector integrated into a holder and wire bonded to signal routing circuit boards attached to the holder. The steps of this process to be covered in this section of the paper are crystal surface inspection, crystal mechanical processing (cutting and lapping), crystal chemical processing (polish etching and surface preparation etching), amorphous semiconductor deposition, metal electrode deposition and lithography, wire bonding, and completed detector inspection and repair. Additionally, the reprocessing of a defective detector is discussed. The starting point assumed is that the equipment and processes have been established, the processing sequence and tools have been validated, and the HPGe crystal is in a state that is typical of the crystal condition when received from a crystal supplier. Such a crystal has been cut into a right cylinder of the dimensions as specified in Figure 2.5, and the flat faces of the crystal have been lapped and possibly polished.

A simplified representation of the detector fabrication sequence used to produce the LBNL strip detectors is shown schematically in Figure 3.1. A design feature of note for this sequence is that all of the HPGe crystal surfaces are chemically prepared (etched) in a single step prior to the complete coating of all surfaces with the amorphous semiconductor layer followed by the deposition of the metal electrodes. An alternative approach would be to separately fabricate each face of the detector. The sequence would consist of the complete fabrication of one face of the crystal including the metal electrodes, masking this face with an acid and solvent resistant coating, and then the complete fabrication of the other crystal face. In comparison to this sequence, the sequence of Figure 3.1 has the benefits of a reduced number of fabrication steps and the elimination of masking and its associated problem of possibly introducing surface contaminants or defects in the deposited films. Another characteristic of the detector processing at LBNL is the practice of minimizing the handling of the crystal surfaces that will eventually be associated with the active volume of the completed detector. Specifically, after the chemical etching of the crystal is completed, all mechanical contact to the crystal is made only with the crystal handles (at least until the crystal is completely coated with amorphous semiconductor). This reduces the risk of introducing surface damage that could detrimentally change the performance of the detector. To implement this practice, custom designed tools and fixtures are required for manipulating and holding on to the crystal during each processing step. In the following subsections, each of the individual steps of the detector fabrication sequence is covered in detail including descriptions or photographs of the crystal handling and processing tools and fixtures.

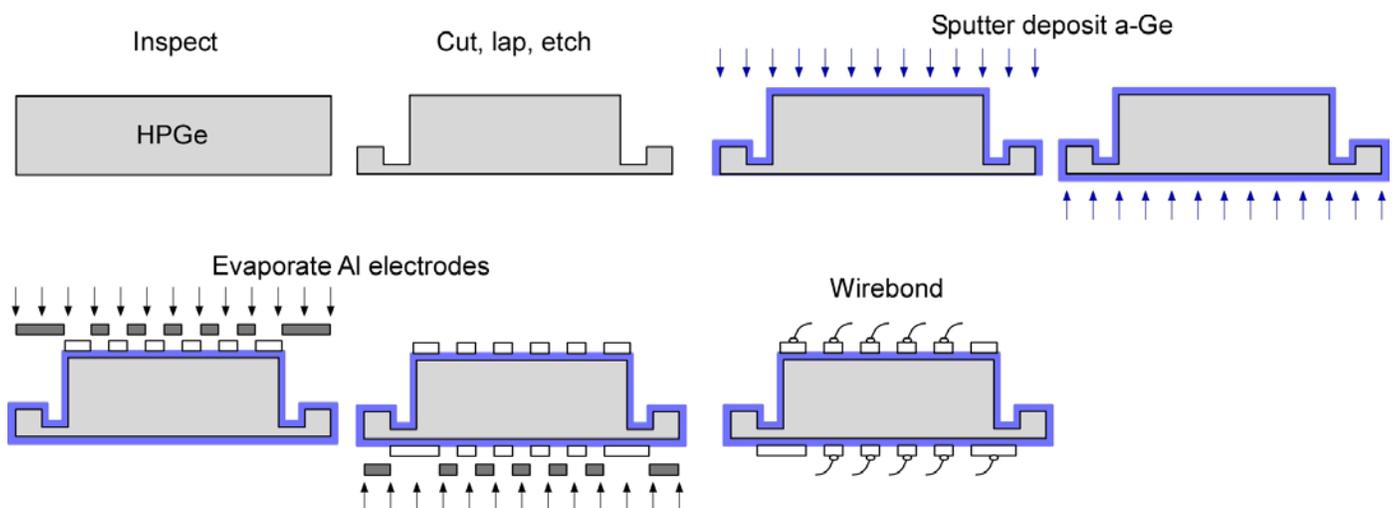

**Figure 3.1** Schematic cross-sectional drawings depicting the strip detector fabrication sequence. This sequence assumes that the metal electrodes are patterned using a thermal evaporation shadow mask technique. The metal electrodes could as well be patterned using photolithography.

## 3.1 Crystal surface inspection

The purpose of this crystal evaluation step is to identify any obvious crystallographic defects appearing on the crystal surfaces prior to significant processing of the crystal. Example surface defects are shown in the photographs of Figure 3.2. These photographs are of a completed strip detector. The strips are oriented horizontally, and the darker horizontal bar in each photograph is the gap between two Al strips. The other lines and surface features appearing in the photographs are the ones of concern and could not be completely eliminated through additional polishing of the surfaces. The crystal shown in the figure was processed into a strip detector and tested three times. Overall, the detector had greater than normal leakage current and electron trapping. The electron trapping necessitated higher operating bias voltages, but, unfortunately, the detector broke down at a detector bias of about 1600 V on two of the three processing runs.

The crystals as received from the manufacturer typically have surfaces that have been only lapped or coarsely polished. Such surface processing produces a rough surface that will likely hide the crystallographic defects of concern. Furthermore, the lapping will likely have left surface scratches that may appear to be critical defects but instead can be readily removed through further surface processing. Consequently, the crystal surface should be carefully inspected after the surfaces have been processed as described below. This detailed surface inspection before the detector fabrication has been found to be of value when previous experience has indicated that the risk of receiving defective material is non-negligible. Additionally, prior to this post-etch surface evaluation, the crystal should be inspected for any obvious damage such as chips or cracks and dimensionally checked to confirm usability for the intended application.

**(a) Lap crystal faces:** If the two flat faces of the crystal are not free from saw marks and significant scratches, lap them as described in section *3.3 Crystal lapping*.
**(b) Polish etch:** Polish etch the crystal as described in section *3.4 Crystal chemical polish etching*. Two etches roughly 3 min each with a change of etchant between the etches will likely be required before smooth surfaces are obtained.
**(c) Inspect:** After the polish etch, all crystal surfaces should be smooth with a possible orange peel texture. Carefully inspect the two faces of the crystal by eye and then with a stereomicroscope. A small number of lines may be visible on the surfaces, but high densities of lines or gross topographic features such as those shown in Figure 3.2 should not be present. If any defects are visible, additional lapping and etching can be done to confirm that they are not simply surface damage. If defects are present and it is decided to continue with the detector processing anyway, it is advisable to fabricate the crystal into a simple planar detector and perform an electrical test.

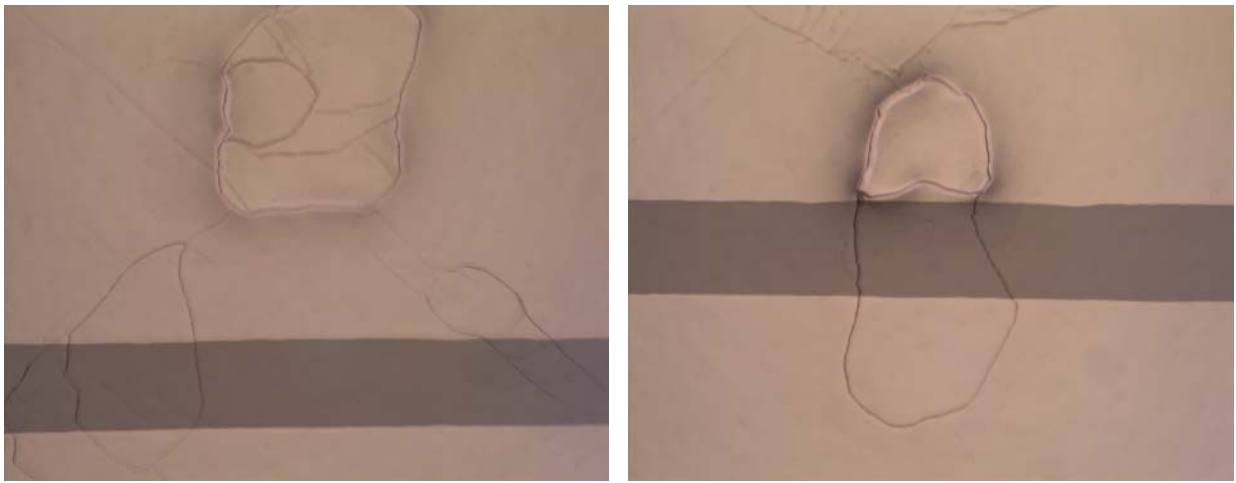

**Figure 3.2** Photographs of surface defects present on a HPGe crystal of the type shown in Figure 1.1. The photographs were taken after the crystal was processed into an orthogonal strip detector. The dark horizontal bar in each photograph has a width of approximately 250 μm and is the gap between two Al strips on the detector.

## 3.2 Crystal cutting

The HPGe crystal as received from the manufacturer is typically a cylindrical slice from a crystal boule. The purpose of the crystal cutting step is to convert this cylinder into the geometry shown in Figures 1.1, 3.3, and 3.4. The primary tool used to accomplish this is an outer diameter (OD) saw equipped with a diamond abrasive impregnated metal blade and a magnetic chuck attached to an x-y positioning stage. The particular saw used at LBNL is a Micromech Micro-Matic Precision Wafering Machine and is many decades old. However, more modern saws are available that could just as easily perform the desired operations. The main processing steps consist of a set of linear cuts. Good crystal mounting, accurate positioning, and slow and constant speed movement of the crystal through the saw blade are needed to achieve smooth, chip-free cuts. A slow feed is particularly important for cutting the grooves that separate the handles from the active volume of the detector. It is also worth noting that a rounded recessed blade is used for these groove cuts to prevent cracking at the bottom of the cuts. A step-by-step procedure for the crystal cutting is provided below. A sequence of photographs of the various steps in the procedure is given in Figure 3.5.

**(a) Measure dimensions of crystal**: With a pair of plastic calipers, measure the diameter and thickness of the crystal. Record these numbers on a cutting diagram log such as those shown in Figures 3.3 and 3.4. The cuts to be made on the crystal will be positioned relative to the crystal edges and will depend on the measured crystal dimensions.

**(b) Mount crystal:** Mount the crystal onto a graphite plate that has been mounted onto a magnetic stainless steel plate (type 410). The magnetic stainless steel plate is required for use with the magnetic chuck of the OD saw, and the graphite plate is necessary for cuts that pass through the HPGe crystal (since the diamond saw blade should not be used to cut into metal). Use a hot plate and red wax (Sticky Wax 1264-671763 by Caulk Company or similar wax product) to attach the crystal to the graphite and the graphite to the steel plate. Completely coat the surfaces with the wax to ensure a strong attachment. Once the stack is assembled and still hot, apply uniform pressure to the top of the crystal to make sure that all parts sit flat against each other. Also, coat the top surface of the crystal with wax. This ensures that the crystal has heated through completely and provides a protective top coat on the crystal. When done, remove the stack from the hot plate and let it cool to room temperature.

**(c) Lay out cuts:** Lay out cuts on top of the mounted crystal using a permanent marker. These lines provide an important visual check indicating whether or not the saw blade has been properly positioned prior to each cut.

**(d) Cut crystal:** Using the OD saw, perform the cuts as indicated in the crystal cutting notes of either Figure 3.3 or 3.4. For each cut, locate the blade carefully, and confirm the blade location relative to the marked lines placed on the crystal in the previous step. Confirm the blade height with a ruler when making the groove cut. During the cutting process, cutting fluid is continually sprayed onto the blade and crystal. This fluid is Silcool-5 diluted in $H_2O$ at a ratio of 15 mL Silcool to 1 gallon of $H_2O$. This fluid is further diluted with $H_2O$ inside the pump unit that supplies the cutting fluid to the saw.

**(e) Demount crystal:** Once all of the cuts have been completed, use a hot plate to heat the crystal-graphite-steel stack and carefully remove the crystal. Wipe the wax off of the crystal with towels while the crystal is still hot. After the crystal has cooled, remove the remaining wax with trichloroethylene.

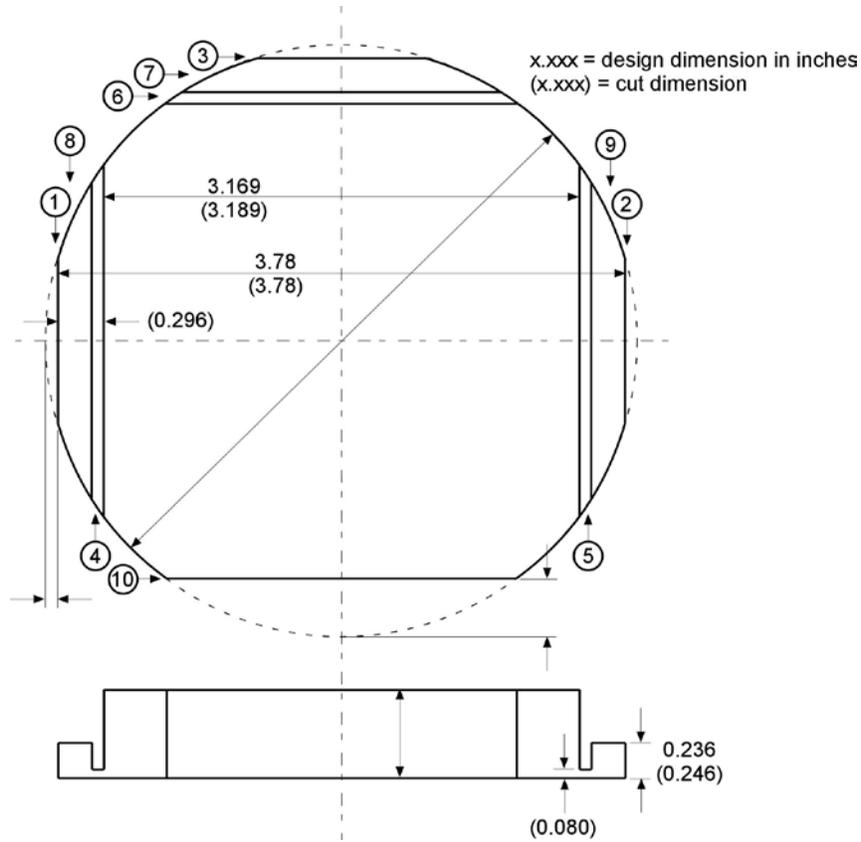

| Cuts | LBNL blade number | Blade part number | Blade diameter (inches) | Blade kerf | Feed setting | Completion date |
|------|------|------|------|------|------|------|
| 1-3 | 1 | M4D180-N100M-1/8 | 5 | 20 mil | 1 | |
| 4-6 | 2 | M4D180-N100M-1/8, rounded, recessed | 5 | 1.5 mm | 0.5 | |
| 7-9 | 3 | M4D400-N50M-1/8 | 4 | 3.8 mm | 0.5, 2 passes | |
| 10 | 1 | M4D180-N100M-1/8 | 5 | 20 mil | 1 | |

**Figure 3.3** HPGe crystal cutting notes specifying the crystal dimensions, saw blades, cut sequence, and cut feeds. Dimensions such as the crystal diameter and thickness vary from crystal to crystal and are added to the notes specific to each crystal. The notes shown here are for the COSI detector design that had three handles rather than the two shown in Figure 1.1. The saw blades listed in the table are all metal bond diamond grinding wheels manufactured by Norton Winter [Norton 2020].

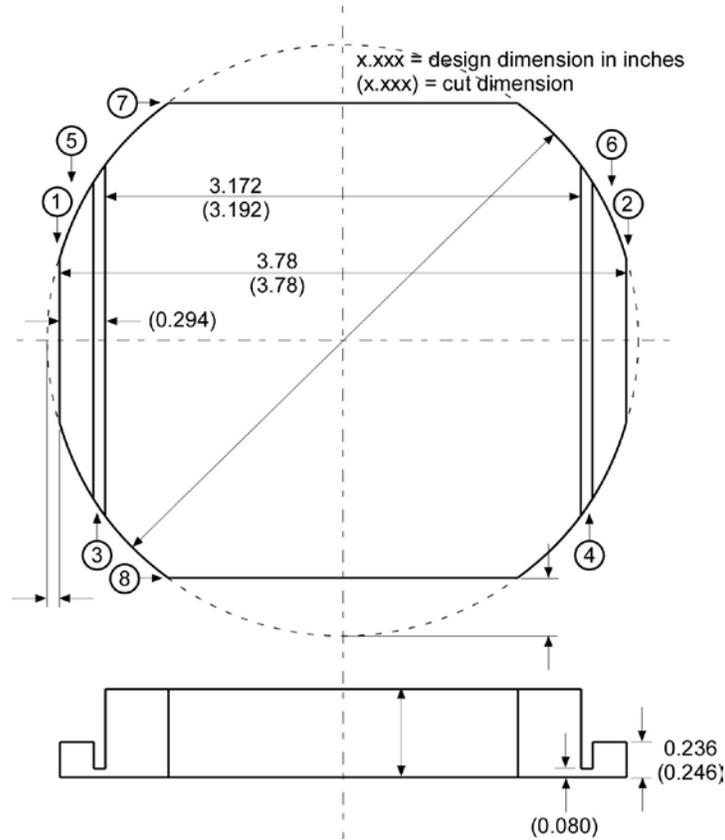

| Cuts | LBNL blade number | Blade part number | Blade diameter (inches) | Blade kerf | Feed setting | Completion date |
|------|------|------|------|------|------|------|
| 1-2 | 1 | M4D180-N100M-1/8 | 5 | 20 mil | 1 | |
| 3-4 | 2 | M4D180-N100M-1/8, rounded, recessed | 5 | 1.5 mm | 0.5 | |
| 5-6 | 3 | M4D400-N50M-1/8 | 4 | 3.8 mm | 0.5, 2 passes | |
| 7-8 | 1 | M4D180-N100M-1/8 | 5 | 20 mil | 1 | |

**Figure 3.4** HPGe crystal cutting notes specifying the crystal dimensions, saw blades, cut sequence, and cut feeds. Dimensions such as the crystal diameter and thickness vary from crystal to crystal and are added to the notes specific to each crystal. The notes shown here are for the GRIPS detector design. The saw blades listed in the table are all metal bond diamond grinding wheels manufactured by Norton Winter [Norton 2020].

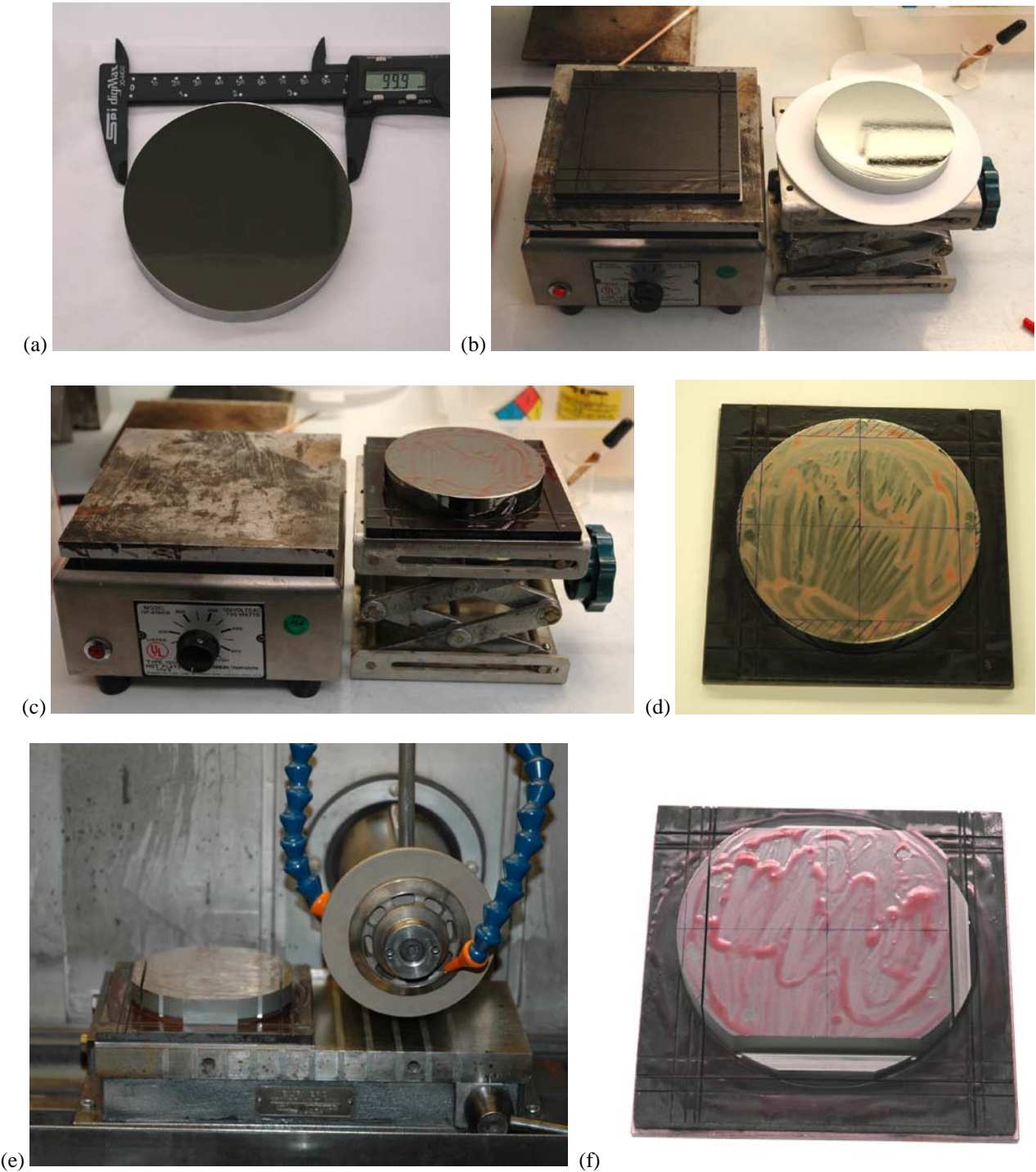

**Figure 3.5** Sequence of photographs showing the crystal cutting process. **(a)** Measuring the crystal dimensions. **(b)** Prior to the attachment of the crystal to the graphite plate and steel plate stack. **(c)** After the crystal, graphite plate, and steel plate have been bonded together with wax. **(d)** Completed stack with the intended cut locations drawn onto the crystal. **(e)** Crystal cutting. **(f)** Crystal, graphite plate, and steel plate stack just after all cuts were completed.

## 3.3 Crystal lapping

Lapping of the HPGe crystal is a process in which Ge material is removed by rubbing the crystal against a flat glass plate with an abrasive slurry placed between the crystal and plate. The purpose of this process is to remove surface features that may have resulted from the mechanical processing done by the crystal supplier or by the shaping of the crystal after it was received. Once the lapping is completed, the planar surfaces of the crystal should have a uniformly matte surface texture, and the crystal edges should be free from cracks and gross chips. The typical abrasive slurry used at LBNL is a mixture of $H_2O$ and fine particles of either SiC or $Al_2O_3$. It has also been found that relatively new glass plates produce a more perfectly flat detector face and can shorten the length of time required to complete the lapping of a large detector face since well-used plates will have developed a concave shape. A detailed procedure for the lapping process is provided below, and photographs of a crystal being lapped are shown in Figure 3.6.

**(a) Perform coarse lapping:** If all planar surfaces of the crystal are relatively smooth, a coarse lapping is not necessary, and only the fine lapping needs to be done. Furthermore, if a preprocessing polish etch was done before the crystal cutting, no lapping (coarse or fine) of the detector faces may be necessary. Prior to lapping, make sure that the glass lapping plate to be used is clean and free of debris. To coarse lap the crystal, squirt $H_2O$ onto the glass plate and place a spoonful of 600 grit lapping powder (Micro Abrasives 15 μm SiC [Micro Abrasives 2020]) onto the wet plate. Mix the two to form a slurry. Carefully place the crystal surface to be lapped onto the plate. With little or no pressure on the crystal and using a figure eight or circular motion, lap the crystal surface until the entire surface has a uniform texture. Repeat this process for all planar surfaces needing the coarse lapping. Thoroughly rinse the crystal in $H_2O$ when done.

**(b) Perform fine lapping:** Repeat the above process using 1900 grit lapping powder (Lapmaster 9.5 μm $Al_2O_3$ [Lapmaster 2020]). No pressure and a small circular motion should be used near the end of the process until a nearly scratch-free surface is obtained. Thoroughly rinse the crystal in $H_2O$ when done.

**(c) Lap away cracks and chips:** Depending on the quality of the mechanical processing at the manufacturer and the cutting performed earlier to shape the crystal, this step may be unnecessary if no defects are identified. Inspect all crystal edges for cracks or large chips (of a size that cannot be chemically etched to a smooth state). All such features should be lapped away. To do this, prepare a glass plate and abrasive slurry as described above. Carefully place the crystal edge containing the defect onto the slurry coated glass plate and rub the crystal edge on the plate until the defect has been abraded away. Repeat this for all identified defects. Note that this step can be performed before the lapping of the planar surfaces and should be done at any point during the detector processing when a crack or large chip is identified. It also may be beneficial to slightly lap all exposed crystal edges (knock off sharp edges) with this technique prior to lapping the planar surfaces so that edge chipping during the lapping of these surfaces is reduced.

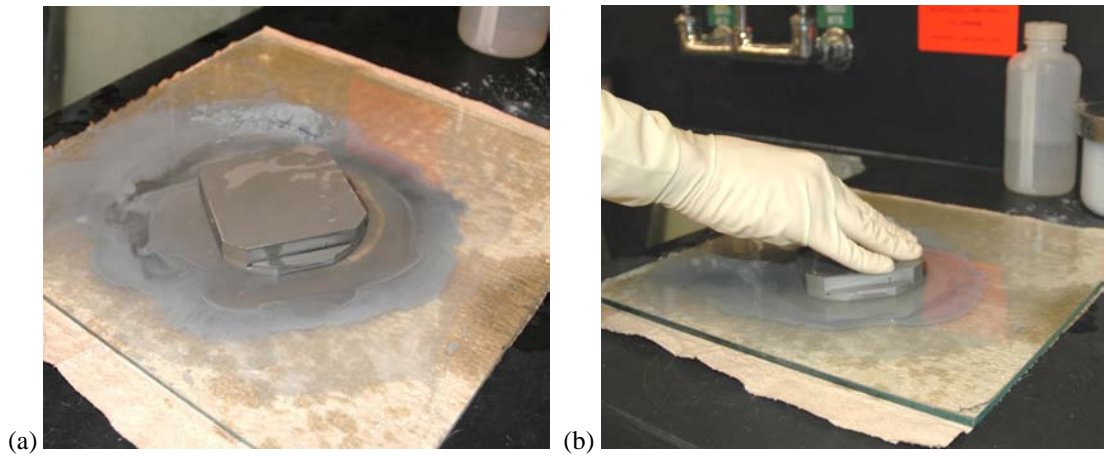

(a)                                         (b)

**Figure 3.6** Photographs of a GRIPS geometry HPGe crystal taken during the lapping of the bottom crystal face.

## 3.4 Crystal chemical polish etching

The purpose of the chemical polish etch is to remove the physical damage done to the HPGe crystal surfaces during the previous cutting and lapping operations. Such damage, if left in place, would negatively impact the performance of the completed detector. This could include electrical contact breakdown or substantial surface leakage current to the extent that the detector cannot be properly operated. The etching of Ge is a well-studied topic, and a wide variety of etchants have been used in the past. For example, see the book **Handbook of Metal Etchants** [Walker 1991]. In the LBNL process detailed below, an acid mixture of $HNO_3$ and HF is used to remove the outer layer of imperfect Ge on the crystal. In addition to the step-by-step procedure below, photographs of the etching sequence are given in Figure 3.7. It is critical to note that the etching is done within a fume hood, and the worker makes use of personal protective equipment. The mixture of concentrated acids used in this process is highly hazardous and could result in great bodily harm including loss of life if not handled properly.

**M. Amman, 2020, "High Purity Germanium Based Radiation Detectors with Segmented Amorphous Semiconductor Electrical Contacts: Fabrication Procedures"**

**(a) Clean crystal:** Remove any wax remaining on the HPGe crystal with trichloroethylene. Remove the lapping residue on the crystal with a deionized $H_2O$ rinse combined with wiping using a wet paper towel. Rinse the crystal with methanol or deionized $H_2O$ and then blow it dry.

**(b) Etch crystal:** Prepare two acid resistant bowls in the sink of the wet bench – one containing a 4:1 volume ratio mixture of $HNO_3$ (69.5% concentration) to HF (49% concentration) and a second one with flowing deionized $H_2O$ (see Figure 3.7). Place a round sheet of PTFE in the bottom of the acid bowl to prevent the crystal from being scratched by any rough surfaces in the bowl. With a multi-gloved hand, place the crystal into the acid mixture. Continuously move the crystal around in the acid, regularly change grip on the crystal (primarily holding it by the crystal handles), and turn the crystal over once or twice during the etching process. Continue this etching with agitation for about 3 min. At this point, the etchant will likely be fuming and can be quite warm. Note that this step describes hand holding the crystal during the etching. For such a process, great care must be given to the selection and preparation of the etch gloves. Multiple layers of acid gloves should be used, and each glove should be leak checked prior to its use. In addition to the acid gloves, an outer glove made of polyethylene is used to reduce the contamination introduced by the submersion of the gloves into the acid and the contact of the outer glove to the crystal. As an alternative to the handheld etching, a crystal holder could be designed and used for manipulating the crystal. Since this etch is long, though, and significant Ge is removed from the crystal, the holder should be carefully designed so that its use during the etching will not leave an undesirable pattern or surface texture on the crystal. The PTFE etch clamps that are used for the shorter etch described in the next section could also be used here, however, when used for this long etch, significant non-flat regions on the crystal handles will be created where the clamp contacts the crystal. These regions can then prevent the masks (shadow or photolithography) from sitting flat on the crystal during the Al electrode patterning and prevent the completed detector from sitting flat in the detector holder.

**(c) Rinse and dry crystal:** After the desired etch time has elapsed, rapidly move the crystal from the acid bowl to the deionized $H_2O$ rinsing bowl. It is best to hold on to the crystal by its handles at this point so that glove marks will not be left on the critical crystal surfaces. Slosh the crystal around in the deionized $H_2O$ bowl and then rinse the crystal and your gloved hand directly in the deionized $H_2O$ stream. If acid is trapped inside the outer polyethylene glove, remove the glove and re-rinse. Place the rinsed crystal on a few sheets of filter paper and then blow the crystal dry with $N_2$.

**(d) Inspect crystal and repeat etching as necessary:** The etched crystal should have surfaces that are smooth with a possible orange peel texture. Pits and scratches from the lapping step should have been etched smooth. If this is not the case, the crystal should be etched further. The etchant can be reused for this additional etching as long as it has not become too warm. It seems, however, that a better surface texture is typically achieved if new etchant is used. The warm fuming etchant sometimes leads to cloudy surfaces. If there are chips or cracks at the crystal edges that have not been etched smooth, lap these edges away and re-etch the crystal.

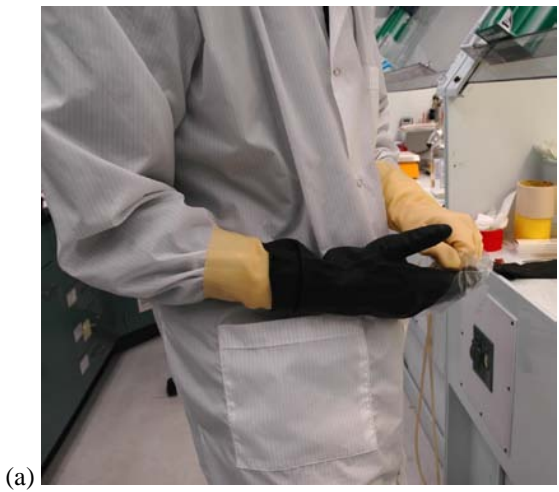

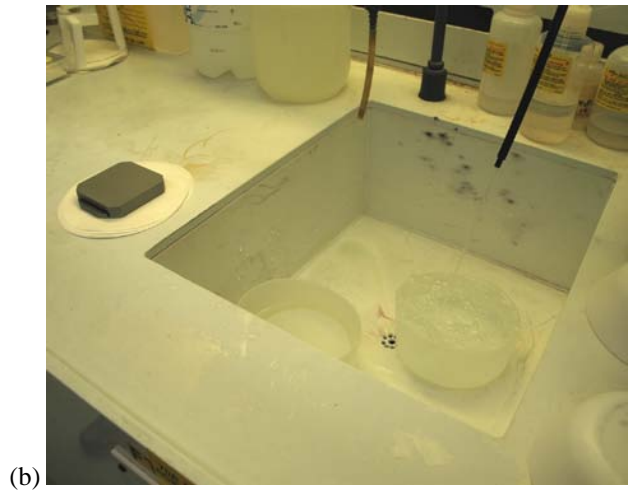

(a)                    (b)

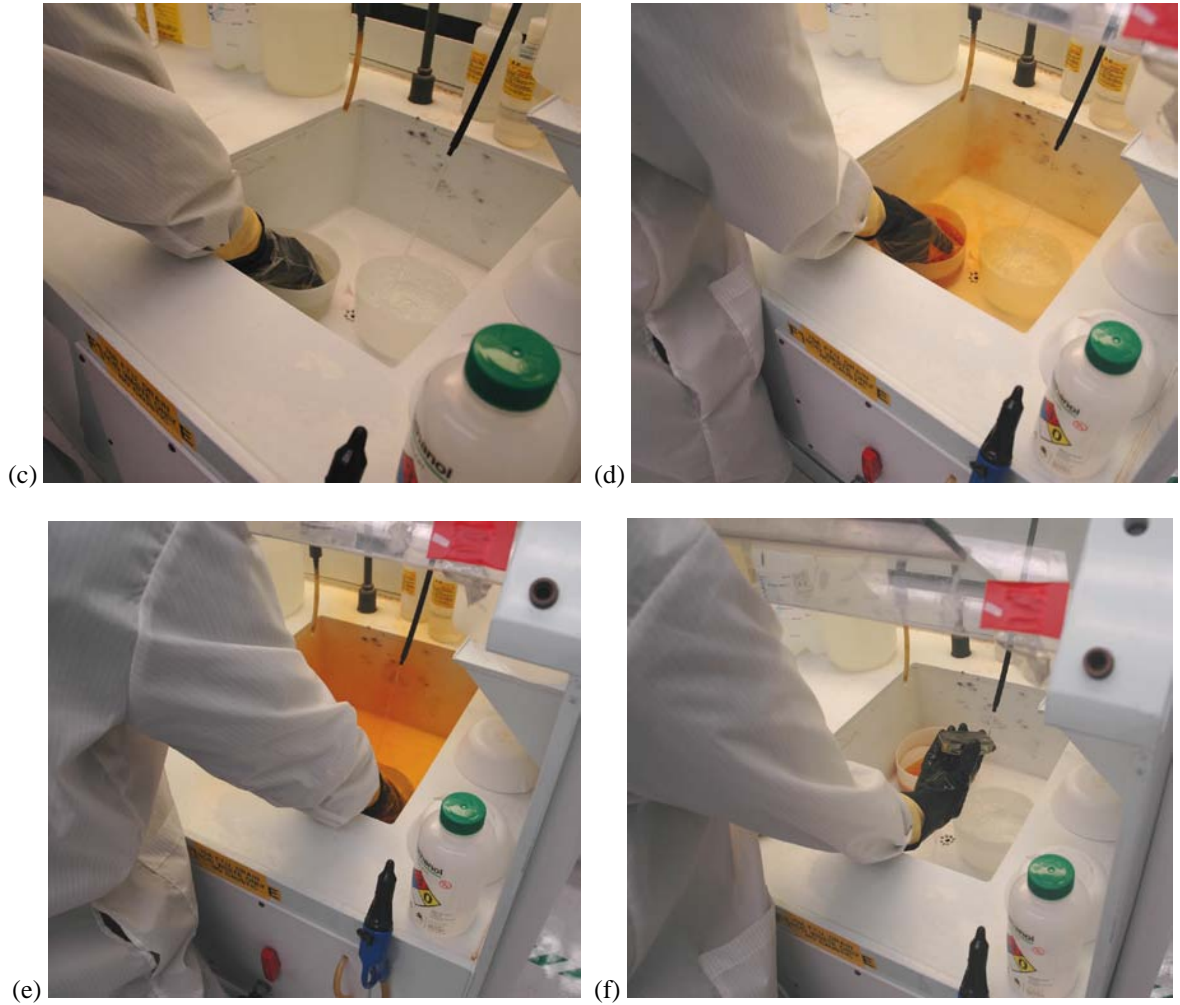

**Figure 3.7** Sequence of photographs showing the HPGe crystal chemical polish etching process. The etching is done within a fume hood, and the worker makes use of personal protective clothing and equipment appropriate for handling the concentrated acid mixture. The sequence shown here is one in which the crystal is handheld during the etching. For such a process, great care must be given to the selection and preparation of the etch gloves. Multiple layers of acid gloves should be used, and each glove should be leak checked prior to its use. A safer and, hence, more desirable process would utilize a crystal holder. The steps shown are: **(a)** glove donning, **(b)** etch and rinse baths preparation, **(c)** etching near the start, **(d)** etching near the end, **(e)** etch quenching in deionized $H_2O$, and **(f)** rinsing in the deionized $H_2O$ stream.

## 3.5 Crystal surface preparation etching

The next step in the detector fabrication sequence is a second chemical treatment of the HPGe crystal surfaces. The purpose of this step is to remove surface contaminants and to adjust the surface chemistry in order to obtain the most desirable properties after the subsequent coating of the crystal surfaces with the amorphous semiconductor and metal electrode layers. As an example, one of the properties of importance is the existence and extent of surface channels associated with the sides of the completed detector (see the relevant discussion in [Amman 2018] and references therein). The early work of Hansen et al. [Hansen 1980] demonstrated that the surface channel behavior on simple planar detectors with a-Ge coatings was strongly impacted by the recipe used to deposit the a-Ge films and also was somewhat dependent on the details of the chemical etching of the crystals done just prior to the a-Ge depositions.

A standard chemical surface treatment used in the past at LBNL to prepare the HPGe surfaces for the sputter deposition of the amorphous semiconductor was a short etch in the concentrated 4:1 $HNO_3$:HF acid mixture followed by a quench in methanol. The pre-sputtering treatment used for the strip detectors is similar to this except that deionized $H_2O$ is used instead of methanol to quench the etch. The methanol quench creates a problem since it results in a hazardous acid-solvent waste mixture that has associated storage and disposal challenges. The less problematic $H_2O$ quench followed by a methanol rinse as described below has been shown to produce good results with the strip detectors. In addition to the detailed procedure given below, photographs of the surface preparation etching steps are provided in Figures 3.8 and 3.9.

**(a) Load crystal into PTFE etch clamp:** If contact has been made to the non-handle crystal surfaces, rinse the crystal with methanol and then blow it dry. While holding the bottom of the crystal with filter paper, clamp onto the crystal with the PTFE etch clamp (see Figure 3.8 below). A gloved hand could instead be used to hold onto the crystal handles and load the crystal into the etch clamp. Secure the clamp with the limit screw so that the clamp cannot be opened during the etch process. Blow any dust off of the crystal with dry $N_2$.

**(b) Etch crystal:** Fill an acid resistant bowl with a 4:1 volume ratio of $HNO_3$ (69.5% concentration) to HF (49% concentration) etchant to a level that will allow the crystal to be fully submerged in the acid. Also prepare a bowl with deionized $H_2O$ flowing into it to be used to quench the etch. Submerge the crystal in the acid and move the crystal rapidly back and forth, and side to side (though not so rapidly that the acid splashes out of the bowl) for approximately 30 s. Take care to not rub the crystal against the bowl and keep the crystal submerged during the agitation.

**(c) Quench etch, rinse and dry crystal:** After the desired etch time has elapsed, rapidly move the crystal from the etch bowl to the deionized $H_2O$ quenching bowl. Slosh the crystal around in the deionized $H_2O$ for at least 30 s, and then rinse the crystal and etch clamp directly in the deionized $H_2O$ stream for another 30 s. Without allowing the crystal to dry, remove the crystal from the deionized $H_2O$ stream and immediately rinse it with methanol from a squirt bottle. The methanol should be collected in a bowl during the rinse and then disposed of properly afterward. Following the methanol rinse, quickly dry the crystal with dry $N_2$. It is very important that this is done rapidly to avoid drying marks/residues. Residues will affect the amorphous semiconductor film morphology, adhesion, etc., and will likely impact detector performance. No drying marks should be present on the active crystal surfaces (all surfaces other than the handles). If some exist, re-rinse the crystal with methanol and re-dry it. It is best to tilt the crystal nearly on edge, with the third handle resting on filter paper (if the crystal is of a three handle design), and then blow the crystal dry starting at the non-handle edge and working down to the opposing handle. The objective is to blow all methanol down to the opposing handle so that any drying marks left will be on this inactive part of the crystal. For a two handle crystal, a similar approach of blowing the methanol down to a handle edge should be used. It is important that once the etching is done that the active surfaces of the crystal not be handled or come into contact with anything. If contact is made with a non-handle surface (particularly if the surface is scratched), the crystal should be re-etched. Following this surface preparation etching, the crystal should be promptly loaded into the sputtering system for the amorphous film deposition. The impact of leaving the freshly etched crystal in air for an extended period of time before loading into the sputterer has not been determined.

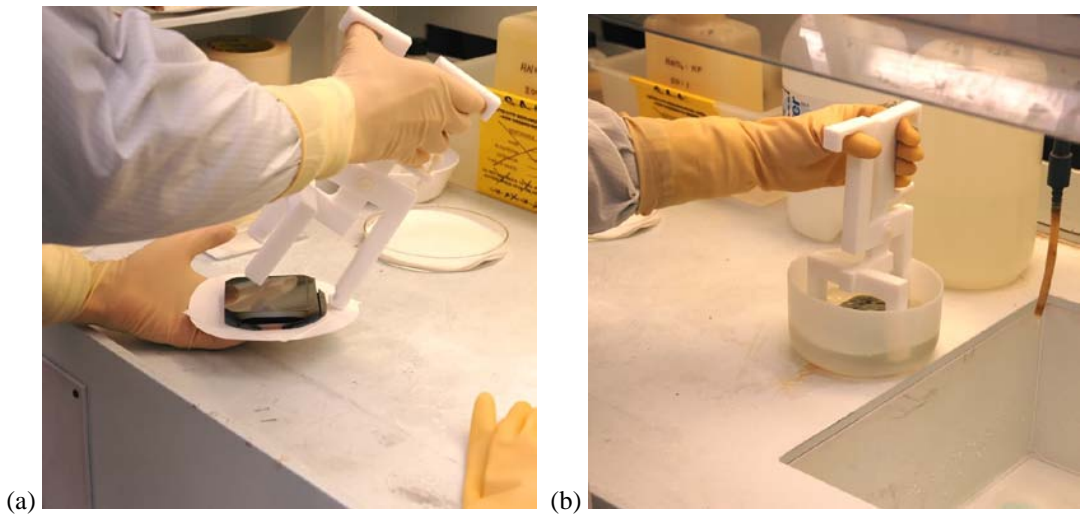

(a)                                        (b)

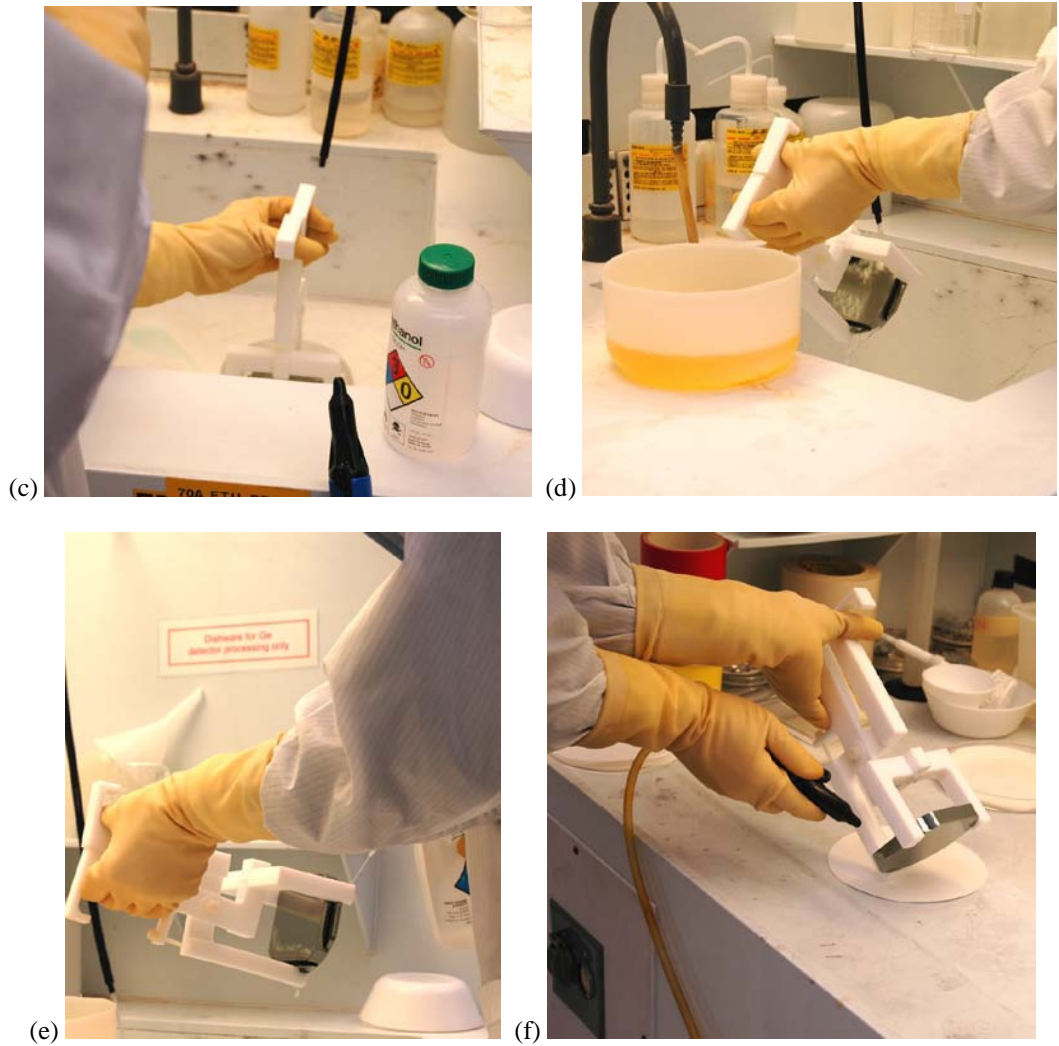

**Figure 3.8** Sequence of photographs showing the crystal surface preparation etching process. The crystal being etched in these photographs is of a three handle design. The steps shown are: **(a)** crystal loading into the custom PTFE etch clamp, **(b)** etching, **(c)** etch quenching in deionized H$_2$O, **(d)** rinsing in a deionized H$_2$O stream, **(e)** rinsing with methanol, and **(f)** drying.

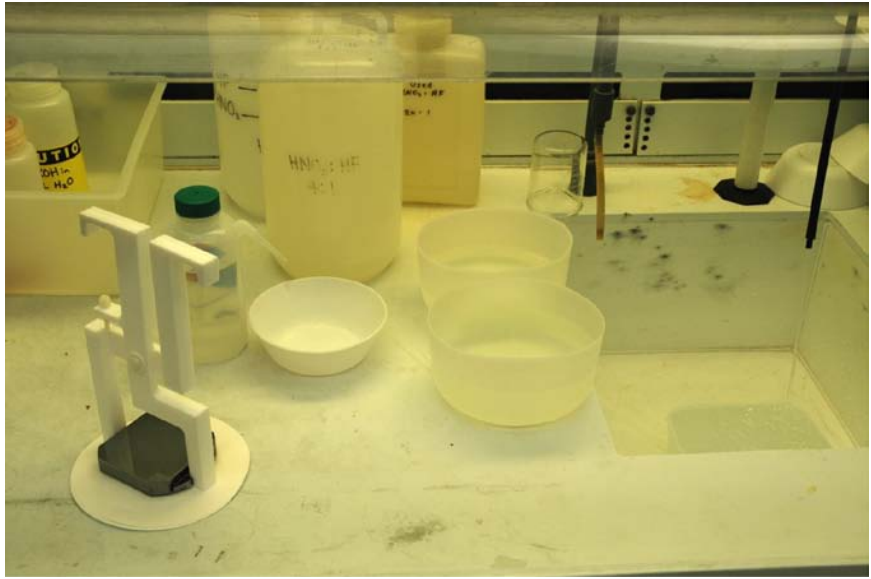

**Figure 3.9** Setup for the crystal surface preparation etching process to be done on a crystal with two handles. The design of the PTFE etch clamp used here is of a similar but different design from that of the clamp used with the three handle crystals.

## 3.6 Amorphous semiconductor depositions

Immediately following the surface preparation etching, all surfaces of the HPGe crystal are coated with an amorphous semiconductor layer (either a-Ge or a-Si). The amorphous film provides the charge injection blocking behavior that is needed for the areas that will be later coated with a metal and used as electrical contacts. In areas without the metal electrodes, the amorphous semiconductor serves as a surface passivation coating. As has been detailed elsewhere [Amman 2018 and references therein], the properties of the amorphous semiconductor affect the performance of the resultant detector, and these properties substantially depend on and are controllable through the process parameters used to deposit the film. The relationships between the process parameters and the detector performance as well as specific deposition recipes will not be covered in this section. The reader should instead refer to the aforementioned paper for such information.

The deposition of the amorphous semiconductor layer is done using RF diode sputtering in a gas mixture of Ar and $H_2$. The sputter deposition system used is described in [Amman 2018] and is referred to as Sputterer 1 in that reference. A two step, two vacuum cycle procedure is used to completely coat the crystal (see Figure 3.10). In the first step, the crystal is placed onto an Al fixture that only makes contact with the crystal handles. The fixture loaded with the crystal is then placed into the sputter chamber onto a rotation stage that is offset from the center of the amorphous semiconductor sputter target. A sputter down deposition of the amorphous semiconductor is done while the offset sample stage is rotated. This coats the top face and sides of the crystal (see the drawing on the left side of Figure 3.10(b)). In the second step, the crystal is flipped over and placed on a different fixture. The crystal and fixture are loaded into the sputter chamber onto a stationary stage that is positioned directly beneath the sputter target. A second sputter deposition is then done that coats the bottom face of the crystal (see the drawing on the right side of Figure 3.10(b)). Ideally, all surfaces of the crystal will become uniformly coated with amorphous semiconductor as shown in Figure 3.10(b). Unfortunately, this is not the case in reality due to the nature of the sputter deposition process. Non-uniform coverage of the crystal sides and possible undercoating of the crystal during the first deposition are typical problems encountered. An incomplete coating of the detector sides and an excessive undercoating (shown in Figure 3.10(c)) have both been observed to be associated with excessive surface leakage currents to the extent of detector failure. To circumvent this problem, the fixtures used to hold the crystal during the depositions must be carefully designed. The first deposition is more critical, and an appropriately designed fixture for this step is shown schematically in Figure 3.10(d). The fixture is designed so as to not block the coating of the sides and to keep only a small gap between the bottom crystal surface and the fixture. As the sputter gas pressure is increased, the problem of undercoating becomes more severe, and a smaller gap under the crystal becomes necessary.

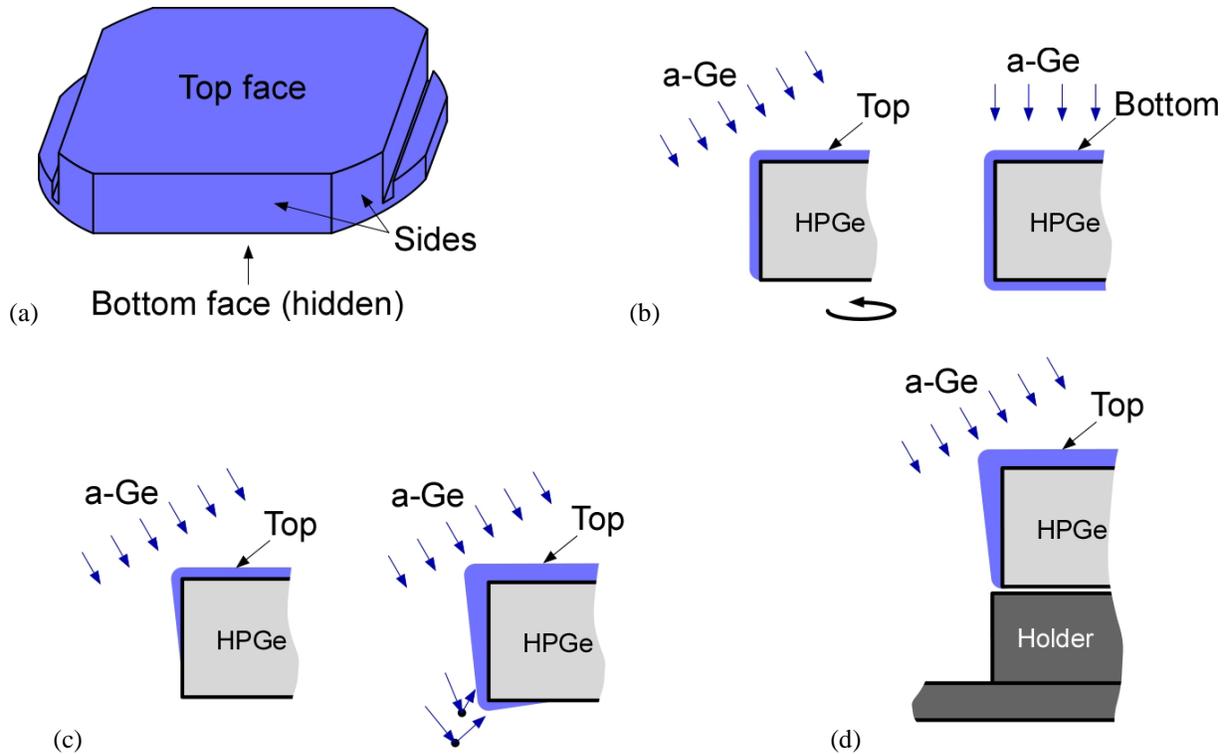

**Figure 3.10** Schematic drawings of the HPGe crystal and the amorphous semiconductor sputter depositions onto the crystal. **(a)** Perspective crystal drawing denoting the crystal surface designations. **(b)** Cross-sectional drawings showing one end of the crystal (a side without a handle) during the two sputter deposition steps. The drawing on the left is of the top and side deposition, which is done with the crystal offset from the center of the sputter target while the crystal is rotating. The drawing on the right is of the bottom deposition, which is done with the crystal centered directly beneath the sputter target. **(c)** Cross-sectional drawings showing one end of the crystal during the top and side deposition. The drawing on the left shows the condition of incomplete side coating while that on the right is of undercoating. Undercoating results from the scattering of the sputtered atoms by the sputter gas and nearby surfaces. **(d)** Cross-sectional drawing showing one end of the crystal during the top and side deposition. In this case, the crystal is placed on top of an appropriately designed fixture that allows the sides to be readily coated while at the same time inhibits undercoating onto the bottom face.

A description of the amorphous semiconductor sputter deposition sequence is provided below. Also provided are photographs of the sequence in Figures 3.11 and 3.12. The sequence of Figure 3.11 is of a COSI crystal that has three handles while that of Figure 3.12 is of a GRIPS crystal that has two handles. The fixture for the top and side deposition shown in Figure 3.12(a) can be used for either the two or three handle crystals and became the preferred fixture for all such depositions.

**(a) Set up for top and side coating of crystal:** Immediately following the surface preparation etch, place the crystal bottom side down onto the Al sputter fixture (see Figures 3.11(a), 3.11(b), and 3.12(a)). The sputter fixture supports the crystal only at the crystal handles and enables easy handling of the crystal with minimal risk of contacting the critical crystal surfaces. The sputter deposition system should have been previously prepared by loading the sample rotation stage and the appropriate sputter targets (Ge and/or Si). Place the mounted crystal on the sample rotation stage in the sputterer. Pump out the sputterer as necessary for the specific amorphous film recipe.

**(b) Sputter coat top and sides of crystal:** While slowly rotating the sample stage, sputter deposit the amorphous film onto the crystal. The sample rotation stage is offset from the center of the target and sample rotation is used to ensure that both the top and side surfaces become completely coated. The rotating sample stage is not water cooled, and a significant temperature increase of the crystal can occur during this deposition. To reduce this heating, a multistep deposition can be and is typically used. This consists of sputtering for 2 min, waiting for a cool down period of at least 5 min, and then repeating the sputtering and cool down until the desired film thickness is achieved.

**(c) Set up for bottom coating of crystal:** Open the sputter chamber, remove the crystal, and load the crystal into the desired sputter fixture (see Figures 3.11(c), 3.12(b), and 3.12(c)). Place the mounted crystal bottom side up on the water cooled plate of the sputterer. Secure the fixture to the plate with screws, and place any sputter shield around the crystal as called for in the deposition recipe (see Figures 3.11(d), 3.12(d), and 3.12(e)). Pump out the sputterer as necessary for the specific amorphous film recipe.

**(d) Sputter coat bottom face of Ge crystal:** Sputter deposit the amorphous film onto the crystal.

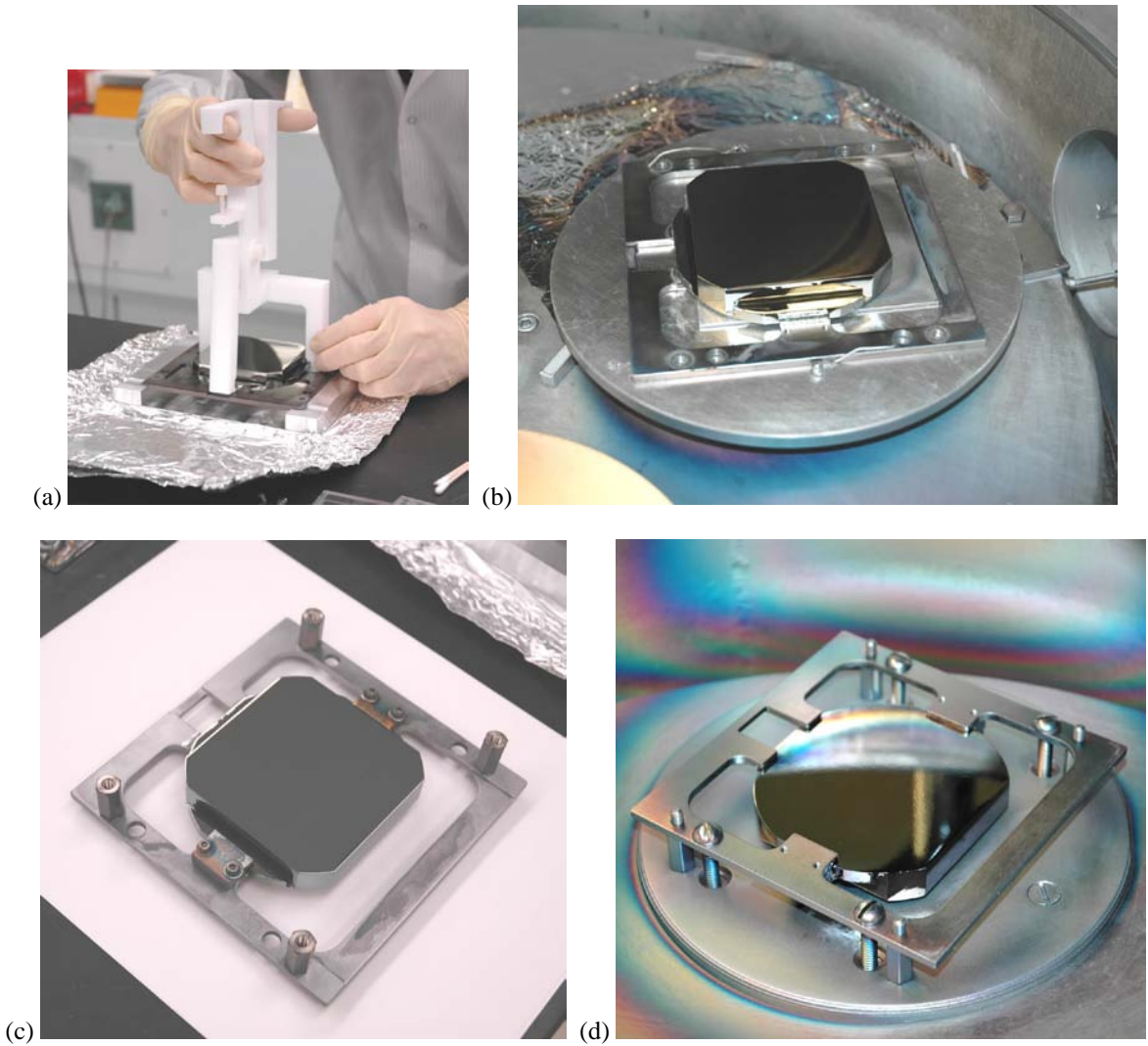

**Figure 3.11** Sequence of photographs showing the amorphous semiconductor sputter deposition process. In this case, the crystal being coated has three handles. **(a)** Crystal being transferred from the etch clamp to the sputter fixture. **(b)** Crystal resting on the sputter fixture that has been placed on the rotating stage inside of the sputter chamber in preparation for the deposition onto the top and sides of the crystal. Note that this sputter fixture is no longer used for the first deposition and has been replaced by that shown in Figure 3.12(a) below for both three handle and two handle crystal designs. **(c)** Crystal mounted in the sputter fixture after the completion of the top and side deposition. The fixture has been reconfigured for the bottom deposition by the addition of standoffs. **(d)** Mounted crystal attached to the water cooled plate of the sputter deposition system in preparation for the bottom deposition. An Al foil shield that is not shown here is added prior to the deposition.

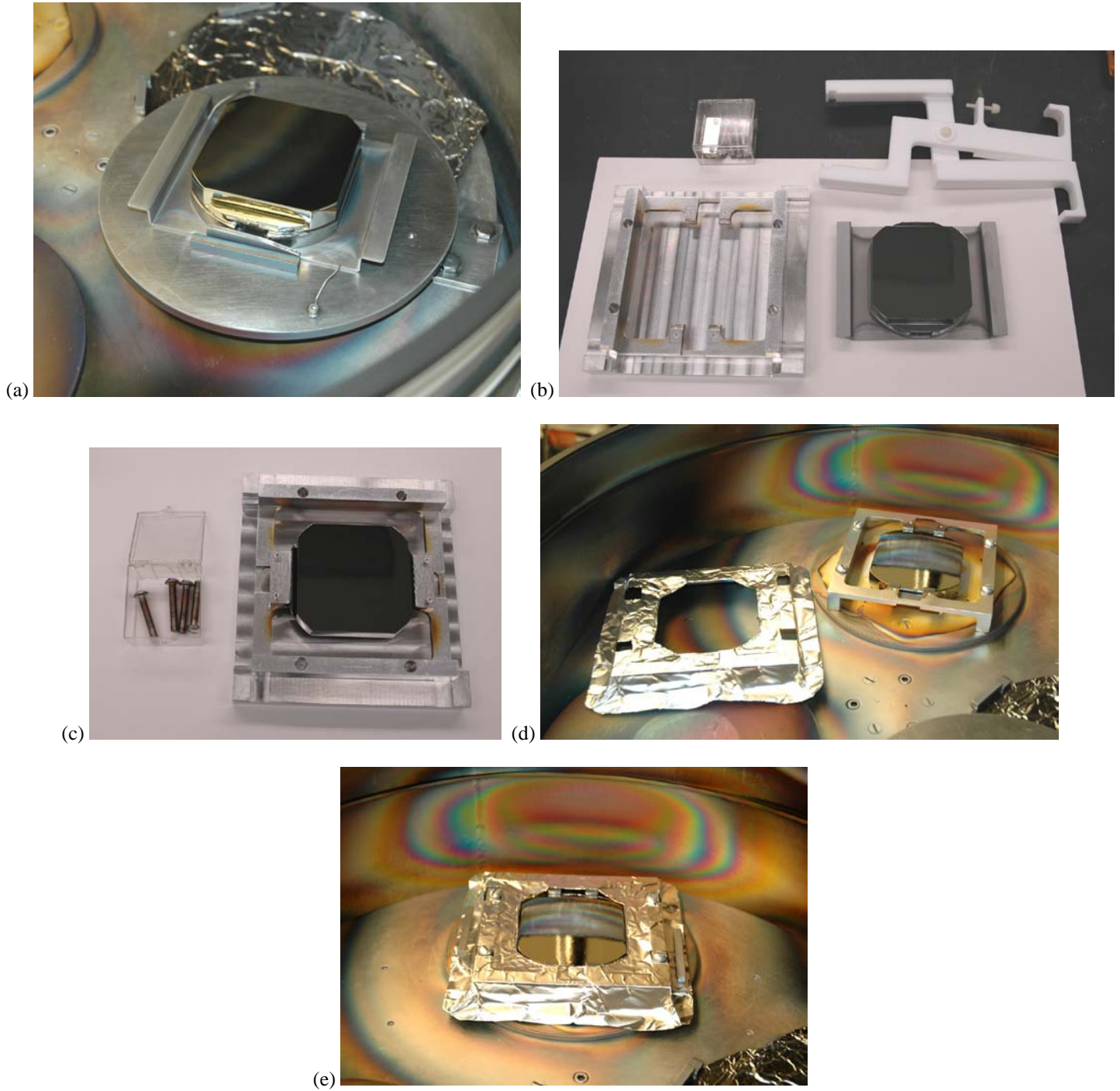

**Figure 3.12** Sequence of photographs showing the amorphous semiconductor sputter deposition process. In this case, the crystal being coated has two handles. **(a)** Crystal resting on the sputter fixture that has been placed on the rotating stage inside of the sputter chamber in preparation for the deposition onto the top and sides of the crystal. **(b)** Sputter fixtures, etch clamp, and crystal immediately after the completion of the top and side deposition. The crystal is resting on the fixture used for the first deposition and is about to be moved using the etch clamp to the fixture on the left that is used for the deposition onto the bottom of the crystal. **(c)** Crystal loaded into the second sputter fixture. **(d)** Mounted crystal attached to the water cooled plate of the sputter deposition system in preparation for the bottom deposition. **(e)** Same as in (d) except with the addition of an Al foil shield that serves to inhibit coating the top and sides during the deposition onto the bottom face.

## 3.7 Metal electrode deposition and lithography

After the HPGe crystal has been completely coated with a layer of amorphous semiconductor, a patterned Al metal layer is fabricated onto each of the two detector faces. The pattern is that of strips with a surrounding guard ring as shown in Figures 1.1 and 1.2. These metalized areas define the physical areas of the electrical contacts on the detector. Though metals other than Al have been successfully used for these electrodes, Al has the advantages of good adhesion to a-Ge and a-Si, being easily and cleanly chemically etched off of a-Ge and a-Si, having a self-passivating oxide layer, and forming an excellent wire bonding pad when deposited in a thick layer.

Two different methods are used at LBNL to produce the patterned Al layer: (1) shadow mask and (2) photolithography. In the first of these two, a custom, commercially produced sheet metal mask containing a cutout of the desired electrode pattern is placed on top of the crystal face. Aluminum is then thermally evaporated through this shadow mask, thereby resulting in the creation of a patterned metal layer. With this shadow mask method, no additional wet chemical processing of the crystal is required. Because of this, the crystal can be loaded into the final detector holder prior to the Al depositions. This eliminates further direct handling of the crystal and reduces the possibility of crystal damage. Note that for this shadow mask process, the mask makes direct contact with the crystal faces. At this stage of the detector fabrication process, the critical interface between the amorphous semiconductor and the HPGe has been completed. Consequently, contact to the crystal surfaces is acceptable as long as the contact does not cause damage completely through the amorphous semiconductor to this interface.

Photolithography, the second method to produce the electrodes, involves more processing steps than the shadow mask method and includes additional wet chemistry steps. The photolithography process begins with the deposition of an Al layer covering both crystal faces. This is done with thermal evaporation but could as well be done with sputtering since, unlike the situation with the shadow mask method, a directional (line of sight) deposition of the Al is not required for the blanket coatings needed for photolithography. Once both faces of the crystal are coated with Al, the crystal has been converted to a simple planar detector. If the Al depositions have been done such that only the crystal faces and not the sides were coated with Al, the detector can be loaded into a test holder and evaluated as a detector. This mid-fabrication evaluation is a valuable tool for identifying detector defects prior to the time-consuming photolithography, wire bonding, and inspection steps. If the detector passes the planar detector performance tests, the process to create the strips and guard ring electrodes is continued. A patterned photoresist mask layer is created on both detector faces. The patterns of these masks are identical to those of the desired electrodes. The mask patterns are then transferred to the Al layers using a wet chemical etchant that readily removes the Al not covered with photoresist yet does not significantly etch the photoresist or the amorphous semiconductor layers. The lithography process is then completed by the removal of the photoresist using an appropriate solvent.

Provided below are detailed procedures for both the shadow mask and the photolithography electrode deposition and patterning methods. Prior to these sections on the electrode fabrication, a section covering the preparation of the Al thermal evaporation source is provided.

### 3.7.1 Evaporation filament preload

The thermal evaporation system used for the deposition of the Al electrodes is a decades-old, cryopumped system that has been customized so that evaporate down or up configurations (HPGe crystal placed below or above the Al source) can be used. The Al evaporation method employed consists first of placing an Al coated W filament source into the evaporator vacuum chamber along with the HPGe crystal. The chamber is then pumped to a high vacuum, and an electrical current is passed through the filament to heat it sufficiently so that the Al melts and then evaporates. All surfaces with a line of sight to the evaporating Al source become coated with Al. A moveable shutter that can be placed between the source and the crystal is contained within the vacuum chamber to control the coating of the crystal. Though the source Al material when in the molten state readily adheres to the W, it also has an undesirable tendency to spit out small molten particles of Al in addition to the flux of evaporated Al atoms. If Al spit lands on the crystal, the performance of the resultant detector can be degraded. A large blob of molten Al landing on a crystal face will cause excessive leakage current in the detector and will ultimately leave a dimple in the crystal that will make the mandatory reprocessing of the crystal more difficult. In the past, detector failure due to Al spitting had been a significant yield reducing problem. Fortunately, this problem can be substantially eliminated through the combined application of several practices. These include: a long Al predeposit with the shutter closed, the use of thin Al when possible, evaporating up rather than down when possible, placing a fine meshed screen between the Al source and the crystal, and carefully preparing the evaporation source (preloading the filament with Al). Depending on the situation, it may not be possible to follow all of these practices. For example, the shadow mask method typically must be done using the evaporate down configuration, and a thick Al layer is required for electrode areas that will receive wire bonds. Provided below is a step-by-step procedure for preparing the Al loaded evaporation sources. This includes information on the W filaments and Al wire used at LBNL.

The W filaments used for Al evaporation are procured from R.D. Mathis Company [RDMathis 2020]. These filaments are a custom part (part number RDM-F-41202) manufactured from multi-stranded W wire (4 x 0.03 inch diameter) and are of a four loop design as shown in Figure 3.13. The overall length of the filaments is 4 inches, which is needed to fit properly in the filament holder contained within the LBNL thermal evaporation system. Other filament configurations have also been used with good results. Important considerations when choosing a filament design include the ability to evaporate down (as is needed for the shadow mask method) as well as up and the need for a large Al holding capacity. The Al source material is obtained from the vendor ESPI Metals [ESPI 2020]. The Al is in the form of 0.04 inch diameter wire, and its purity is 5N (ESPI stock number K234L). Aluminum material from other vendors and of a lower impurity would also likely be suitable. Prior to loading a crystal into the thermal evaporator for Al coating, the evaporation filaments are prepared by cleaning and then melting the Al material onto them. The procedure for doing this is the following.

**(a) Detergent clean filaments:** Ultrasonically clean the as received W filaments in a solution of $H_2O$ and detergent (for example, International Products' Micro-90 or Surface Cleanse 930). Thoroughly rinse the filaments in $H_2O$ and then dry them in an oven.

**(b) Load filament with Al wire:** Cut an approximately 400 mm length of Al. Using a portion of this length, loosely wrap Al wire around each loop of the filament (see the top of Figure 3.13).

**(c) Melt and wet Al onto filament:** Load the Al wire wrapped filament into the thermal evaporator and pump out the system. Somewhat quickly heat the filament to a glowing state. Since the Al wire only weakly makes thermal contact to the filament, it will not immediately melt. This delay allows the filament to be substantially heated and cleaned prior to the melting of the Al. Once the Al melts, it should immediately wet and adhere to the filament. As soon as this happens, quickly reduce the filament current to zero. Let the Al coated filament cool in vacuum for about 15 min before venting. Note that Al readily alloys with W, and, as a result, the structural integrity of a filament rapidly degrades when the Al is in the molten state. This alloying is ultimately what limits the useful lifetime of a filament. During the Al preloading procedure described here, the length of time this alloying is occurring should be minimized so as to not take away from the evaporation lifetime of the filament. If during preload, the filament is held at a high temperature for an extended period of time (several minutes), the filament will likely suffer a premature break and not fully evaporate all of its Al when it is later used to coat the HPGe crystal with Al.

**(d) Fully load pre-wetted filament with Al:** Vent and open the evaporator chamber. Using the remaining Al wire, load the filament with Al wire such that the majority of the Al is located between the center two loops (see the third from the top filament in Figure 3.13). Place the loaded filament back into the chamber and pump out the chamber. Somewhat quickly heat the filament to a glowing state. The newly added Al should melt and rapidly wet the filament since the filament has already been coated with Al. As soon as the Al wets, quickly reduce the filament current to zero. The filament is now ready for use as an Al thermal evaporation source.

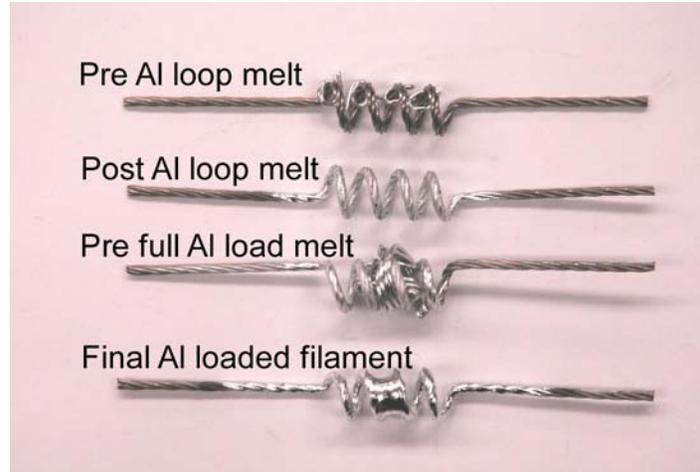

**Figure 3.13** Tungsten thermal evaporation filaments shown at various stages of the Al loading process.

### 3.7.2 Shadow mask electrode definition

With the shadow mask method, the openings in the mask produce the metalized electrode areas on the HPGe crystal face. The desired electrode pattern for the COSI and GRIPS detectors consists of a set of strips surrounded by a guard ring. There are a number of challenges encountered when producing such a pattern with shadow masks. Since the guard ring would translate in the mask to an open area completely surrounding the strips, a single mask cannot be used to create the entire electrode pattern. Furthermore, the narrow gap between the strips translates to a long thin wire in the mask. If a straightforward translation of the electrode pattern to the mask is done, the resultant long wires in the mask will tend to meander and bow away from the crystal surface thereby distorting the strip pattern. Also, there are small areas at the end of the strips that will eventually receive wire bonds. These areas must have a thick Al coating, whereas the rest of the electrode pattern should only receive a thin coating in order to minimize the possibility of an Al spit defect. Consequently, multiple Al evaporations and masks are used to create the electrode pattern. Drawings for an example set of masks are given in Figure 3.14. Corner portions of the three masks needed to deposit the electrodes on a COSI detector are shown in parts (a) through (c). The dark areas represent the open regions in the masks. The mask shown in Figure 3.14(a) is used to partially form the guard ring and strips, and only a thin layer of Al is needed. The deposition of a second thin Al layer through the mask of Figure 3.14(b) completes the electrode pattern. Finally, a thick Al deposition is made through the mask of Figure 3.14(c) to add Al to the areas that will eventually receive a wire bond. The combination of the three masks produces the completed pattern of Figure 3.14(d).

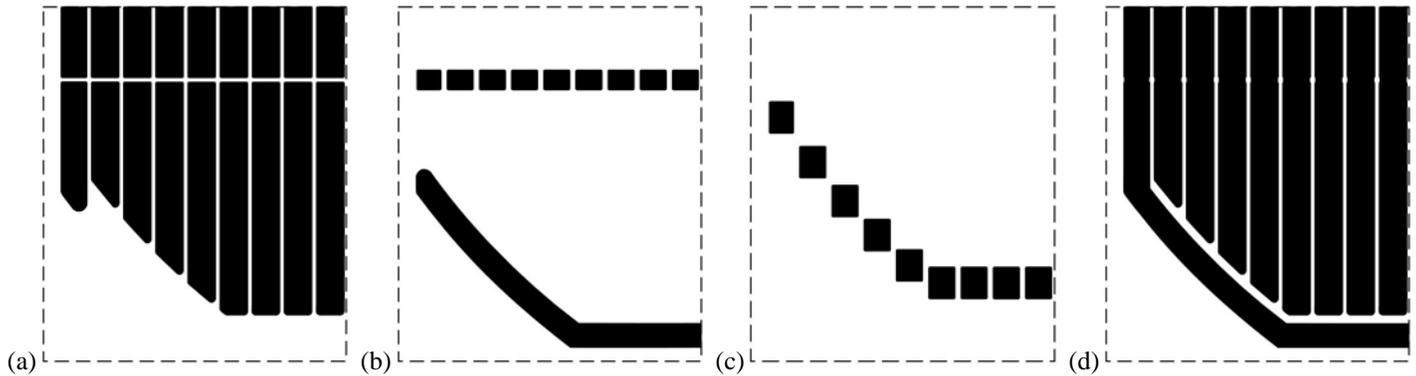

**Figure 3.14** Drawings of a set of shadow masks used to define the COSI detector electrodes. Only a portion near one corner of each mask is shown. The black areas in the drawings represent the open areas of the mask that when evaporated through will create the metalized areas on the detector. **(a)** Mask used to create nearly all of the strip areas and a portion of the guard ring (strip mask). **(b)** Mask used to interconnect the strip segments and to complete the guard ring (strip connector mask). **(c)** Mask used to produce the wire bonding pads (bonding pad mask). **(d)** Addition of all three masks.

Alignment of the masks to the HPGe crystal is made through the use of temporary pins installed in the detector holder along with corresponding holes in the shadow masks. As noted previously, the crystal is loaded into the final detector holder prior to the Al depositions. The holder is designed to accommodate alignment pins, and the shadow masks are designed to fit within the detector holder and make use of the pins. Two pins are used and are positioned diagonally opposite each other across the detector. In the masks, one pin passes through a clearance hole and the other through a slotted hole. The slotted hole is intended to accommodate for any dimensional inaccuracies.

The shadow masks are commercially produced from thin (typically 0.003 inch thick) sheet metal. Both hard tempered BeCu and 302 stainless steel have been used for the LBNL masks. Shadow masks must have no sharp edges that could possibly scratch the amorphous semiconductor layer. Masks produced through chemical etching have the desired smooth edges. Most of the masks used at LBNL were produced by either Vaga Industries [Vaga 2020] or Photofabrication Engineering, Inc. [PEI 2020], but other suitable manufacturers exist.

The step-by-step procedure to create the strip and guard ring electrodes on an amorphous semiconductor coated HPGe crystal using shadow masks is given below. Also provided (in Figure 3.15) is a sequence of photographs that were taken at various stages during the execution of the procedure.

**(a) Transfer crystal to final detector holder:** At this point in the detector fabrication process, the HPGe crystal has just received the amorphous semiconductor coatings and is under vacuum in the sputter deposition system. Experience indicates that the Al adhesion to the a-Ge and a-Si layers is best if the exposure of these films to air is minimized. Therefore, air exposure should be minimized until after the Al depositions are completed. This includes minimizing air exposure between the Al depositions. If the detector processing must be delayed before the Al electrodes are completed, the crystal should be stored under vacuum. To proceed with the processing, remove the crystal and sputter fixture from the sputter deposition system. Place the crystal and fixture on the sputter fixture holder. Using the etch clamp or gloved hand, move the crystal from the sputter fixture to the PTFE transfer fixture (Figure 3.15(a)). This transfer fixture has been designed to support the crystal by its handles and has features intended to align the crystal with the detector holder. Place In foil (improves thermal conductivity at the interface) and BN pieces (electrically isolate the crystal from the detector holder yet still provide good thermal conductivity) on the two opposing handles of the crystal (Figure 3.15(b)). Attach two shadow mask alignment pins to the detector holder as appropriate for use with the bottom shadow masks. Indium foil should have already been applied to the detector holder areas that will make contact with the BN pieces. If this has not been done, do so now. Carefully lower the detector holder onto the crystal handles (Figure 3.15(c)). Place the u-shaped Al plate on top of the detector holder. While squeezing the Al plate together with the transfer fixture, flip the entire assembly over (Figure 3.15(d)). Carefully lift off the transfer fixture. Place three PTFE shims between the detector holder and the crystal handle edges to prevent the crystal from sliding in the holder. Lay the first of the bottom strip masks on the crystal using two pieces of filter paper (Figure 3.15(e)). Take care to not slide the mask along the crystal surface. Make sure the mask sits flat on the crystal. If it does not, place small weights (such as nuts) on the mask to hold it down (Figure 3.15(f)).

**(b) Deposit bottom strips:** Great care must be taken during this deposition of Al since nearly the full crystal face is exposed to the evaporation source and could be damaged by Al spitting. Load the thermal evaporation system with the preloaded evaporation filaments created previously. Place a wire mesh screen beneath the filaments to provide additional protection from Al particles. Good results have been obtained with a 180 x 180 mesh size screen constructed from 0.0018 inch diameter 304 stainless steel wire forming 0.004 inch openings and providing an open area of 46% (McMaster-Carr part number 85385T107 [McMaster-Carr 2020]). Place the mounted crystal covered by the shadow mask directly beneath the filaments on the vibration isolation sample stage in the evaporator. Close the evaporation shutter. The setup as just described is shown in the photograph of Figure 3.15(g). Close and pump out the evaporator chamber to the $10^{-7}$ Torr range. Predeposit Al at the target deposition rate (between 0.1 and 0.5 nm/s) with the shutter closed for approximately 3 to 4 min in order to reach the stage where there is little chance of Al spitting. Open the shutter and deposit between 30 and 50 nm of Al. After allowing the parts to cool under vacuum for about 15 min, open the evaporator chamber and remove the mounted crystal. Note that due to the Al alloying with the W filament, each filament is typically used for only a single evaporation in which most all of the loaded Al is evaporated away.

**(c) Deposit bottom strip connectors:** Carefully remove the strip shadow mask using two pieces of filter paper. Add the bottom strip connector mask. Deposit another Al layer following the procedure described in (b).

**(d) Deposit bottom bonding pads:** Replace the strip connector mask with the bonding pad mask. Since only a small area of the crystal is exposed and a thick layer of Al is required, remove the wire mesh screen beneath the filaments. On top of the bonding pad mask, add an additional Al foil shield (with the appropriate openings) so that heating of the mask is minimized (Figure 3.15(h)). Otherwise, the mask will distort during the long deposition. Deposit Al as described in (b) except predeposit for about 1.5 min and use a deposition rate between 0.5 and 2 nm/s. Deposit a minimum of 500 nm. Typically, two filaments will be required to get this thickness.

**(e) Prepare crystal and holder for top electrode depositions:** Remove the mask. Remove the PTFE shims. Add the detector clamps to fix the crystal in the detector holder. Relocate the mask alignment pins. Flip the mounted crystal over.

**(f) Deposit top electrodes:** Perform the Al depositions for the three top shadow masks using the procedures (b) through (d).

**(g) Prepare detector assembly for wire bonding:** Remove the shadow mask and the mask alignment pins. Attach the detector circuit boards to the detector holder. The detector assembly is now ready for wire bonding.

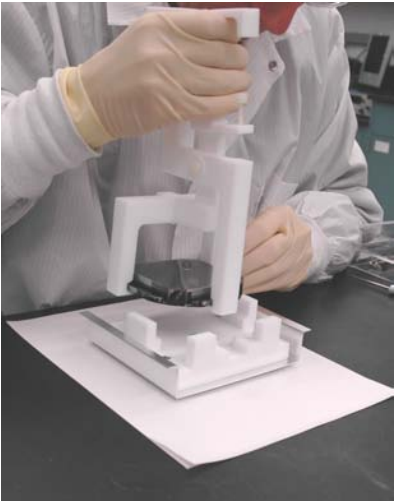
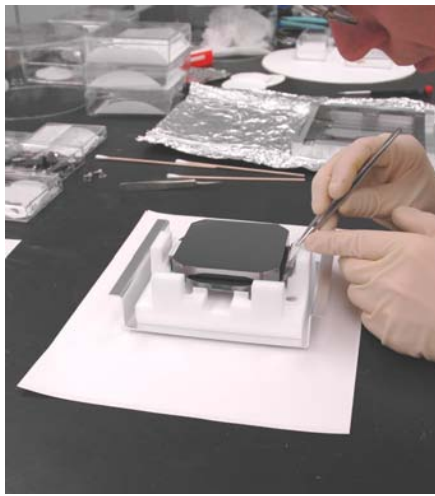
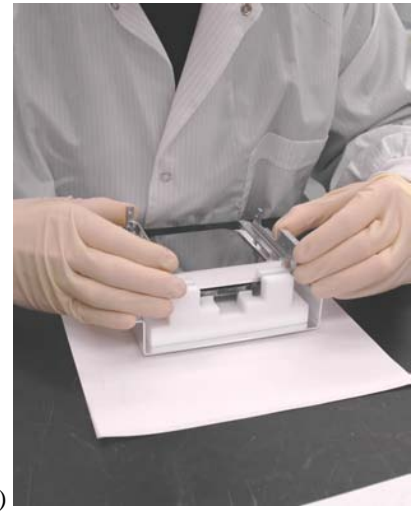

(a)                               (b)                               (c)

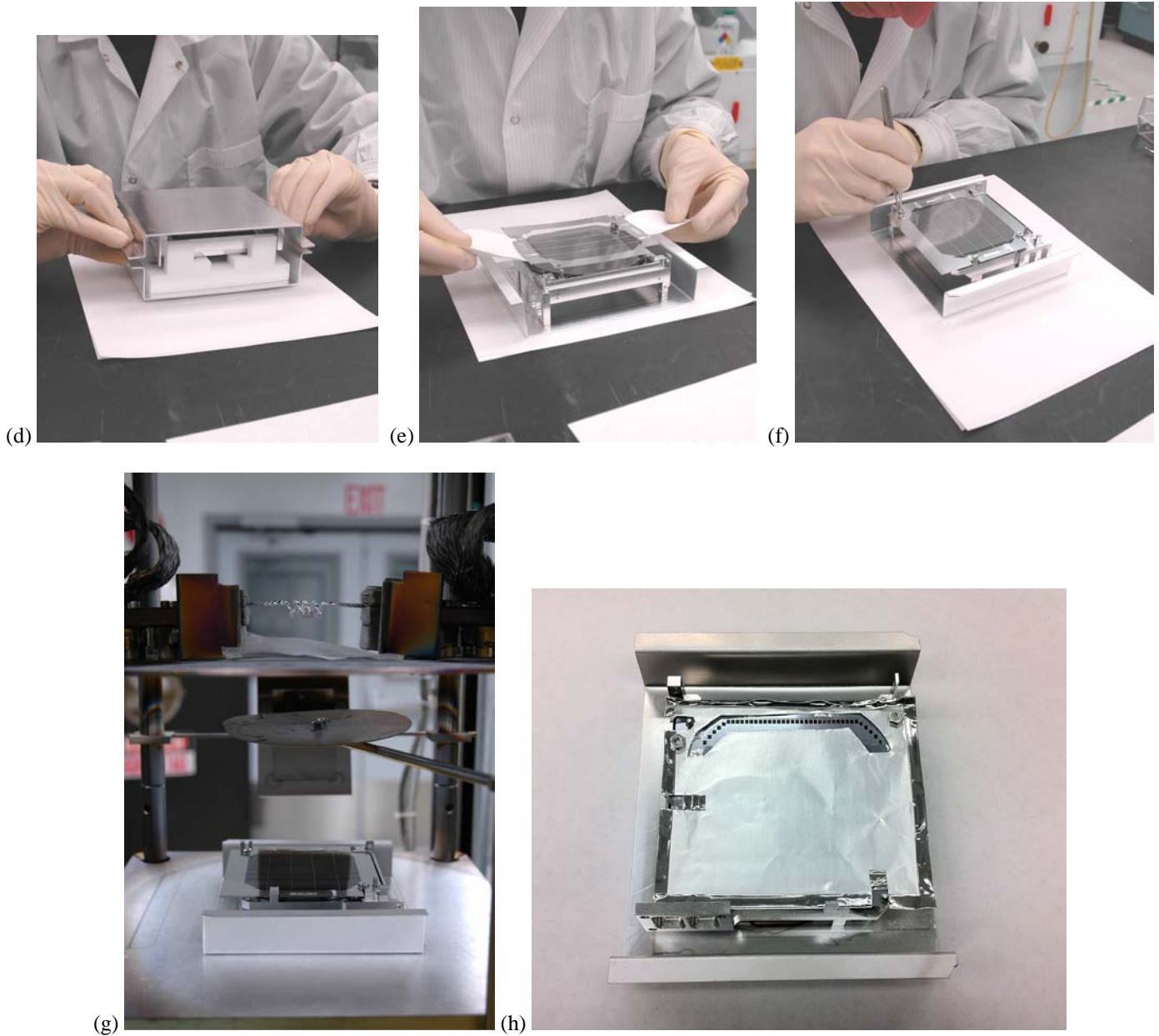

**Figure 3.15** Sequence of photographs showing the shadow mask method for creating the Al electrodes on a COSI detector. **(a)** Crystal being transferred from the sputter fixture to the transfer fixture. In the photograph, the etch clamp is used to move the crystal. A gloved hand also works well for this operation. **(b)** Indium foil and BN pieces being added to the crystal handle areas that will later be clamped to the detector holder. **(c)** Detector holder being placed onto the crystal. A proper alignment between the crystal and the detector holder is ensured by the transfer fixture. **(d)** A u-shaped Al plate has been placed on top of the detector holder in preparation for flipping the entire stack over so that the crystal will then be resting in the detector holder. The bottom side of the crystal will be accessible once the transfer fixture is removed. **(e)** The transfer fixture has been removed, and the bottom side strip mask is being lowered onto the crystal using two pieces of filter paper. **(f)** Small nuts are being placed onto the mask to force the mask to sit flat on the crystal surface. **(g)** The crystal in the detector holder with the bottom side strip shadow mask covering the crystal has been placed into the thermal evaporation system. Near the top of the photograph are two evaporation filaments loaded with Al. Directly beneath the filaments is a fine mesh screen, and beneath this screen is a shutter. **(h)** Shadow mask setup for the deposition of the wire bonding pads. An additional Al foil shield is placed on top of the mask and acts as a heat shield preventing the mask from becoming excessively warm and distorted.

If the quality, impurity concentration, or some other critical property of the starting HPGe crystal is unknown or a new fabrication process is being tested, it is a good idea to convert the crystal into a detector that has a simple electrode configuration rather than the strip one and then evaluate the performance of the simple detector. A simple planar configuration in which the two electrode faces of the planar crystal are completely metalized (full area electrodes) is one possibility. Another is a guard ring configuration in which a perimeter guard ring electrode and central electrode are fabricated onto the bottom face of the crystal, and a full area electrode is fabricated onto the top face (see Figure 3.16(b)). The use of these simple detector configurations eliminates the complications associated with wire bonding, detector boards, and reading out a large number of strip channels. The simple full area electrode and guard ring detector configurations can both be produced using the shadow mask method. To do this, follow steps (a) and (b) presented earlier in this section except that, rather than the bottom strip mask, a perimeter mask should be placed on the bottom face of the detector. This mask is open over nearly all of the crystal face and is meant to shield the sides of the crystal from the Al deposit. If the crystal is positioned directly beneath the Al source and there is no chance of coating the sides of the crystal, the perimeter mask can be left off. If a simple full area electrode detector is desired, no other mask is needed. For a guard ring detector, a wire ring mask is added to form the gap between the center electrode and the guard ring. The masking for a guard ring detector is shown in Figure 3.16(a). After the bottom electrodes have been deposited, the detector is completed by following steps (e) and (f) given earlier in this section except that only a single deposition with a perimeter mask is done. Note also that steps (a) and (b) of the photolithography section below are also a procedure for creating a simple detector with full area electrodes. Once the detector electrodes are finished, remove the mask alignment pins and attach spring steel wire clips to make electrical contact to the detector center, bottom guard ring (if one is present), and top electrodes. Each wire clip makes use of an In ball to make a soft, non-damaging connection to a detector electrode. A completed guard ring detector including the spring wire electrical connection attachments is shown in Figure 3.16(b). With the electrical connections to the detector electrodes made, the detector should then be evaluated as instructed in section *4.1 Evaluation of full area electrode and guard ring detectors*.

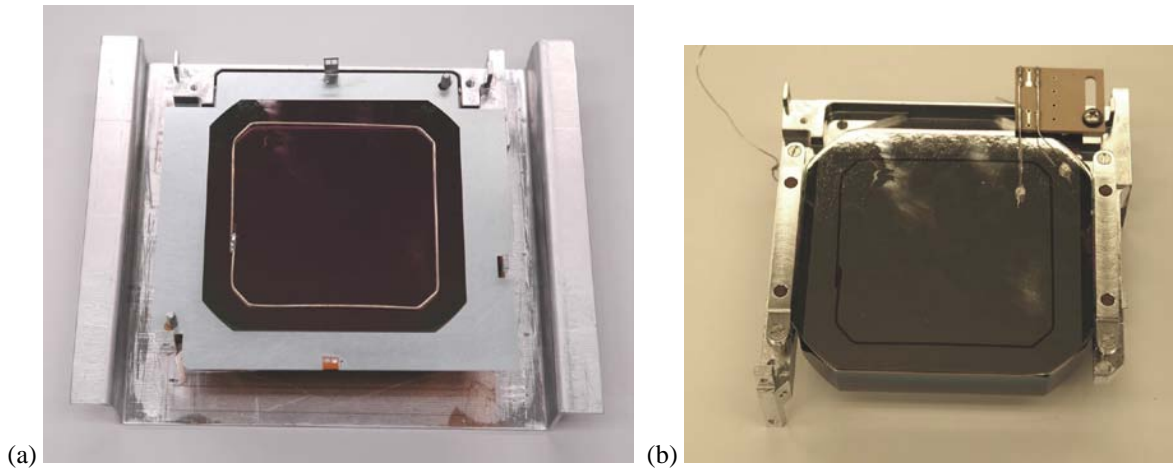

**Figure 3.16 (a)** Photograph of the shadow masking used to produce a detector whose bottom electrodes consist of a large area central electrode surrounded by a perimeter guard ring. **(b)** Photograph of a completed guard ring detector loaded into a detector holder. Spring steel wire clips with In balls at their ends have been attached to the detector holder and make electrical contact to the electrodes on the detector but are electrically isolated from the detector holder. The clips will eventually be wired out to measurement electronics when the detector is evaluated.

### 3.7.3 Photolithography electrode definition

Though the shadow mask method just described is simple and can be highly reliable, it has its disadvantages. These include being limited to larger electrode feature sizes, requiring multiple masks and Al depositions to implement designs such as that on the strip detectors, needing line of sight deposition methods such as thermal evaporation and a deposit down configuration, and not providing a convenient mid-processing check of the detector functionality. In contrast, the photolithography method is capable of producing fine features of complex patterns through a single mask process. Furthermore, since the electrode layer is first deposited as a full area coating, a wider variety of deposition methods can be used, and, once these full area electrodes have been deposited, the resultant simple planar detector can be tested for basic functionality. The main drawbacks of photolithography are the additional processing steps needed to produce the segmented electrodes, which can lower the yield of good detectors, and the need for additional processing equipment. The additional equipment needed includes a spinner for the application of the photoresist onto the crystal, an oven to bake the photoresist, and a mask aligner to provide a UV light source for exposing the photoresist. The photoresist spinner must be capable of spinning the combined mass of the HPGe crystal and the custom fixture used to attach the crystal to the spinner spindle. The unit used at LBNL is a G3P Spincoat from Specialty Coating Systems [SCS 2020]. The mask aligner must have sufficient headroom between its sample stage and exposure head in order to accommodate the thick crystal. The aligner used at LBNL is a decades-old unit from Quintel [Quintel 2020] that was designed for research and development type of work. In addition to these two pieces of processing equipment, custom fixtures and a photolithography mask are required. One of the fixtures is used to hold onto and flip over the crystal (flipping fixture), and another is the photoresist spinner fixture referred to earlier. Both of these fixtures were designed so that they only make contact with the handles of the crystal. The photolithography mask used for the patterning of the electrodes on the GRIPS detectors is a 4 inch x 4 inch piece of 0.06 inch thick glass with a patterned iron oxide mask coating. The choice of iron oxide for the mask layer facilitates aligning the mask to the edges of the crystal since the iron oxide is optically transparent. Many companies are capable of making such a mask. The masks for the LBNL strip detectors were produced by Photo Sciences, Inc. [Photo Sciences 2020].

The step-by-step procedure to create the strip and guard ring electrodes on an amorphous semiconductor coated HPGe crystal using photolithography is given below. Also provided (in Figures 3.17 through 3.22) is a sequence of photographs that were taken at various stages during the execution of the procedure as well as other information pertinent to the procedure.

**(a) Deposit full area electrodes:** At this point in the detector fabrication process, the HPGe crystal has just received the amorphous semiconductor coatings and is under vacuum in the sputter deposition system. As noted previously in the shadow mask section, it is best to minimize air exposure of the just deposited amorphous semiconductor film until after the Al electrode layer has been deposited. To begin, the thermal evaporator should be prepared. To do this, load the evaporator with two preloaded Al evaporation filaments. Place a wire mesh screen (180 x 180 mesh as specified previously in the shadow mask section) above the filaments. Install the evaporate up shutter if it is not already in place. When all is ready for the Al depositions, remove the crystal and sputter fixture from the sputter deposition system. Depending on the sputter fixture used, either leave the crystal in the fixture or move the crystal from the fixture to a holder that can be attached to the evaporate up plate of the thermal evaporator (Figure 3.17(a)). Attach the mounted crystal to the evaporate up plate. Load the evaporate up plate with the mounted crystal into the evaporator (Figure 3.17(b)). Close the evaporation shutter. Close and pump out the evaporator to the $10^{-7}$ Torr range. Using the first filament, predeposit Al at the target deposition rate (~ 1 nm/s) with the shutter closed for approximately 2 to 3 min to reach the stage where there is little chance of Al spitting. Open the shutter and deposit the entire filament load onto the crystal. Close the shutter and then repeat the evaporation sequence with the second preloaded filament. A target minimum total thickness for the Al is 500 nm. If this has not been achieved, reload the evaporator with new preloaded filaments and then deposit additional Al. After allowing the parts to cool under vacuum for about 15 min, open the evaporator and remove the evaporate up plate holding the mounted crystal. Remove and then reattach the mounted crystal to the plate with the other side of the crystal facing outward. Repeat the Al deposition sequence in order to coat the other crystal face with at least 500 nm of Al. Remove the mounted detector from the evaporator and then from the evaporate up plate. Note that due to the effect of gravity, the evaporate up configuration is less prone to Al spitting problems than evaporating downward. Because of this, a shorter pre-deposition time, a higher deposition rate, and no screen can be used when evaporating up. However, to achieve a near-perfect yield, the screen should still be used. Since the recommended screen only passes about 46% of the deposit, four evaporation filaments are often required for each face of the crystal. For the GRIPS detectors, the actual Al film thickness was typically about 700 nm, and four Al filaments were used.

**(b) Electrically test as planar detector:** Since the remaining steps required to complete the detector are quite involved, it is best to evaluate the detector at this point to make sure that it has sufficiently low leakage current at the desired operating voltage bias. To do this, load the detector into a detector holder following the procedure given in part (a) of the shadow mask section. Attach spring steel wire clips with In balls at least their ends to make electrical contact to the bottom and top electrodes. With the electrical connections to the detector electrodes made, the detector should then be evaluated as instructed in section *4.1 Evaluation of full area electrode and guard ring detectors*. If the detector electrically breaks down or has excessive leakage current, the detector fabrication process should be restarted as described in section *3.10 Detector reprocessing*.

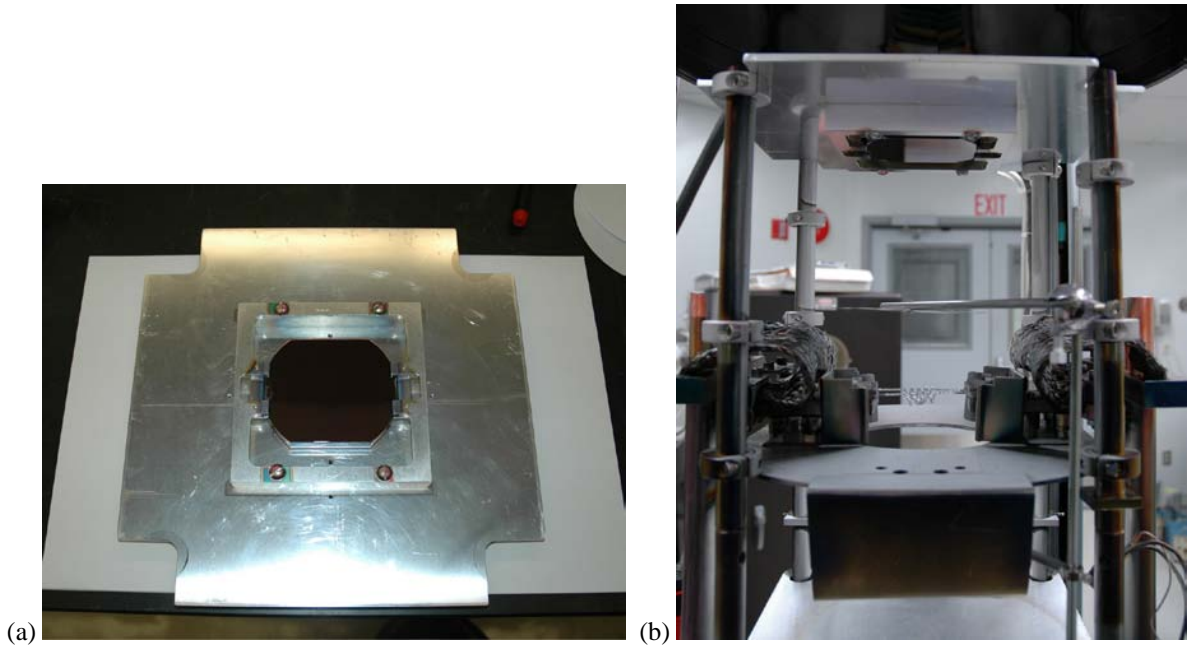

(a)                                    (b)

**Figure 3.17** Photographs showing the setup for the deposition of full area Al electrodes. **(a)** HPGe crystal loaded into the sputter deposition fixture that is attached to the evaporate up plate. This assembly will later be loaded into the upper part of the thermal evaporation system. **(b)** View of the inside of the thermal evaporation system after it has been configured for evaporating upward onto an installed crystal. Note that in this photograph there is no Al spit screen between the Al filaments and the crystal, but one should be used to improve the yield of good detectors.

**(c) Apply photoresist:** With a successful test of the planar detector completed, remove the detector from the test holder by reversing the loading procedure. After this is done, the detector should be sitting in the transfer fixture. Using either a gloved hand or the etch clamp, move the detector from the transfer fixture to the PTFE flipping fixture. Use the flipping fixture to transport the crystal between processing stations and for flipping the detector over when necessary (see Figure 3.18(c) for a photograph of the two piece flipping fixture). Set the photoresist baking oven to 90°C and wait for the oven temperature to stabilize. Attach the custom holder to the spindle of the photoresist spinner (Figure 3.18(a)). Set the spinner program to 1 krpm with a 5 s ramp up, a 60 s spin at the full speed, and a 5s ramp down. Test spin the setup without the detector loaded. Transfer the detector from the flipping fixture to the spinner holder. Orient the detector such that the top is facing up. Clamp the detector in place with the side screws and the top PTFE clamps (Figure 3.18(b)). Test spin the setup. Nearly completely fill a disposable pipette (23 mL size) with Shipley Microposit S1818 photoresist or any other similar photoresist. Without the spinner running, dispense most of the photoresist onto the detector face. This should nearly cover the entire surface. Do not dispense bubbles onto the detector surface during this process, and do not let the photoresist spill over the edges of the face. Start the spinner and let it complete the spin process. When the spin is finished, remove the detector from the spinner holder. Flip the detector over using the flipping fixture (Figure 3.18(c)). Reload the detector into the spinner holder with the bottom of the detector facing up (Figure 3.18(d)). Repeat the photoresist application sequence to coat the bottom with photoresist. Remove the detector from the spinner holder and place it on one of the flipping fixture halves. Bake the detector in the photoresist oven at a temperature of 90°C for 30 min. Remove the detector and flipping fixture from the oven, and then let them cool on the countertop for 15 min. At this point, it may be necessary to remove any photoresist that has spilled onto the sides of the detector. This would be needed if there is a chance that Al has been deposited onto the detector sides rather than just the two faces. Removing the photoresist from the sides ensures that any Al on these surfaces will not be masked and will be removed during the later Al etch step. To complete this step, use a cleanroom swab and acetone to wipe off any unwanted photoresist. To do this, dip the swab in acetone, blot off any excess acetone, carefully wipe the detector sides, and then repeat until the sides are clean (Figures 3.18(e) and 3.18(f)). Note that this side cleaning is unnecessary if the Al depositions were done in the thermal evaporator as instructed above and the planar detector electrical testing was successfully completed. A good electrical test of the detector indicates that there must not be any significant Al coating on the sides of the detector.

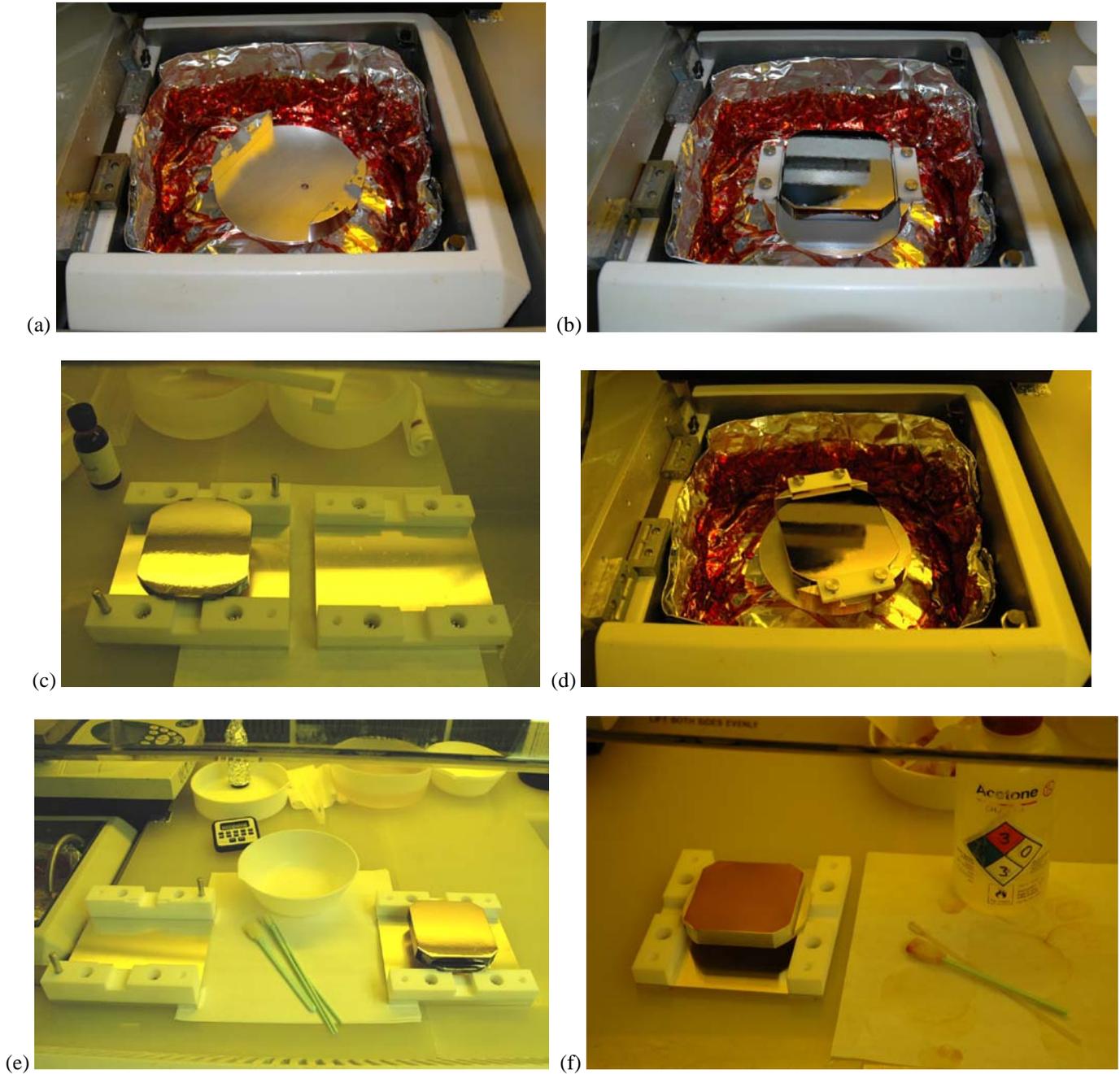

**Figure 3.18** Sequence of photographs showing the double sided photoresist application process. **(a)** Photoresist spinner with the custom detector holder installed. **(b)** Detector installed top face up in the photoresist spinner. **(c)** Flipping fixture with a detector sitting bottom face up on one side of the fixture. **(d)** Detector installed bottom face up in the photoresist spinner. **(e)** Setup for removing photoresist from the sides of a detector using acetone and cleanroom swabs. **(f)** View after the completion of the acetone swabbing.

**(d) Expose photoresist:** Turn on the mask aligner and start the UV lamp. Let the lamp stabilize for at least 15 min before doing an exposure. This startup can be done during the photoresist bake step. Remove the mask holder and the wafer vacuum chuck from the mask aligner. These components are not needed, and removing them is critical since, otherwise, the flipping fixture loaded with the detector will not fit under the mask aligner head. Once these parts have been removed, confirm that there will be sufficient clearance. Place the detector loaded onto one side of the flipping fixture (use the thinner side of the fixture) into the mask aligner (Figure 3.19). Place the photoresist mask on top of the detector, and adjust its position by eye. Make sure that the strip orientation is appropriate for the detector face being exposed. Note that the top strips run from one detector handle to the other, and the bottom strips, which are perpendicular to the top ones, run parallel to the length of each handle. Lower the mask aligner head. Manually expose the detector for the desired length of time (see recipe parameters in Figure 3.20). Flip the detector over and similarly expose the other detector face.

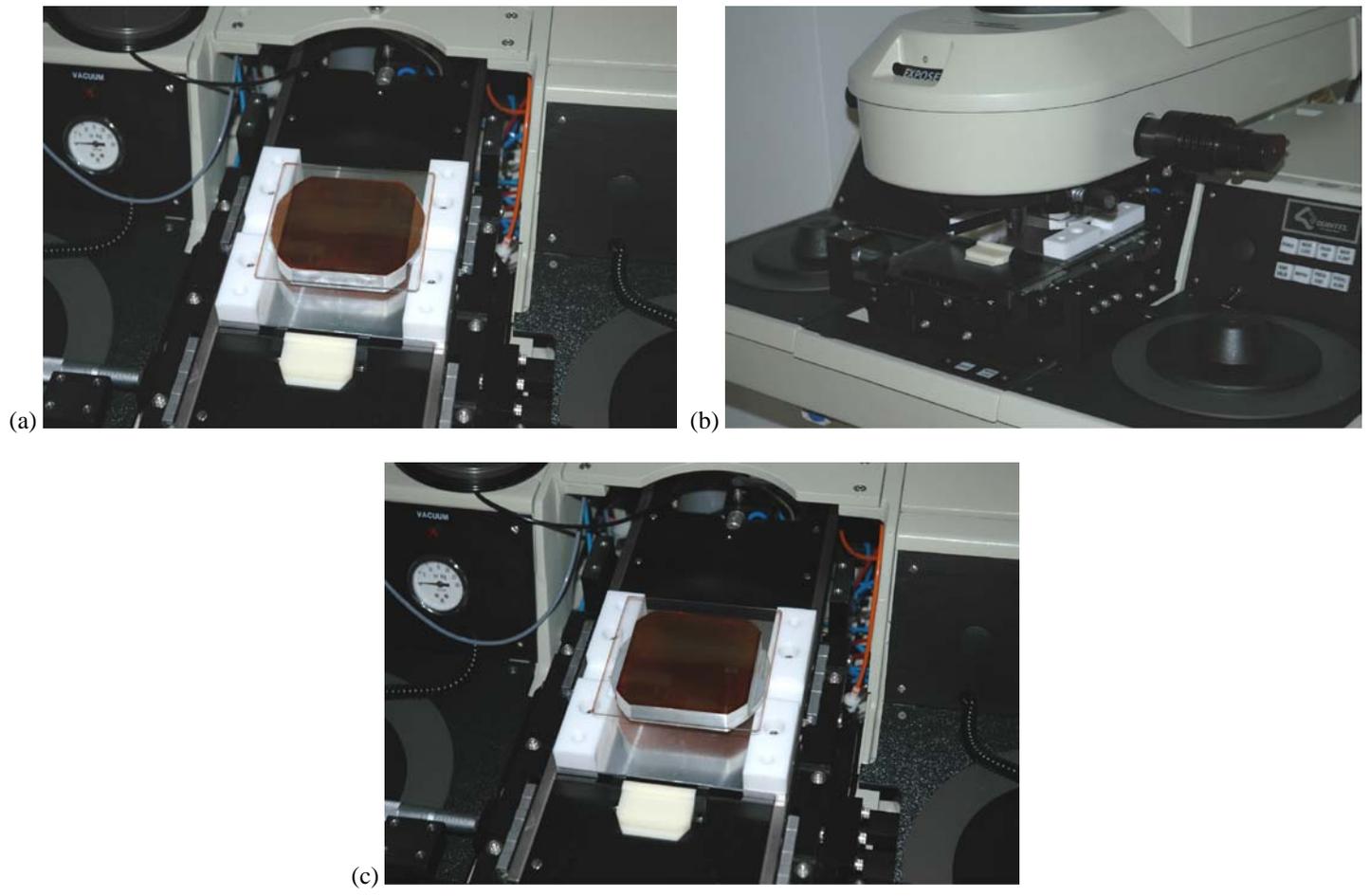

(a)                    (b)

(c)

**Figure 3.19** Sequence of photographs showing the double sided photoresist exposure process. **(a)** Detector positioned bottom side up on the flipping fixture that has been loaded onto the mask aligner sample stage. A photolithography mask has been placed on the detector. The mask aligner exposure head has been raised to provide access to the sample stage. **(b)** Mask aligner with the exposure head lowered and ready to expose the photoresist coating on the detector. **(c)** Detector positioned top side up on the flipping fixture that has been loaded onto the mask aligner sample stage. Note that the flipping fixture consists of two pieces or sides and that one side is thinner than the other. The thinner of the two must be used for this exposure process. Otherwise, the fixture, detector, and mask stack will be too tall and will come into contact with the mask aligner head when it is lowered for an exposure.

**(e) Develop photoresist:** The setup used for developing the pattern in the UV exposed photoresist is shown in the photographs of Figure 3.21. To start, fill a bowl with Microposit developer to a level that will allow the detector to be fully submerged. Also prepare a bowl with deionized $H_2O$ flowing into it for rinsing after developing. While holding the detector with the etch clamp, submerge it in the developer and move it slowly back and forth until the pattern fully develops, and then slightly overdevelop to make sure that the underside of the detector has also fully developed (see recipe parameters in Figure 3.20). Take care to not rub the detector against the bowl. Also make sure air bubbles are not trapped under the detector by tilting the detector and periodically removing it from the developer to allow the liquid to flow off of the surfaces. When the developing is done, remove the detector from the developer and immediately rinse it in the $H_2O$ bowl. Following the rinse, blow dry the detector with $N_2$. Inspect both faces of the detector to confirm that the photoresist mask is fully developed and defect free. If the mask is not fully developed, additional developing can be done. If unacceptable defects in the mask exist, the photoresist must be removed and the process repeated. If this is necessary, the photoresist can be removed with an acetone rinse followed by a methanol rinse.

| Photoresist | Spin speed | Oven bake | Exposure time | Developer | Develop time |
|---|---|---|---|---|---|
| Microposit S1818 | 1000 rpm | 90°C, 30 min | 20 s @ 18 mW/cm$^2$ <br> 23 s @ 16 mW/cm$^2$ | Microposit MF-24A or Microposit MF-314 | ~ 60 s |

**Figure 3.20** Recipe parameters for one possible photoresist process.

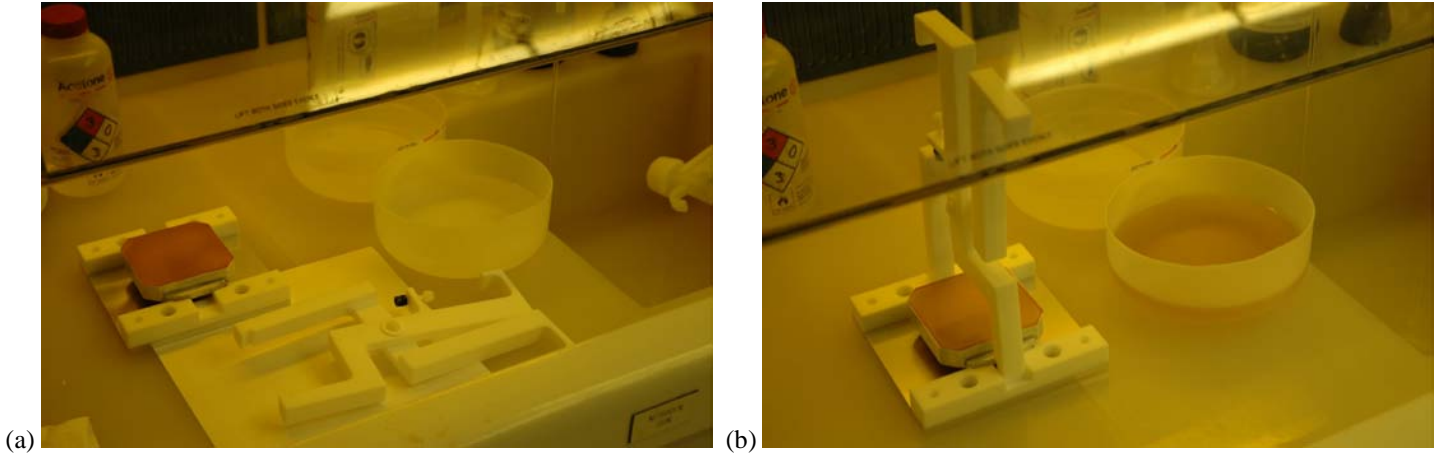

(a)  (b)

**Figure 3.21** Photographs of the double sided photoresist developing process. **(a)** Setup prior to the start of the process. **(b)** View after the completion of the developing process.

**(f) Etch Al:** Fill a flat bottomed bowl with a 1:1 volume ratio mixture of 1% HF dip and deionized $H_2O$ to a level that will allow the detector to be fully submerged in the etchant. Also prepare a bowl with deionized $H_2O$ flowing into it to be used for rinsing the detector. Using the etch clamp, submerge the detector in the etchant and continually slosh the detector back and forth allowing the etchant to flow off of the surfaces. A pendulum type motion should be used so that bubbles formed on the underside do not become stationary and mask the etching process. Periodically pull the detector completely out of the etchant to remove any gas bubbles that may have formed at the Al surface. A periodic rinse under the $H_2O$ stream should also be used as a means to eliminate bubbles. Continue this process until the unmasked Al areas have been completely removed (approximately 10 to 15 min). Rinse the crystal in deionized $H_2O$, and then dry it with $N_2$. Inspect both faces to confirm that the Al has been completely etched away from all unmasked areas. If this is not the case, perform additional etching.

**(g) Remove photoresist mask:** While holding the detector with the etch clamp, rinse and/or soak the detector in acetone until the photoresist mask is gone. Without letting the crystal dry, follow this with a methanol rinse, and then dry the detector with $N_2$. Additionally, a deionized $H_2O$ rinse can follow the methanol one.

**(h) Prepare detector assembly for wire bonding:** Load the detector into its holder using the transfer fixture as described previously in the shadow mask section, but do not add the alignment pins, PTFE shims, or shadow mask. Add the detector clamps to fix the detector in the detector holder. Attach the detector circuit boards to the holder. The detector assembly is now ready for wire bonding. A sequence of photographs capturing this sequence of operations is provided in Figure 3.22.

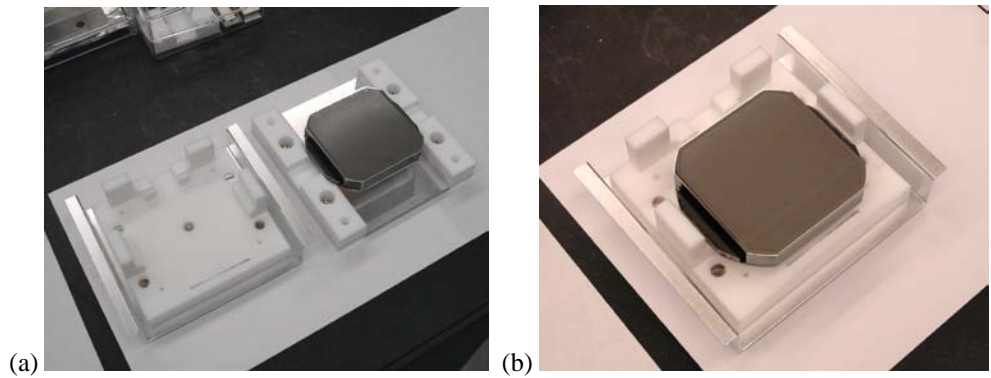

(a)  (b)

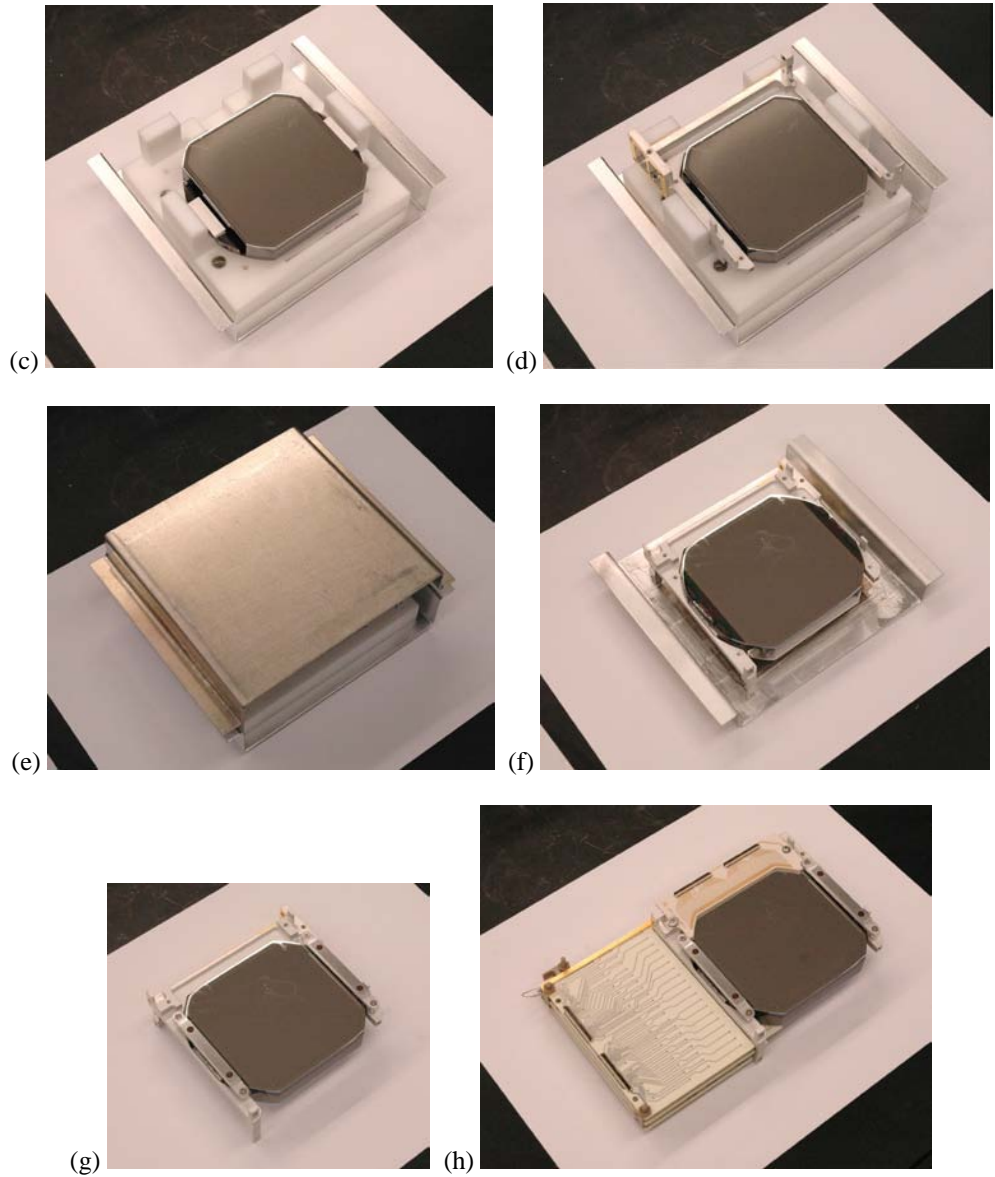

**Figure 3.22** Sequence of photographs showing the loading of a two handle GRIPS detector into a detector holder. **(a)** Completed detector sitting on the flipping fixture in preparation for moving the detector to the transfer fixture. Note that a more convenient transfer fixture than the one shown in the photographs is available for the GRIPS detectors. The one in these photographs was actually designed for the three handle COSI detectors. **(b)** Detector located on the transfer fixture. **(c)** Detector after the In foil and BN pieces have been added to each of the two handles. **(d)** Detector holder added to the detector and transfer fixture stack. **(e)** Al plate added to the detector holder, detector, and transfer fixture stack. **(f)** Detector loaded into its holder sitting on the Al plate after the stack pictured in (e) was flipped over and the transfer fixture was removed. **(g)** Detector assembly with the addition of the clamping bars that fix the detector in its holder. **(h)** Detector assembly completed with the attachment of the detector circuit boards.

## 3.8 Wire bonding

The final step needed to complete a detector assembly is to electrically connect the detector electrodes to the electrical traces on the detector circuit boards. This is done using ultrasonic wedge wire bonding. Making interconnects using this technology is routinely done on Si based devices. The wire commonly used when bonding on these devices is Al with 1% Si. Wire bonding to Ge, though, can be more challenging than it is for Si, since Ge is more brittle. Depending on the quality of the wire bonding machine, Al wire with 1% Si may not be the best choice due to its hardness, which likely contributes to a greater tendency for the bonding process to damage Ge.

**M. Amman, 2020, "High Purity Germanium Based Radiation Detectors with Segmented Amorphous Semiconductor Electrical Contacts: Fabrication Procedures"**

Two different wire bonders have been used at LBNL. One is a Kulicke & Soffa 4523 Manual Wedge Bonder (to be referred to as the manual bonder), and the other is a higher end fully automatic wire bonder. With the manual bonder, the use of standard 0.001 inch diameter Al with 1% Si wire along with bonding parameters chosen to produce reasonably strong bonds has led to a seemingly random occurrence of detector damage in the form of small divots or chips in the HPGe crystal. However, larger diameter (0.002 inch) pure Al wire produces excellent results with this bonder. Strong bonds on both the detector and circuit boards can consistently be achieved, and many thousands of bonds have been made on detectors with no visual damage or apparent degradation in the detector leakage currents. All of the COSI detectors were wire bonded with the larger diameter pure Al wire and this manual bonder. The downside of this bonder is that manual wire bonding becomes tedious when a large number of bonds are needed. A COSI detector requires 76 wire bonded connections, which is a reasonable job for a manual bonder. A GRIPS detector, however, requires 300 connections, and manually bonding such a detector takes more than two hours to complete when all goes well with the process. In contrast, the automatic wire bonder can complete the job in less than 15 min once the parameters have been optimized and the bonding program written. The consistency of the automatic bonder is also apparently superior to that of the manual one. This is concluded from the fact that the automatic bonder can use 0.001 inch diameter Al with 1% Si wire on the detectors without causing visual damage or performance degradation. Many thousands of bonds have been made with the automatic bonder on GRIPS and other HPGe detectors with no apparent issues.

It was stated earlier in this paper that properly choosing the metallization on both the detector and the circuit boards is critical for good wire bonding. It is worth restating that point here. The Al metallization on the detector must be adherent and reasonably thick. A minimum of 500 nm of thermally evaporated Al deposited onto freshly sputtered a-Ge or a-Si allowed strong wire bonds to be consistently formed on the detector. For the circuit board, a multilayer metallization finish capped with soft bondable Au is appropriate, and a specific example is the electroless Ni/electroless Pd/immersion Au (ENEPIG) surface finish [IPC-4556 2013]. The bonding areas of the board traces should also be thoroughly cleaned before the boards are added to the detector assembly. Swab cleaning with acetone and alcohol is appropriate. High quality wire bonding also requires a mechanically solid setup. Characteristics of a good setup include a firm base supporting the detector, tightly fastened detector and circuit boards, and a minimization of the circuit board cantilever distance. Essentially it is desirable to minimize the average loss and variability in the loss of the bonding energy in components other than the location of the wire bond.

Provided below is a procedure for the wire bonding of a strip detector. Supporting information and photographs are provided in Figures 3.23 through 3.25.

**(a) Prepare detector assembly:** For the GRIPS detectors, a holder was designed and fabricated specifically for rigidly supporting the detector assembly during wire bonding and protecting the sides of the detector from incidental contact. The holder provides access to both detector faces and allows the detector assembly to sit flat on a solid Al plate. Photographs of the wire bonding holder loaded with a GRIPS detector are provided in Figure 3.23. This detector assembly preparation step consists of loading the detector assembly into the wire bonding holder. Note that a custom holder was not used with the COSI detectors and the manual wire bonder. Also, at this step, a dummy detector and circuit board setup should be assembled if one does not already exist. This dummy setup should closely resemble the actual detector assembly configuration including replicating the detector and circuit board metallizations and mechanical support structures. The dummy setup will be used for bond parameter optimization and bond consistency checks.

**(b) Prepare wire bonder:** Load the wire bonding wedge tool and wire, and set the bonding parameters as determined previously. If the bonding parameters have yet to be optimized, this will be done as part of step (c). An optimized set of parameters as well as other relevant information for the manual wire bonding are given in Figure 3.24.

**(c) Perform test bonds:** Using the dummy setup, perform test bonds and adjust bonding settings so that strong bonds are consistently produced.

**(d) Wire bond detector:** Wire bond each strip and the guard ring to its corresponding trace on the circuit board using a single wire. Do this for both faces of the detector. Make sure that the wire bond loops have sufficient slack to accommodate any relative shifting between the detector and circuit board that may take place during detector cooling. Photographs of the manual bonder setup and a GRIPS detector with manually bonded interconnects are shown in Figure 3.25.

**(e) Verify connections:** Each wire bond should be visually inspected. This is naturally done during the wire bonding process, but an additional survey after the job is completed is recommended since wires may get disturbed during the detector assembly handling. Next, verify each connection through a resistance or continuity check. Using a digital multimeter and fine tipped probes, measure between the guard ring circuit board trace and each strip circuit board trace. Any open connection indicates an incomplete wire bond that will need to be removed and then remade.

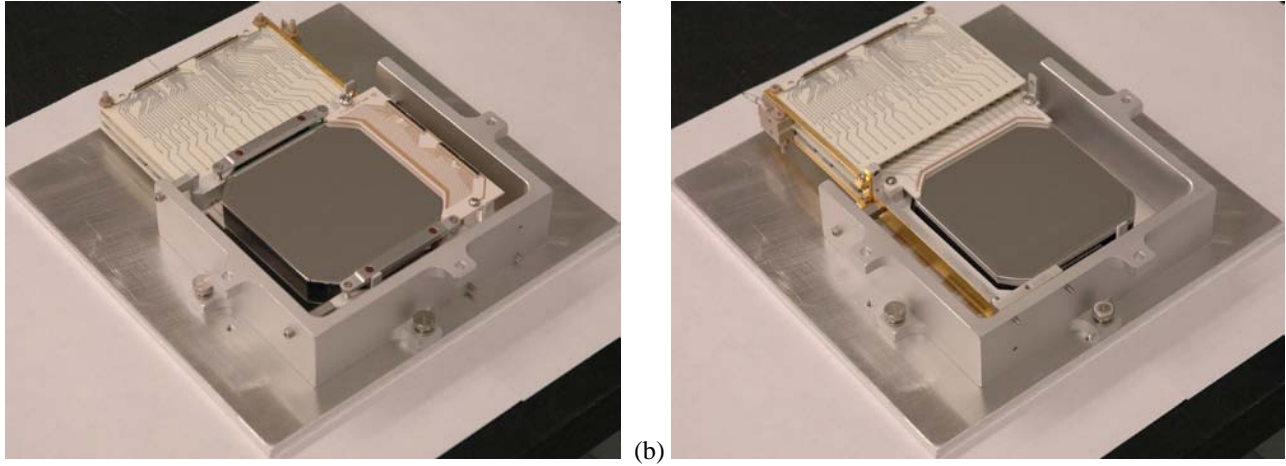

(a)  (b)

**Figure 3.23** Photographs of a GRIPS detector assembly loaded into a wire bonding holder. This holder was used for bonding the detectors with the automatic wire bonder. **(a)** Detector assembly oriented bottom face up. **(b)** Detector assembly oriented top face up.

| Detector bond parameters | Circuit board bond parameters | Other parameters |
|---|---|---|
| Search 1: 10 | Search 2: 10 | Loop: 7 |
| Power 1: 3.5 | Power 2: 7 | Tail: 10 |
| Time 1: 4 | Time 2: 6 | Tear: 5 |
| Force 1: 2 | Force 2: 2 | |

**Figure 3.24** Wire bonding parameters for the Kulicke & Soffa 4523 Manual Wedge Bonder. These are the parameters that produce strong bonds with 0.002 inch diameter pure Al wire. The bonding wedge used is a Deweyl [Deweyl 2020] part number CKSVO-1/16-750-45-C-3540-M.

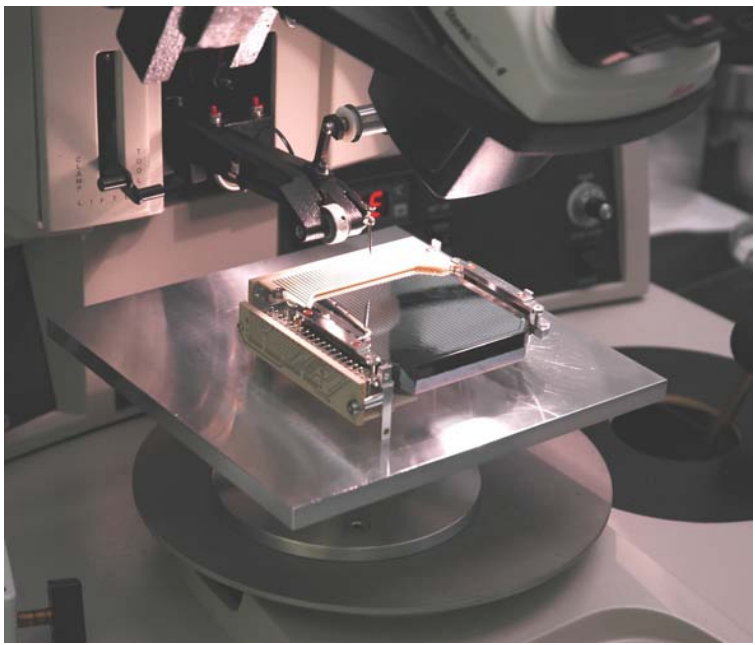

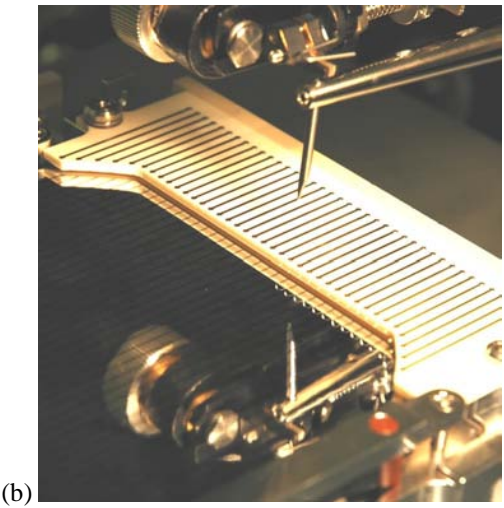

(a)  (b)

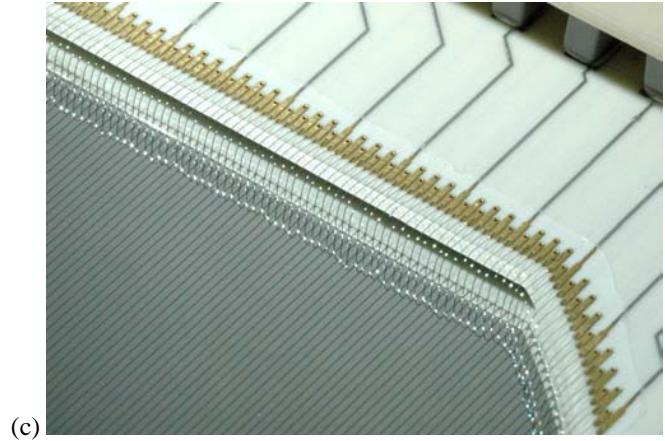

**Figure 3.25 (a)** Manual wire bonder with a COSI detector assembly positioned for bonding. **(b)** Close up view of a COSI detector during the wire bonding of its bottom face using the manual wire bonder. **(c)** Close up view of the manually produced wire bonds on a GRIPS detector assembly.

## 3.9 Detector defect inspection and repair

Visual inspection at each step of the fabrication sequence should be done to identify defects. If any are discovered, a decision must be made as to whether the fabrication should be continued or restarted. A fatal flaw will necessitate detector reprocessing as described in section *3.10 Detector reprocessing*. For the GRIPS detectors, which have finely segmented electrodes, additional detector flaw detection work should be done after the detector assembly is complete. The aim is to identify common photolithography imperfections with an emphasis on strip interconnection defects. A common defect on the GRIPS detectors is a short between adjacent strips. A photoresist mask defect or a bubble formed and left in place during the Al etch can create such a short. Since a GRIPS detector is read out by an ASIC that is intolerant to having input channels shorted to each other, the inter-strip shorts must be identified and repaired for the detector to be functional. The shorts themselves cannot be removed from the detector, but one of the two wire bonds connected to the shorted pair can be removed to ameliorate the readout electronics problem. It is convenient to identify shorts after the wire bonding since, depending on the amorphous semiconductor recipe, they can often be efficiently detected through a resistance or continuity check at room temperature. The procedure for the short identification and repair is described below. Note that interconnected strips have not been an issue with the COSI detectors due to their coarser electrode structure and the use of the shadow mask method described previously. Furthermore, GRIPS inter-strip shorts can be substantially eliminated through a careful inspection of the photoresist etch mask (followed by the remaking of the mask when defects are identified) and a well-implemented Al etching procedure focused on removing gas bubbles as they form on the detector surfaces during the etch.

**(a) Visually inspect detector:** Using a stereomicroscope, carefully scan the entire detector face looking for shorts. Any identified shorts should be marked by placing a small dot on the corresponding circuit board traces using a fine permanent marker. During this inspection, also check that all wire bonds are good. If any have come off, they should be rebonded, and the detector should be rechecked electrically.

**(b) Probe test detector boards:** Using a digital multimeter set for a continuity check and fine tipped probes, measure between each pair of adjacent circuit board traces on the detector boards (see Figure 3.26 for the basic setup). Set the meter so that it will sound an audio alert when a connection is made. This way, the meter does not need to be viewed during the testing, and the visual focus can be on probing the fine circuit board traces. Depending on the amorphous semiconductor recipe used to coat the detector face, the inter-strip resistance at room temperature for properly separated strips may be high enough to not trip the continuity audio alert. This should be true for the high resistivity a-Si that has been used on the bottom face of the GRIPS detectors. For this detector face, any audio alert indicates a shorted pair of strips. In contrast, the a-Ge used for the top face of the GRIPS detectors produces an inter-strip resistance of only about 15 $\Omega$ at room temperature. This is likely not high enough to prevent the audio alert from sounding. For this case, a series resistance can be added so that the sum of a good inter-strip resistance and the series resistance will not trip the audio alert, but the series resistance itself will trip the alert. For example, the threshold for the audio alert on a meter used at LBNL is about 50 $\Omega$. Adding a series resistance of just above 35 $\Omega$ will prevent a good inter-strip measurement from tripping the audio alert, yet an inter-strip short will sound the alert. As was done during the visual inspection, any identified shorts should be marked by placing a small dot on the corresponding circuit board traces using a fine permanent marker.

**(c) Remove bond wire(s):** One of the two bond wires from each shorted pair of strips should be removed. While viewing the detector under a stereomicroscope and using a fine tipped pair of tweezers, scrape the target bond wire off of the circuit board bonding pad.

Detach the wire from the detector side by grabbing the wire with the tweezers and then wiggling it back and forth until the wire breaks at the foot of the bond on the detector. In the documentation for the detector, record all shorted pairs and the bond from each pair that was removed.

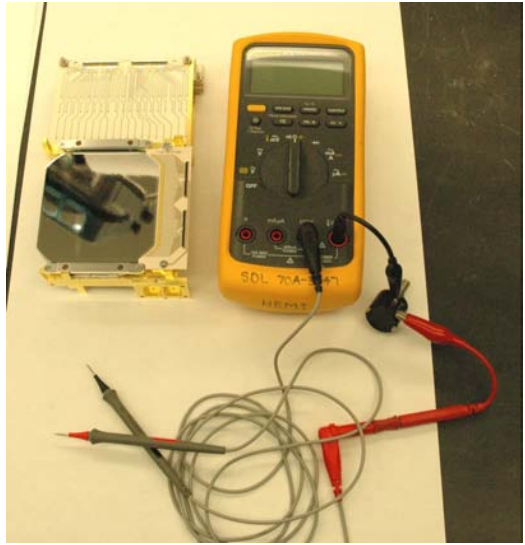

**Figure 3.26** Setup used to identify inter-strip shorts on the GRIPS detectors. A continuity check at room temperature is used to identify the shorts. For convenience, the audio alert feature of the meter is turned on so that the operator only needs to visually focus on probing the neighboring circuit board traces and not on the meter readout. A ten-turn potentiometer is added in series with the meter for the case when the amorphous semiconductor layer is not high enough in resistivity at room temperature to prevent the meter audio alert from sounding.

## 3.10 Detector reprocessing

Detectors do not always function properly despite the use of high quality HPGe crystals and a robust fabrication process. Typical causes for less than desirable performance include inadequate mechanical and/or chemical processing of the crystal or a defect introduced at one of the subsequent processing steps. When a detector fails electrical testing or does not meet performance expectations, it should be reprocessed. A detector may also need to be reprocessed if it has first been made into a simple planar detector (such as the simple guard ring detector of Figure 3.16) and now must be converted into a strip detector. The procedure provided below can be used to convert the detector into a bare HPGe crystal that will be in an appropriate state for the reapplication of the fabrication procedures described earlier in this paper.

**(a) Remove and prepare detector circuit boards for reuse:** Remove the detector circuit boards from the detector assembly. Pull off the Al bonding wires still attached to the boards with a pair of tweezers. Clean the board traces with methanol and a cotton swab. Rinse the boards with methanol and then blow them dry.

**(b) Prepare crystal for etching:** Pull off any Al bonding wires still attached to the HPGe crystal with tweezers. Remove the crystal from the holder using the PTFE detector transfer fixture. Clean off any In remaining on the crystal wings using a razor blade, wooden stick end of a cotton swab, and/or a cotton swab (Figure 3.27(a)).

**(c) Etch off Al electrodes and remaining Al bonding wire:** Hold onto the crystal with the etch clamp and submerge the crystal in 1% HF (Figure 3.27(b)). Periodically pull the crystal out of the acid and resubmerge it to remove bubbles that form on the crystal. After the crystal has soaked for at least 1 hour, rinse it in deionized $H_2O$ and then blow it dry. This long etch is required to ensure that the thick bonding wire is completely etched away. If this is not done, any remaining Al could mask the subsequent Ge etch.

**(d) Etch off a-Si if necessary:** Amorphous Si can be resistant to the concentrated 4:1 $HNO_3$:HF etchant used for the a-Ge and Ge etching. This is particularly true of a-Si sputtered at higher gas pressures. If the a-Si is not removed prior to the 4:1 etch, a rough and poorly etched surface can result. If the detector has any a-Si coatings, remove them with alternating 2 min soaks in concentrated HF and KOH solution (50 g KOH in 100 mL deionized $H_2O$). Rinse in deionized $H_2O$ when switching between the two soaks. When the a-Si is visibly gone, rinse the crystal in deionized $H_2O$ and then blow it dry.

**(e) Etch off amorphous semiconductor layer and any underlying damage to crystal:** Follow the surface preparation etch procedure described previously in section *3.5 Crystal surface preparation etching*, except in this case a methanol rinse is not necessary, and the crystal can be placed on filter paper when it is blown dry. If the crystal has significant surface damage, the polish etch procedure of section *3.4 Crystal chemical polish etching* should be followed except that the etch time can be reduced depending on the extent of the damage.

**(f) Inspect crystal:** The crystal surfaces should be inspected (using a microscope if necessary) to make sure all scratches, pits, etc., have been etched smooth. If any damage remains on the active detector surfaces, the crystal should be re-etched. If the crystal is free of defects, it is ready to be fabricated into a detector starting at the *3.5 Crystal surface preparation etching* step.

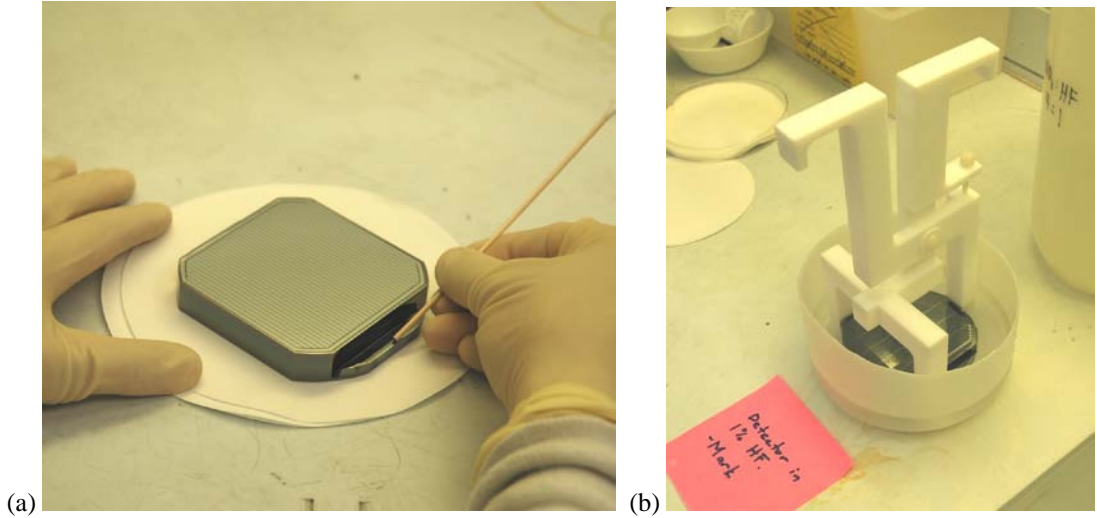

(a)                                                             (b)

**Figure 3.27** Photographs of some steps done during detector reprocessing. **(a)** Removal of In foil that has adhered itself to the handles of the HPGe crystal. **(b)** Al electrode and bonding wire removal in a 1% HF soak.

# 4. Detector evaluation and troubleshooting

As a HPGe crystal is converted into a detector through the procedures presented above, it is natural to note and lament over visible defects that appear on the crystal during the conversion sequence. These can include smoothed out lapping scratches, surface haze or drying marks from the chemical etching, or cloudiness in the Al electrodes. These may appear to be problems but instead are often only cosmetic defects. What ultimately matters is the performance of the detector. In this section, the procedures for evaluating a completed detector are provided. This detector testing is primarily intended to be a check of basic functionality and not a detailed evaluation. The detailed testing is often application specific and is typically done by the end-user after the detector has been installed in the scientific instrument. These instruments contain all of the electronics and data processing needed to fully exercise the detector.

Two general types of detectors need to be tested. The first consists of the HPGe crystal with a simple electrode configuration of either full area electrodes on both crystal faces or a full area electrode on one face and a guard ring electrode structure on the other face (see Figure 4.1 for a drawing of a guard ring detector). As explained earlier in this paper, it is advantageous to convert the crystal into one of these simple detectors prior to using the crystal for a strip detector. This allows for a more efficient determination of the crystal properties and the suitability of a particular fabrication process. Furthermore, a simple planar detector is produced as an intermediate step in the photolithography method, and it is beneficial to test the detector at that point in order to determine if the processing should be restarted due to poor electrical performance. Both electrical and spectral testing are done with these simple detector configurations. The electrical testing consists of the measurement of detector capacitance and leakage current as a function of applied detector high voltage. The capacitance measurement is used to determine the voltage at which the detector becomes fully depleted, which in turn dictates the suitable operating voltage for the detector. The leakage current measurement provides an indication as to whether or not the crystal and fabrication process will be capable of meeting the leakage current and energy resolution specifications. The leakage current of a detector must be low enough so that the detector energy resolution is not significantly impacted. The spectral testing entails the operation of the detector as a gamma-ray spectrometer when exposed to a variety of standard radioisotope based gamma-ray sources. These measurements can be used to determine the typeness of the crystal and, to some extent, the quality of its charge transport.

The other type of detector that must be evaluated is the one that is the end product of this work and consists of the HPGe crystal with strips and a perimeter guard ring on both crystal faces (see, for example, Figure 1.1). As with the simple detectors, the strip detectors are first evaluated through electrical measurements. Typically, the strips on each side of the detector are tied together, and their summed leakage current and that on the guard ring are measured as a function of the applied detector voltage. The measurement of the detector capacitance is normally not done with the strip configuration since the crystal would typically have previously been characterized as a simple detector, and a capacitance versus voltage characteristic would have been acquired at that time. Following a successful electrical evaluation, select strips on the detector are connected to pulse processing electronics, and energy spectra are acquired with the detector exposed to a variety of standard radioisotope based gamma-ray sources.

To perform the testing just described, the detectors must be cooled and are typically cooled to a temperature near that of the boiling point of $N_2$. To accomplish this, two different custom cryostats are used. One is configured for the simple detectors and the COSI strip detectors, and the other one is set up for the GRIPS strip detectors. The two cryostats are of a similar design and make use of components designed and fabricated many decades ago at LBNL [Miner 1967]. Each consists of a rectangular vacuum chamber with two large area flanges providing access to both faces of the installed detector (for example, see Figure 4.2). Inside each vacuum chamber is a hollow cold finger that is filled with liquid $N_2$ during detector operation. The liquid $N_2$ is supplied by a removable gravity feed dewar. Such a design allows the detectors to be quickly cooled and warmed back up, which is a valuable attribute to have in a detector production test setup. The vacuum in the cryostats is achieved and maintained through continuous pumping with a turbopump. All of the electronics needed to test and operate the detectors are located outside of the vacuum enclosures, and electrical connections between these electronics and the detectors are made through both custom and commercial electrical feedthroughs integrated into the vacuum chamber walls. The electronics include standard commercial NIM based and stand-alone modules from Ortec [Ortec 2020] and Mirion Technologies (Canberra) [Mirion 2020] for the high voltage supplies, shaping amplifiers, pulsers, and multichannel analyzers. The preamplifiers used with the detectors are custom, low power, discrete component based units similar in design to that of the MaFaYa preamplifier [Fabris 1999]. The leakage current measurements are made using either picoammeters from Keithley such as the model 6485 [Keithley 2020] or transimpedance amplifiers from Portable Technologies (model PTA-100) [Portable Technologies 2020].

To conclude this detector evaluation introduction, some precautions will be given. When testing a detector, a simple mistake can damage the detector and destroy months of work. To help avoid this, the instructions provided below should be followed.

(1) All of the equipment needed to operate the detector should be thoroughly checked for functionality prior to use with the detector. This includes the cryostat, vacuum pump, liquid $N_2$ dewar, and electronics. Of particular importance is the high voltage system of the electronics, which includes the high voltage supply, high voltage cabling, and high voltage filters. If not designed and tested correctly, the high voltage system could be susceptible to electrical breakdown events. If such events occur during detector testing, they are difficult to distinguish from detector breakdown failure and, therefore, can complicate detector debugging. Furthermore, high voltage breakdown within the high voltage system can cause damage to and failure of the detector. Essentially, a good detector can be damaged by faulty testing hardware, and, if this does occur, it would be difficult to determine that the detector was originally good.

(2) Meticulous housekeeping should be maintained throughout the detector testing. All cabling should be neatly arranged prior to powering the electronics. The high voltage cabling must not be disconnected and should not be rearranged after the high voltage has been applied to the detector. Inadvertently breaking the high voltage connection with the high voltage applied can result in costly damage to the detector and the readout electronics.

(3) All electrical connections to the detector should be checked. High voltage should not be applied to the detector unless all electrodes of the detector are either grounded or connected to electronics (signal readout electronics, picoammeters, etc.) that are powered. This is particularly important for the high voltage side of the strip detectors, which will be discussed further in the section concerning the testing of these detectors.

(4) After all of the electrical connections are properly made and the readout and testing electronics are powered, the detector high voltage can slowly be applied. This can be done by slowly increasing the voltage in steps of no more than 100 V and then waiting at least 1 min between each step. The removal of the high voltage must also be done using this slow adjustment process. It is important not to switch off the high voltage supply power or disconnect the high voltage cable unless the detector voltage is at zero.

The procedures for evaluating the detectors are provided below in two separate sections. The first section covers the simple planar detectors and the other the strip detectors. A final section provides help on troubleshooting a poorly performing detector.

## 4.1 Evaluation of full area electrode and guard ring detectors

The assumed starting point for this testing procedure is that the completed detector has been loaded into a holder and spring wire contacts have been added. Each spring wire contact is fixed to an insulator (PTFE or Rogers RO4003 circuit board) that is screwed down onto the detector holder. One end of each spring wire is terminated with an In ball that makes a gentle electrical connection to one of the detector electrodes. The other end of each spring wire is eventually solder connected to the wiring inside the test cryostat. The electrical connections and test setup for the detector testing are shown in the schematic diagram of Figure 4.1.

In addition to the instructions provided below, the paper [Amman 2018] is a source of information on the testing of simple HPGe detectors.

**(a) Load detector into COSI test cryostat:** A sequence of photographs illustrating the loading of a guard ring detector into the COSI cryostat is provided in Figure 4.2. All simple planar detectors are tested in this cryostat. To load a detector, first, remove both vacuum plates on the cryostat. Since the flexcircuits will not be used in the testing of a simple detector, loop each back towards the feedthrough plate and tie it with wire to keep it out of the way. There should be a layer of In foil on the L-shaped cold bracket covering the area that the detector holder will contact. If this In foil is missing or damaged, replace it. Carefully place the detector assembly onto the cold bracket without handling or making contact with the detector itself (Figures 4.2(a) and 4.2(b)). The assembly should only be handled by the holder. Screw the assembly to the cold bracket. Attach the IR shield associated with the bottom detector face (Figure 4.2(c)). Note that the detector could be tested without an IR shield, but a good measure of the detector leakage current would not be obtained. Without the shield, the IR radiation impinging on the detector would lead to leakage currents in excess of ~ 1 nA. The bottom shield contains an electrically isolated high voltage wire. With the shield installed, solder the interior end of this wire to the spring wire contacting the top face of the detector. Great care should be taken when performing this task since the detector is exposed. The detector should be protected from the spatter of solder or solder flux as well as from the solder flux vapor. A piece of filter paper temporarily placed near the detector can be used as a shield. With this electrical connection made, the top IR shield can be attached and screwed into place. With the detector now fully covered by the IR shields, solder connections should be made between the wires from the detector assembly (which should now be passing through the IR shields) and the cryostat wiring (Figure 4.2(d)). At this point, a digital multimeter can be used to confirm that the electrical connections have been properly made through resistance checks. The wiring should also be checked for unwanted shorts to ground. If all looks good, the vacuum sealing o-rings should be wiped clean and then the vacuum plates reinstalled.

**(b) Pump out cryostat:** Attach a dry pump backed turbopump to the vacuum pumpout port of the cryostat. Check that the cryostat pumpout valve is open. Pump on the cryostat until a pressure in the $10^{-6}$ Torr range (as measured near the pump) is attained. This should be achieved within a day. If not, a vacuum leak may exist and should be investigated. If all is good, the pump will typically be left on to actively pump the cryostat for the duration of the detector testing.

**(c) Cool the detector:** Once the cryostat has been pumped out, add a gravity feed dewar to the cryostat and partially fill the dewar with liquid $N_2$ (~ 8 L is sufficient for measurements that will be completed within a day). Wait approximately 3 hours for the detector to cool. A thermometer attached to the cold bracket within the cryostat can be monitored to determine when the detector is cold. However, keep in mind that there is some time lag (~ 1 hour) and temperature difference between the cold bracket temperature and that of the detector. A temperature near 80 K should eventually be achieved.

**(d) Make electrical measurements:** Wire the detector as shown in the Figure 4.1 circuit diagram. The passive electrical components connected to the top, high voltage face of the detector (side facing down in the figure) form a high voltage filter that is contained inside an electrical box attached to the COSI cryostat. This filter is designed so that a step voltage pulse can be applied across the detector for the measurement of the detector capacitance. A standard NIM based pulser provides the voltage step signal for this purpose. Attached to the opposing face of the detector are the circuits for leakage current and induced charge signal pulse measurements. The passive components and preamplifier are housed in a preamplifier electronics box attached to the cryostat. The picoammeters and pulse processing electronics are commercially procured units as described previously in the introduction to this section. After making the necessary connections, it should be confirmed that the polarity of the high voltage supply is set as desired and that the voltage is set to zero prior to powering the electronics. The polarity should be chosen as appropriate for the detector design. Both the COSI and GRIPS detectors typically operate with a positive high voltage applied to their top faces. To proceed with the measurements, set the detector bias $V_d$ to 50 V. Turn the shaping amplifier (shaper) to a low gain and the pulser to the highest amplitude allowed that does not produce a saturated pulse out of the shaper. Record the leakage currents $I_{gr}$ and $I_c$ and the pulse height of the pulses from the shaper $V_p$. Note that $V_p$ is directly proportional to the detector capacitance and serves as a measure of its value. Slowly adjust the detector high voltage $V_d$ in steps of about 100 V. At each step setting, wait approximately 3 min for the current readings to settle. Record $I_{gr}$, $I_c$, and $V_p$. Once $V_p$ no longer changes with increasing $V_d$, full depletion has been reached. Continue the measurements until $V_d$ is at least 500 V above full depletion and typically to at least about 1500 V, but do not go above 2000 V. Components of the high voltage system (such as the high voltage capacitors) are not rated for use above 2000 V. If at any point the leakage currents begin to rise rapidly or the signal out of the shaper becomes noisy, stop increasing $V_d$. If either of these conditions occurs, the detector will likely need to be reprocessed. Example measurements obtained from a properly functioning detector with full area electrodes covering both the top and bottom faces are plotted in Figure 4.3. In this example, the detector fully depletes at about 600 V. Since this detector is about 15 mm thick, a 600 V depletion voltage indicates that the average net impurity concentration of the HPGe crystal is about 5 x $10^9$ cm$^{-3}$ [Amman 2018].

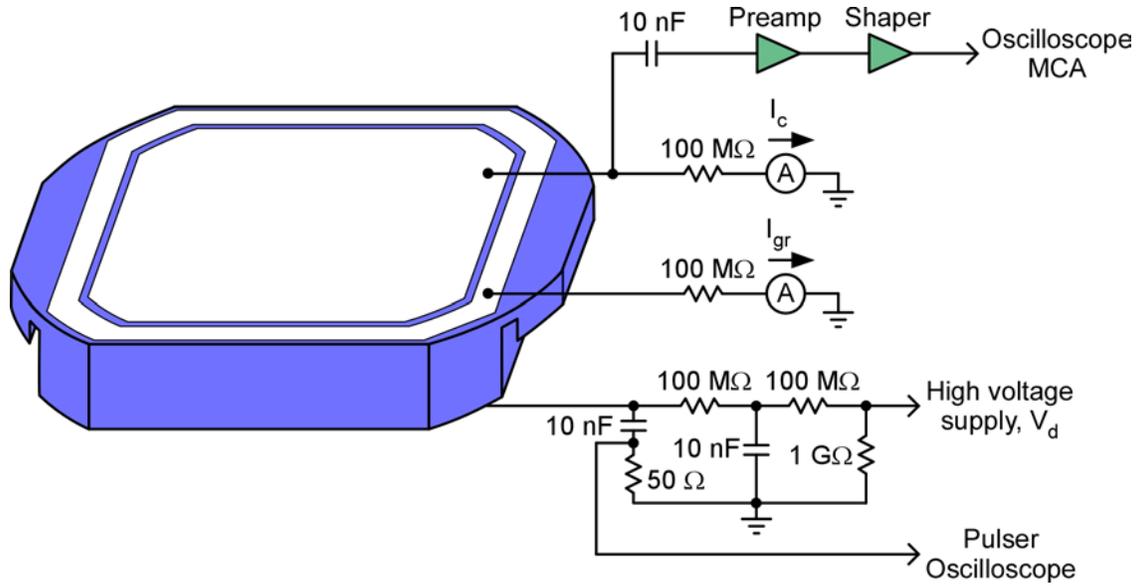

**Figure 4.1** Schematic circuit diagram of the setup used for the testing of a guard ring detector. The detector is shown with its bottom facing upward. The diagram for a simple full area electrode planar detector is similar with the only difference being that there is no guard ring and associated current measurement electronics.

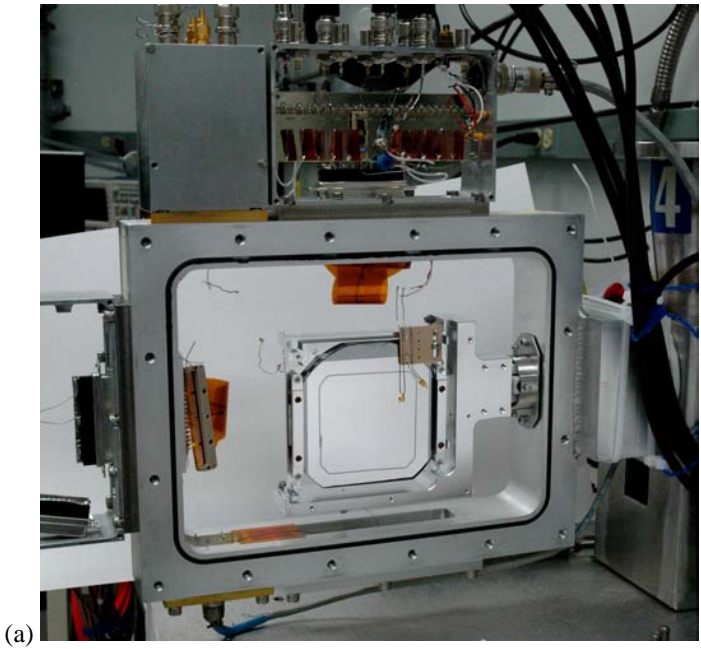

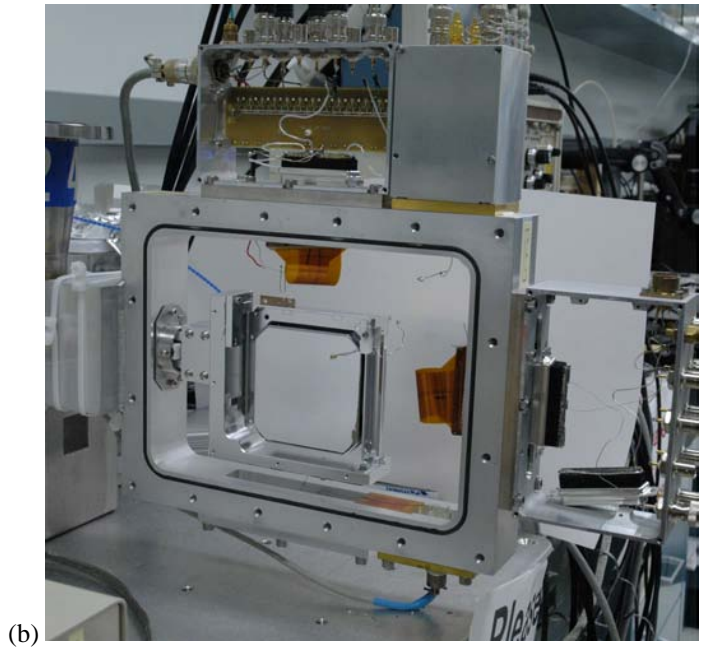

(a)                                                                     (b)

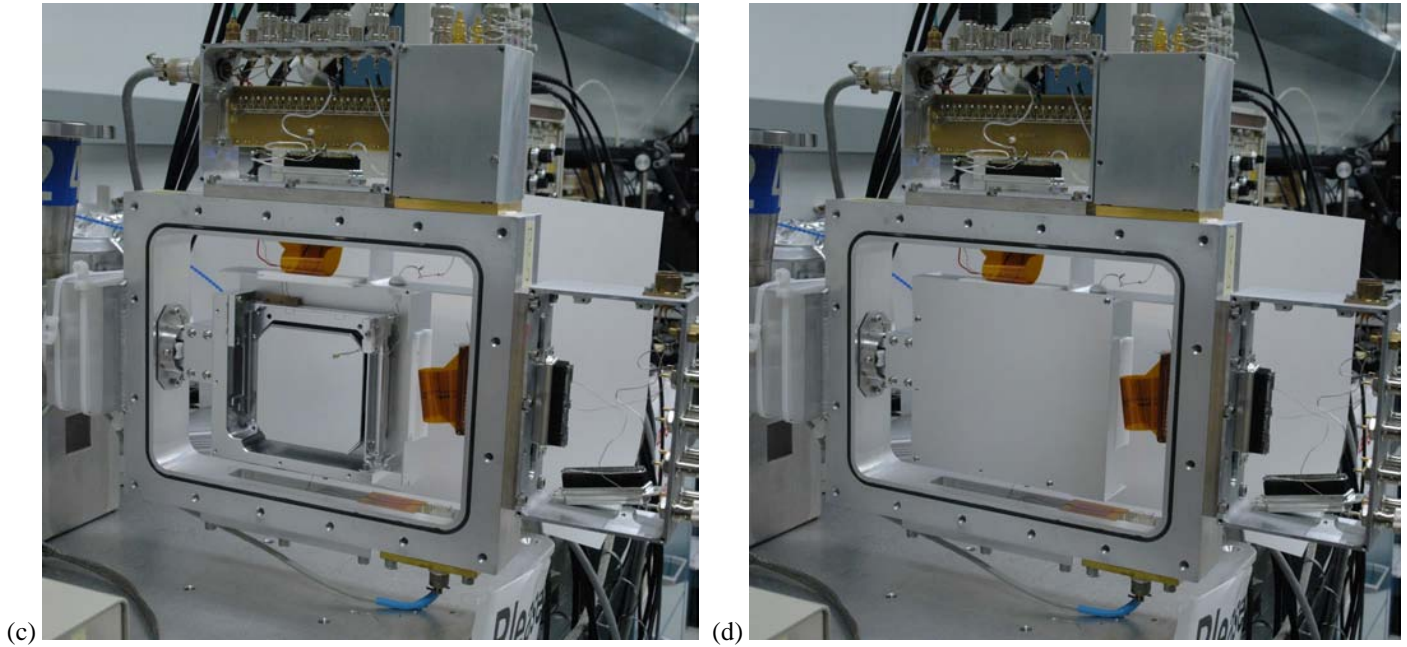

(c)          (d)

**Figure 4.2** Sequence of photographs showing the loading of a guard ring detector into the COSI test cryostat. **(a)** View of the bottom face of the detector after the detector assembly has been attached to the cryostat cold bracket. Flexcircuits can be seen within the cryostat chamber but are not used with the simple planar detectors. These flexcircuits are instead used for the testing and operation of the COSI strip detectors. **(b)** View of the top of the detector. **(c)** View of the top of the detector after the bottom IR shield has been added and the electrical connections to the detector have been completed. **(d)** View of the cryostat after all electrical connections have been made and the IR shields have been completely installed.

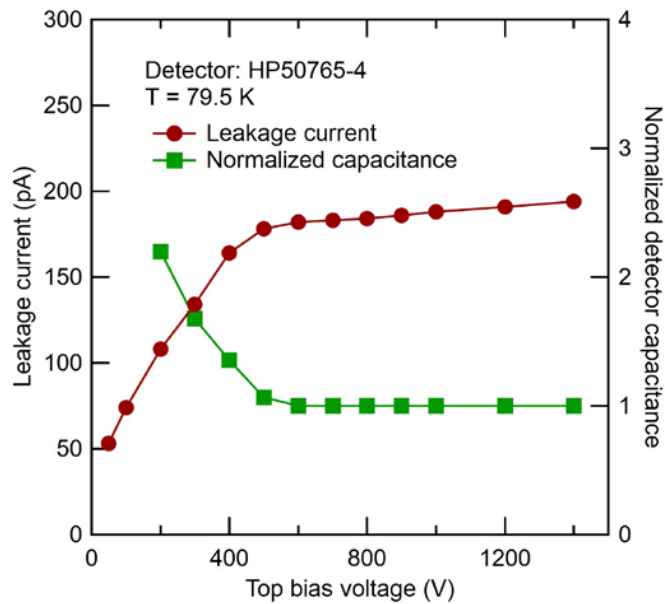

**Figure 4.3** Example electrical measurements from a simple planar detector with full area electrodes. The plotted data consist of the leakage current and detector capacitance measured as a function of the detector voltage. The capacitance has been normalized to the value measured at full depletion. The measurements were made using a setup similar to that of Figure 4.1 (there was no guard ring, so the guard ring current measurement channel was not necessary). The starting HPGe crystal used for the detector was of the typical dimensions for the COSI and GRIPS detectors.

**(e) Make gamma-ray spectroscopy measurements:** These measurements would normally be made immediately following the electrical testing, so the detector should already be loaded into the cryostat and cold. If this is not the case, follow the procedures described above for loading and cooling the detector. Wire the detector as shown in Figure 4.1 except do not connect the picoammeters. Instead, the 100 MΩ resistor attached to the center electrode and the preamplifier should be connected to ground. The guard ring (if the detector has one) should be directly connected to ground (bypassing its 100 MΩ resistor). At this point, the typeness of the HPGe of the detector can be determined with an Am-241 gamma-ray source. This is done by setting the detector voltage bias well below its full depletion voltage (for example, 100 V for the detector of Figure 4.3) and then observing the detected event count rate when the top and then the bottom faces of the detector are illuminated by the source. Since the Am-241 source emits low energy gamma rays (60 keV) and since gamma rays that interact in the undepleted Ge do not contribute to the count rate, the source position resulting in the higher count rate indicates the side of the detector that is depleted. If, for example, a higher count rate is observed when the top is illuminated and a positive bias voltage is applied to the top, then it can be concluded that the depletion starts at the top, and, since the applied voltage is positive, the HPGe must be p-type.

The quality of the charge collection in the detector can also be roughly assessed with gamma-ray spectroscopy measurements. For this task, a normal operating voltage is applied to the detector, and higher energy gamma-ray sources such as Ba-133, Cs-137, or Co-60 are used to illuminate the detector. Ideally, it is desirable to apply a voltage that is about 1000 V above the full depletion value so that the charge carriers travel at near their saturation velocities. However, to achieve a higher detector yield, the general practice for these detectors is to set the voltage to at least 1000 V and at least about 500 V above the full depletion value. With the electronics powered, the high voltage set at zero, and no sources near the detector, the signal out of the shaping amplifier should be inspected using the oscilloscope. Any non-white noise present on the signal should be debugged and eliminated by modifying the shielding, grounding, filtering, etc. With this done, the detector voltage should be slowly increased to the desired operating point. While this is being done, the shaper signal should be monitored for any appearance of non-white noise. Any significant increase in the noise likely indicates a problem with the detector or the high voltage system. With the detector at its operating voltage, place a single gamma-ray source near the cryostat wall and accumulate a pulse height (energy) spectrum with the multichannel analyzer (MCA). Note that the gain of the shaping amplifier will need to be adjusted from that used previously for the detector capacitance measurement. This gain should be set to place the spectral peaks of the source appropriately within the MCA channel range. It is also important to optimize the pole-zero setting of the shaping amplifier. Once the energy spectrum has been accumulated long enough for good statistics to be obtained in the spectral peaks, the acquisition can be stopped and the source removed. At this point, it is common to add a pulser peak to the accumulated spectrum. The width of such a peak is a measure of the electronic noise. To add a pulser peak, set the pulser amplitude so that it will produce a peak in the accumulated spectrum that is away from the existing spectral peaks. Restart the acquisition and then stop it once a peak with sufficient counts has been obtained. The measurement procedure can then be repeated for other gamma-ray sources. From these energy spectra, the quality of the HPGe material can roughly be determined. The spectral peaks should be sharp and have no low energy tailing. Reasons for energy peaks being broader than expected include excess detector noise (faulty fabrication) and charge trapping. A noisy detector would produce a broad pulser peak. Excessive charge trapping is normally identifiable by low energy tailing on the spectral peaks. Note that due to the large capacitance of the simple planar detector, the electronic noise is somewhat high in even a perfectly made and properly functioning detector. For this reason, the quality of the charge transport in the HPGe material can only roughly be assessed. A more accurate characterization must wait for the testing of the HPGe crystal as a strip detector.

## 4.2 Evaluation of strip detectors

The assumed starting point for this testing procedure is that the completed strip detector has been loaded into a holder, the circuit boards for interfacing between the detector and the cryostat flexcircuits have been attached to the holder, and wire bonds have been made between the circuit boards and the detector. This completed detector assembly should also have been visually inspected for defects and electrical connections to the detector confirmed through continuity checks between circuit board traces. The electrical connections and test setup for the detector testing to be performed are shown in the schematic diagrams of Figure 4.4.

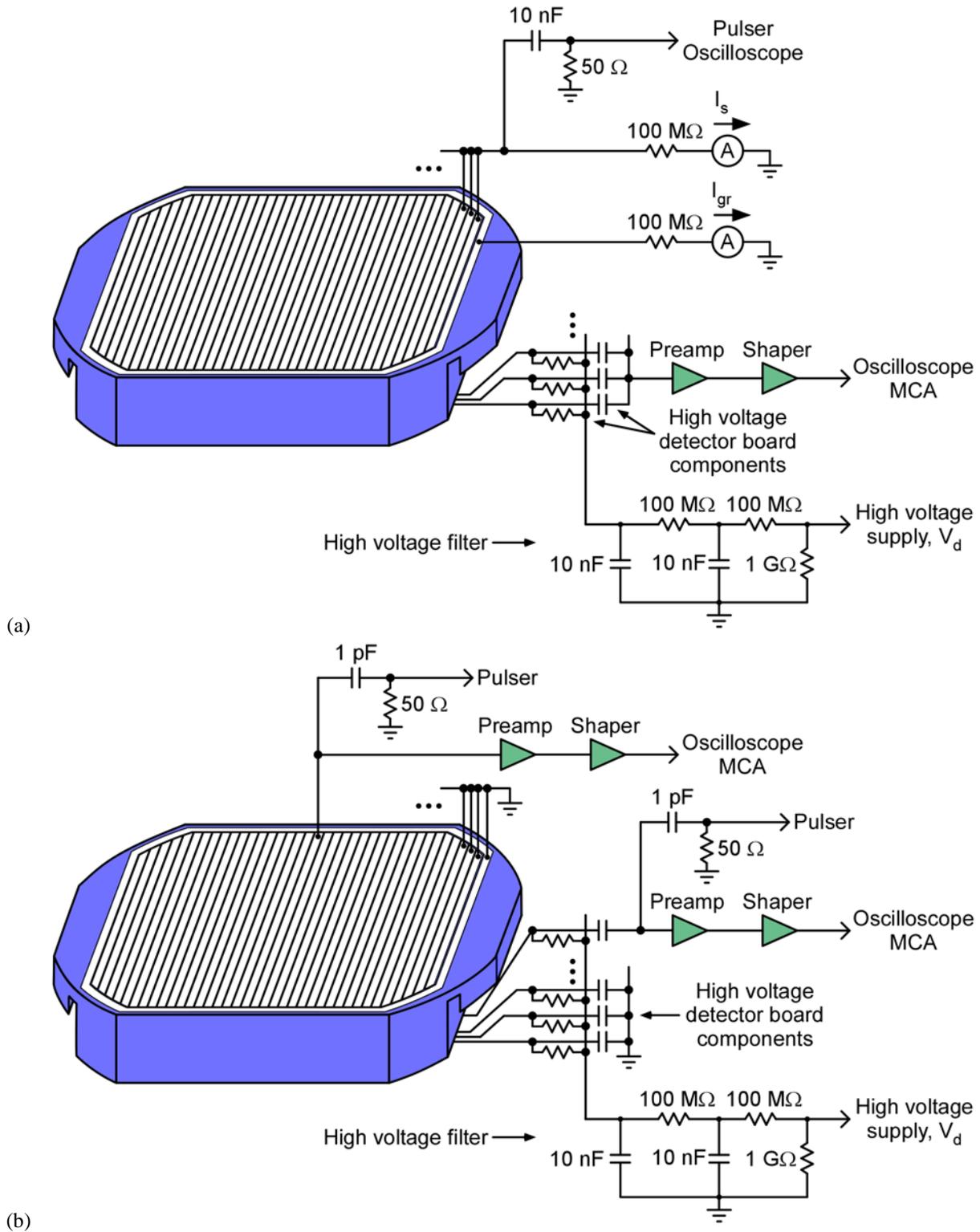

**Figure 4.4** Schematic circuit diagrams of the setups used for testing COSI and GRIPS strip detectors. **(a)** Setup for the electrical measurements. For the measurement of the leakage currents, all strips on the bottom detector face are interconnected and then tied to a single picoammeter, thereby giving the summed strip leakage. The guard ring is connected to a second picoammeter in order to separately measure its leakage current. **(b)** Setup for the spectroscopy measurements. When the spectroscopy measurements are made, one strip from each face of the detector is connected to a separate chain of pulse processing electronics. All other detector channels are connected to ground.

In the introduction to this detector evaluation section, a list of precautions that should be followed when testing a detector was provided. For a COSI or GRIPS strip detector, it is worthwhile to further emphasize the importance of making proper electrical connections to the detector assembly before applying high voltage to the detector. Specifically, all channels from the high voltage detector board must be either grounded or connected to electronics (signal readout electronics, picoammeters, etc.) that are powered. Otherwise, the detector could be damaged during testing. To see why this is critically important, consider the situation depicted in the circuit diagram of Figure 4.5. The diagram is of two strips on the high voltage side of a detector along with the biasing and signal output circuitry. A high voltage of 1000 V is assumed to be applied to the strips through the 1 GΩ resistors on the detector board. Strip 2 is wired properly with the output connected to ground (either directly or through a preamplifier), while the output from strip 1 has been left disconnected. The voltage at point B on the output side of strip 2's coupling capacitor is at 0 V as it should be, but the voltage at point A on the strip 1 channel will have come up to near 1000 V. To see that this is true, consider the circuit path from the high voltage source through the 1 GΩ resistor for strip 1, the 10 nF coupling capacitor for strip 1, the small ~ 10 pF parasitic capacitance between points A and B, and then to ground. Assuming little to no current flow (at least initially), there will be almost no voltage drop across the 1 GΩ resistor. That means the applied 1000 V will be shared between the series connected 10 nF and ~ 10 pF capacitors. Due to the orders of magnitude difference in capacitance values, the ~ 10 pF capacitor will have almost all of 1000 V dropped across it. Therefore, point A will be near 1000 V in potential, and that is a problem. The high voltage detector circuit boards, flexcircuits, and connectors were not designed to hold off that high of a voltage difference between adjacent channels. Because of this, there will be a momentary high voltage breakdown current flow between points A and B. That will bring the potential at A down to a low value (for simplicity, assume 0 V). The potential at strip 1 will follow this and drop to near 0 V since the 1 GΩ resistor prevents the high voltage source from rapidly supplying the current necessary to maintain the voltage at a high value. Strip 1 is now near 0 V and strip 2 at 1000 V. This voltage difference between adjacent strips will cause a high voltage breakdown current flow between the two that will likely damage both the strip metallization and the underlying amorphous semiconductor layer. A cascade of inter-strip breakdown events will then proceed to take place across the face of the detector. As a result, the detector will become damaged and will likely be leaky.

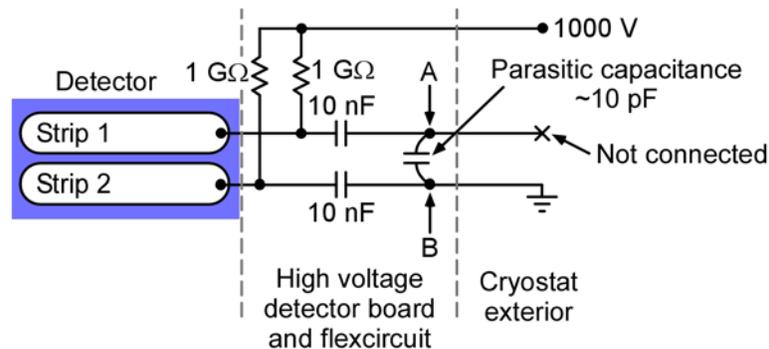

**Figure 4.5** Schematic circuit diagram used to explain the problem of leaving a connection to the high voltage detector board floating when high voltage is applied to the detector board.

The step-by-step procedures for evaluating COSI and GRIPS strip detectors are provided below.

**(a) Load detector into test cryostat:** The strip detector must first be loaded into the appropriate cryostat. COSI detectors are tested in the COSI test cryostat, which is the same cryostat as that used for the evaluation of the simple planar detectors. The GRIPS detectors are tested in a separate though similar cryostat that is dedicated to that purpose. Photographs of the loading sequence for a COSI detector are provided in Figure 4.6 and those for a GRIPS detector in Figure 4.7. To load the detector, the procedure described previously for the simple planar detectors should be followed except that the flexcircuits in the cryostat are used to make the electrical connections to the strip and guard ring channels of the detector assembly. After the detector assembly has been attached to the cold bracket in the cryostat, the flexcircuits should be connected to their corresponding detector boards (Figures 4.6(a) and 4.7(a)). The COSI detectors have a single flexcircuit for each detector face, and the GRIPS detectors have four for each face. The COSI detectors make use of a custom connector assembly (attached to the end of each flexcircuit) that must be screwed to the detector boards in order to make the flexcircuit connections. The GRIPS detectors, however, instead rely on commercial connectors soldered to the detector boards as the means for the connections. With the attachment of the flexcircuits completed, the electrical connection to each channel of the detector should be confirmed by probing with a digital multimeter. The bottom connections can be confirmed with a simple continuity measurement. The connections to the top (high voltage) face of the detector, however, may require a capacitance measurement depending on whether or not the high voltage coupling capacitors on the high voltage detector board are part of the circuit being probed. As noted previously, it is critically important to verify all connections to the high voltage detector board. After a successful verification, attach the bottom IR shield and carefully solder the high voltage wire from the shield to the high voltage trace or wire stub on the high voltage detector board (Figure 4.6(b)). Attach the top IR shield and then solder the IR shield high voltage wire to the cryostat high voltage wire (Figures 4.6(c) and 4.7(b)). Note that the COSI IR shield fully covers the detector, but the one for the GRIPS detectors is only a partial shield. This means that the leakage currents measured from the GRIPS detectors will partially consist of that caused by the absorption of IR radiation in the detector. If all looks good with the detector installation, the vacuum sealing o-rings of the cryostat should be wiped clean and then the vacuum plates reinstalled (Figures 4.6(d) and 4.7(c)).

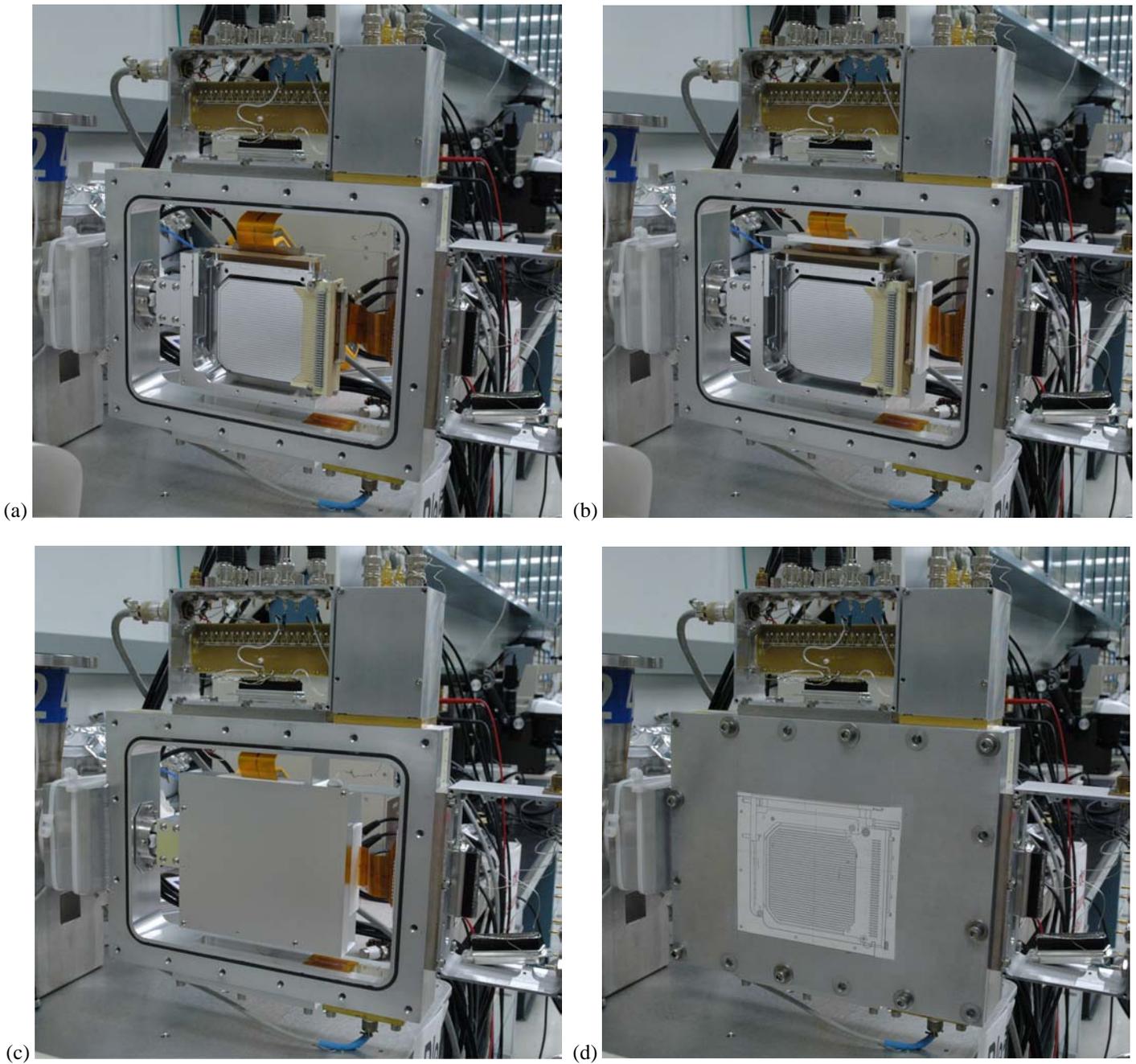

(a)

(b)

(c)

(d)

**Figure 4.6** Sequence of photographs showing the loading of a COSI strip detector into the COSI test cryostat. **(a)** View of the top face of the detector after the electrical connections to the strips and guard rings have been made through the attachment of the flexcircuits. **(b)** View of the top of the detector after the bottom IR shield has been added. **(c)** View of the cryostat after all electrical connections have been made and the IR shields have been completely installed. **(d)** View of the closed cryostat.

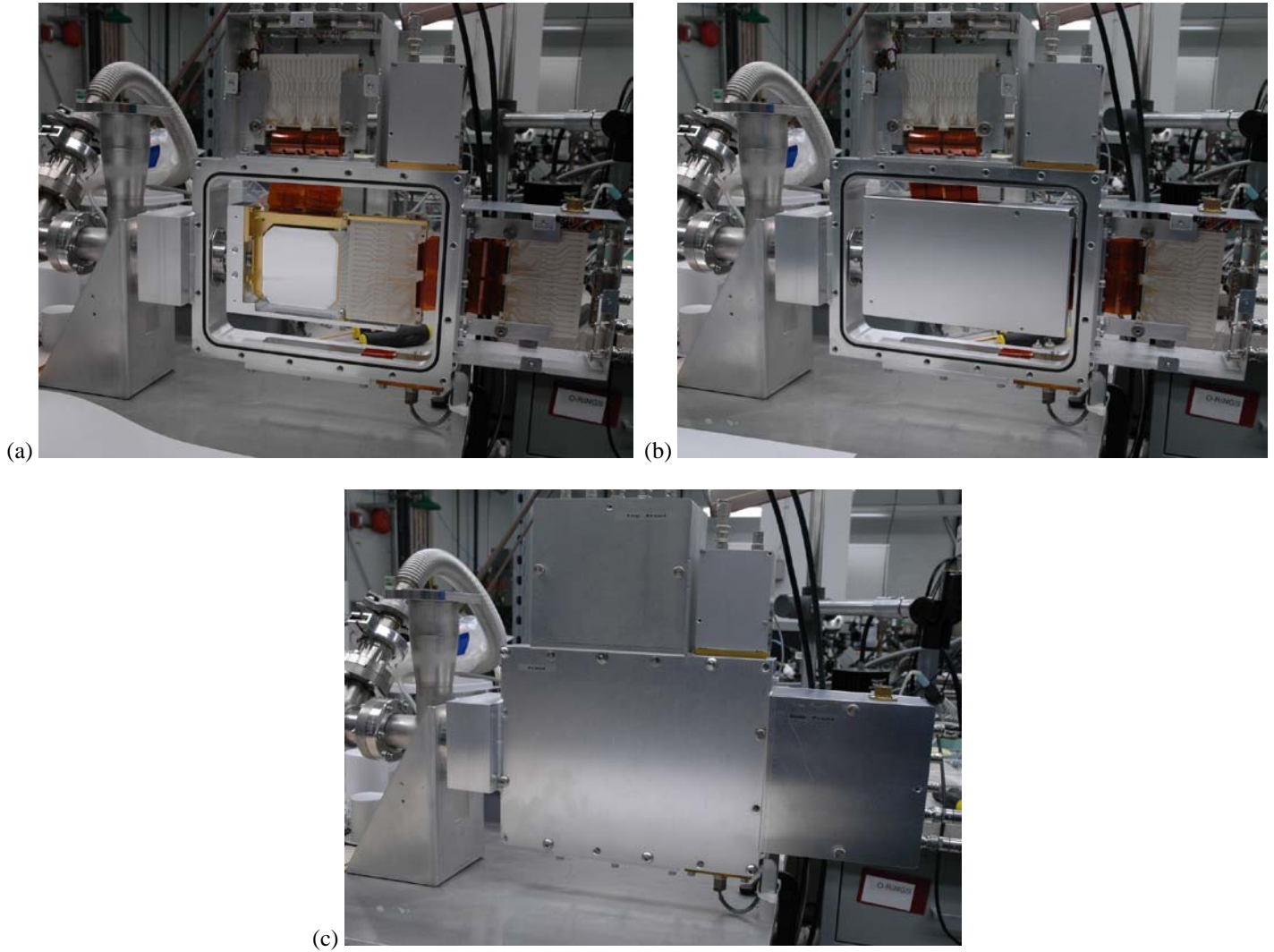

**Figure 4.7** Sequence of photographs showing the loading of a GRIPS strip detector into the GRIPS test cryostat. **(a)** View of the top face of the detector after the electrical connections to the strips and guard rings have been made through the attachment of the flexcircuits. **(b)** View of the cryostat after all electrical connections have been made and the IR shields have been completely installed. **(c)** View of the closed cryostat

**(b) Pump out cryostat:** Follow the pump out procedure described previously in the section on the evaluation of simple planar detectors.

**(c) Cool the detector:** Follow the detector cooling procedure described previously in the section on the evaluation of simple planar detectors. Both the COSI cryostat and the GRIPS cryostat make use of the gravity feed dewars to supply liquid $N_2$ to the cryostat cold fingers. The cold finger temperature should be monitored to determine when the detector is ready for testing. A temperature near 80 K should eventually be achieved.

**(d) Make electrical measurements:** Wire the detector as shown in the Figure 4.4(a) circuit diagram. For these measurements, the strips on the bottom detector face (strips facing up in the figure) are interconnected and tied to a picoammeter for the measurement of the summed strip leakage current. The bottom guard ring is connected to a second picoammeter for the measurement of its leakage current. The high voltage connection to the top of the detector is made through a high voltage filter (contained in a shielded box attached to each cryostat) that was wired previously during the detector loading step to the 1 G$\Omega$ bias resistors on the high voltage detector board. The signal output channels from the top of the detector are interconnected and then tied to a chain of pulse processing electronics if the detector capacitance is to be measured or tied to ground if only the leakage currents will be measured. For the capacitance measurement, a step voltage pulse is applied to the bottom of the detector through a large value capacitor (~ 10 nF). With a strip detector, it is normally unnecessary to measure capacitance as part of the electrical characterization since such a measurement would likely have already been done with the HPGe crystal when fabricated previously into a simple planar detector. The crystal properties of net impurity concentration (and full depletion voltage) and material type should already be known. To proceed with the electrical testing, follow the procedure given previously in the section on the evaluation of simple planar detectors. Acquire the summed strip leakage current $I_s$, guard ring leakage current $I_{gr}$, and shaping amplifier output pulse height $V_p$ (if the detector capacitance is desired) as a function of the detector high voltage $V_d$. As before, these measurements should be done up to a value of $V_d$ that is at least 500 V above full depletion but not above 2000 V. Example electrical characteristics obtained from properly functioning COSI and GRIPS detectors are shown in Figure 4.8.

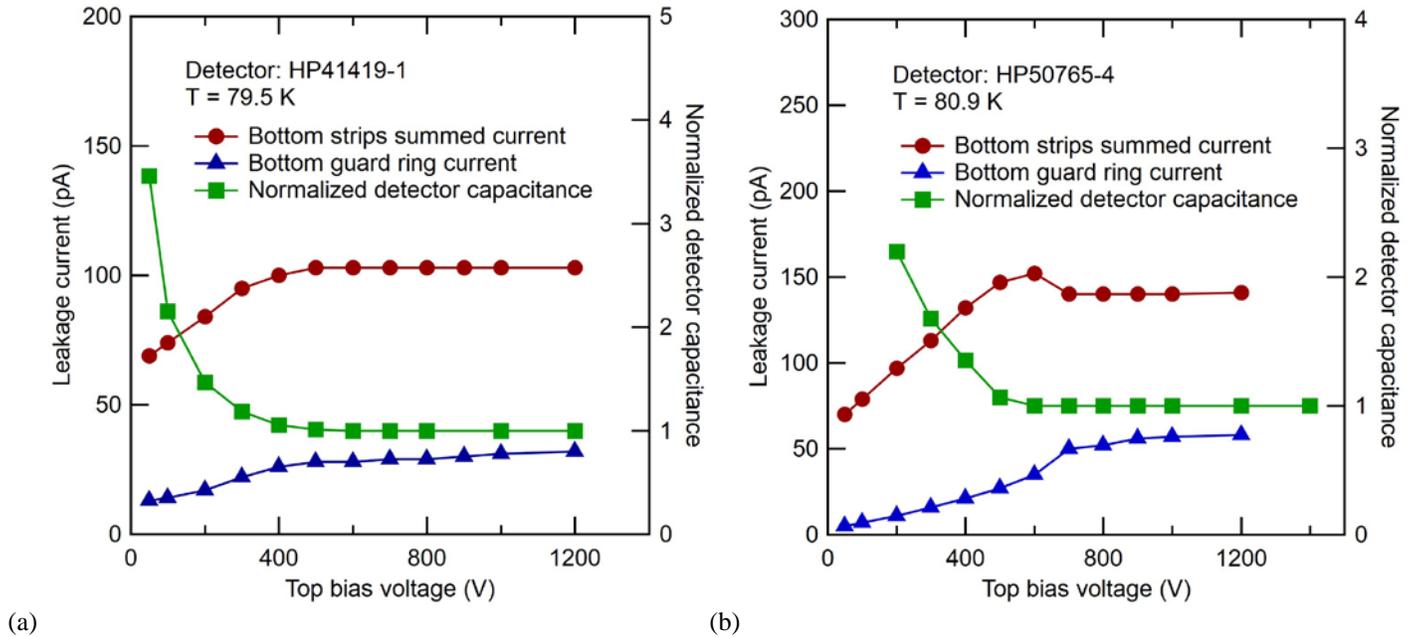

(a)                                                                                            (b)

**Figure 4.8** Example electrical measurements obtained with two different strip detectors. For each detector, the summed leakage current from all bottom strips and the leakage current from the bottom guard ring are plotted as a function of the voltage applied to the top of the detector. Also plotted for reference is the normalized detector capacitance that was obtained with the HPGe crystal when tested previously as a simple planar detector. **(a)** COSI detector. **(b)** GRIPS detector.

**(e) Make gamma-ray spectroscopy measurements:** These measurements would normally be made immediately following the electrical testing, so the detector should already be loaded into the cryostat and cold. If this is not the case, follow the procedures described above for loading and cooling the detector. Wire the detector as shown in Figure 4.4(b). For these measurements, a single strip on each face of the detector is instrumented with a standard chain of pulse processing electronics. All other bottom strips and the bottom guard ring are interconnected and tied to ground. All other readout channels from the top of the detector including the guard ring are also interconnected and tied to ground. The preamplifiers and passive components associated with the pulser inputs to the preamplifiers are contained in shielded boxes attached to the cryostats. The high voltage connection to the top of the detector is made through a high voltage filter that is also housed in a shielded box attached to each cryostat. Due to a lack of a full set of readout electronics to conveniently test all strips on the detector, only a sampling of the strips is typically checked. Using Am-241, Ba-133, and Cs-137 sources, acquire pulse height spectra from select strips on both sides of the detector. At a minimum, edge strips should be assessed for surface channel problems using the Am-241 source, and charge trapping (particularly electron trapping as measured by the anode strips) should be assessed based on Cs-137 spectra acquired from strips near the center of the detector face. For the testing of the COSI detectors, a flight preamp electronics box is available for instrumenting an entire side of the detector. This can be used to more easily make the strip measurements. Example spectra obtained from properly functioning COSI and GRIPS detectors are shown in Figure 4.9.

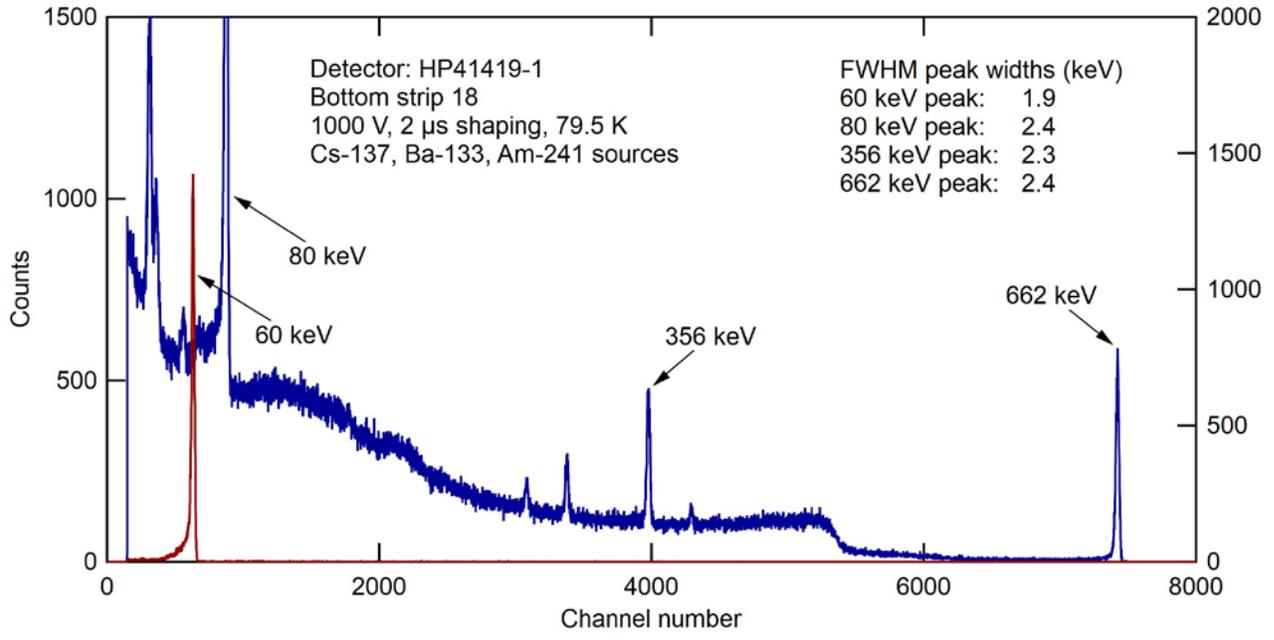

(a)

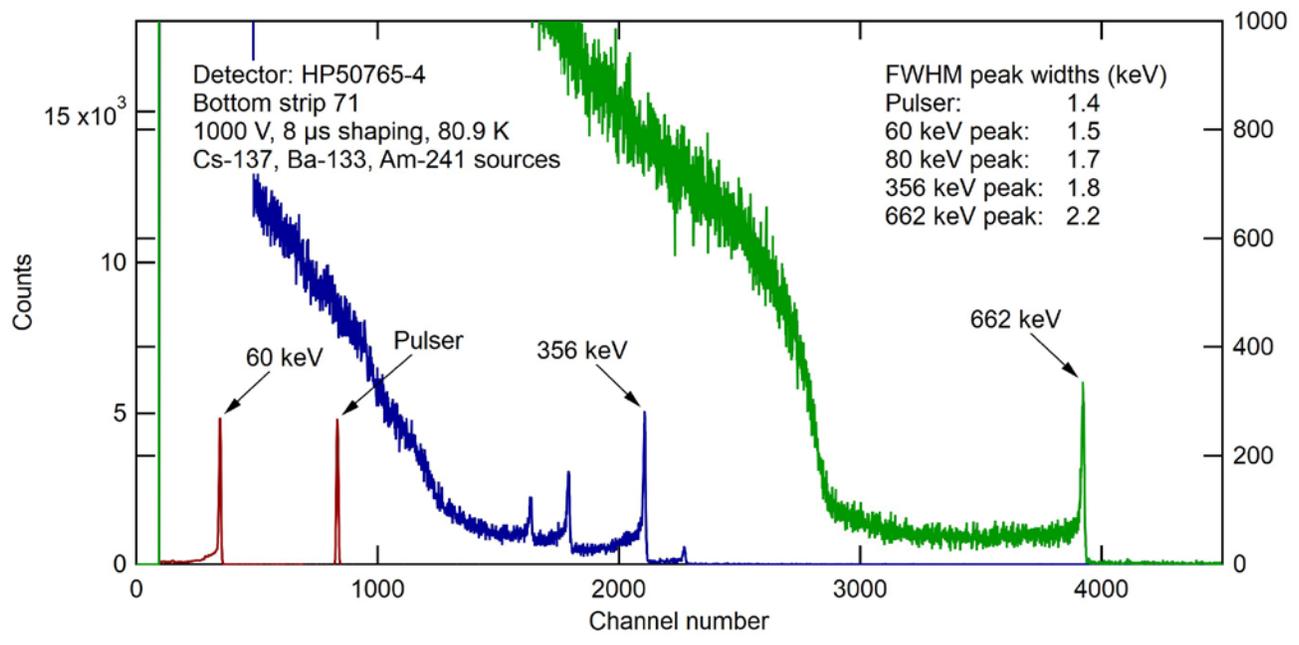

(b)

**Figure 4.9** Example gamma-ray spectra acquired with two different strip detectors. Am-241, Ba-133, and Cs-137 gamma-ray sources were used. The spectra were measured using a bottom strip near the center of the detector face. Additional details such as the detector operating voltage, temperature, and shaping amplifier shaping time are listed in the figures. **(a)** COSI detector spectra. The spectrum plotted in red is from an Am-241 source. The spectrum plotted in blue is from a combination of Ba-133 and Cs-137 sources. **(b)** GRIPS detector spectra. The spectrum plotted in red is from an Am-241 source and includes a pulser peak. The spectrum plotted in blue is from a Ba-133 source. Finally, the spectrum plotted in green is from a Cs-137 source.

## 4.3 Detector troubleshooting

Many things can go wrong during the fabrication and testing of a HPGe strip detector of the COSI or GRIPS type. In this final section of the paper, some of the more common problems and possible causes are listed.

**Symptom:** High guard ring leakage current
**Possible causes:**
(a) HPGe crystal mechanical damage that has not been fully etched away. Such damage should be identifiable through careful inspection of the crystal surfaces using a stereomicroscope.
(b) Improper (incomplete or non-uniform) coating of the detector sides with the amorphous semiconductor thin film. When depleting from the top of the detector, the guard ring leakage will often run away near or just above the full depletion voltage.
(c) Partially undercoating the detector (coating the bottom face near the edges) with amorphous semiconductor during the coating of the top and sides. When depleting from the top of the detector, the undercoating appears to be less of a problem. When depleting from the bottom of the detector, the leakage current can become high at a low voltage.
(d) Coating of the sides of the detector with a relatively low resistivity amorphous semiconductor film. Amorphous Ge sputtered in pure Ar is an example of a side coating that will produce a relatively high guard ring leakage current. See the paper [Amman 2018] for more details.
(e) High detector temperature.
(f) Partial or no IR shield.

**Symptom:** High strip leakage current
**Possible causes:**
(a) HPGe crystal mechanical damage that has not been fully etched away. Such damage should be identifiable through careful inspection of the crystal surfaces using a stereomicroscope.
(b) Improper selection of the electrical contact process (amorphous semiconductor sputter deposition recipe, see [Amman 2018]).
(c) Low inter-strip resistance. This would not cause leakage current through the detector. However, if the strips are dc coupled to preamplifiers, each strip will then sit at a slightly different voltage. This voltage difference between adjacent strips will lead to a current. This current would cause a change in the dc offset at the output of the preamplifiers and, if large enough, could push them into saturation. A low inter-strip resistance can be caused by an improper selection of the electrical contact process (amorphous semiconductor sputter deposition recipe, see [Amman 2018]).
(d) High detector temperature.
(e) Partial or no IR shield.

**Symptom:** High strip noise
**Possible causes:**
(a) High strip leakage current (see above for more information on the causes of high strip leakage current).
(b) Low inter-strip resistance (see above for more information on low inter-strip resistance).
(c) Microphonic noise caused by movement of the wiring (or surrounding structure) between the strips and readout electronics.

**Symptom:** High rate of negative going pulses from strips
**Possible cause:**
Guard ring not properly tied to ground. The guard rings should have low impedance connections (through a low resistance on the grounded detector side or a large capacitor on the high voltage detector side) to ground or to the input of a preamplifier.

**Symptom:** Anomalous spectral peaks
**Possible cause:**
Electrodes may be floating or only weakly connected to either ground or the readout electronics.

# Acknowledgments

The author thanks Paul Luke for useful discussions throughout this project and Julie Lee for help in producing this paper. The author also thanks the numerous collaborators who were involved in projects that contributed to the knowledge needed to write this paper. They include Steve Boggs and the NCT/COSI team, Albert Shih, Pascal Saint-Hilaire and the GRIPS team, Alan Poon and the Majorana team, Dongming Mei and the University of South Dakota team, Steve Asztalos, Craig Tindall, Quinn Looker, Anders Priest, Ren Cooper, Paul Barton, and Kai Vetter.

This work was supported by the U.S. Department of Energy, Office of Science, under contract number DE-AC02-05CH11231.

# References

[Amman 2000a] M. Amman, P. N. Luke, "Three-Dimensional Position Sensing and Field Shaping in Orthogonal-Strip Germanium Gamma-Ray Detectors," Nuclear Instruments and Methods in Physics Research A 452, 155 (2000).

[Amman 2000b] M. Amman, P. N. Luke, "Position-Sensitive Germanium Detectors for Gamma-Ray Imaging and Spectroscopy," Proceedings of SPIE 4141, 144 (2000).

[Amman 2007] M. Amman, P. N. Luke, S. E. Boggs, "Amorphous-Semiconductor-Contact Germanium-Based Detectors for Gamma-Ray Imaging and Spectroscopy," Nuclear Instruments and Methods in Physics Research A 579, 886 (2007).

[Amman 2018] M. Amman, "Optimization of Amorphous Germanium Electrical Contacts and Surface Coatings on High Purity Germanium Radiation Detectors," arXiv: 1809.03046, DOI: 10.13140/RG.2.2.34748.08327/1 (2018).

[Bertolini 1968] G. Bertolini, A. Coche, Editors, **Semiconductor Detectors**, American Elsevier Publishing Company, New York, 1968.

[Chiu 2017] J.-L. Chiu, S. E. Boggs, C. A. Kierans, A. Lowell, et al., "The Compton Spectrometer and Imager," Proceedings of Science, ICRC2017, 796 (2017).

[Deweyl 2020] Deweyl Tool Inc., https://www.deweyl.com/.

[Dinger 1975] R. J. Dinger, "Dead Layers at the Surface of p-i-n Detectors – A Review," IEEE Transactions on Nuclear Science 22, 135 (1975).

[ESPI 2020] ESPI Metals, http://espimetals.com/.

[Fabris 1999] L. Fabris, N. W. Madden, H. Yaver, "A Fast, Compact Solution for Low Noise Charge Preamplifiers," Nuclear Instruments and Methods in Physics Research A 424, 545 (1999).

[Goulding 1961] F. S. Goulding, W. L. Hansen, "Leakage Current in Semiconductor Junction Radiation Detectors and Its Influence on Energy-Resolution Characteristics," Nuclear Instruments and Methods 12, 249 (1961).

[Haller 1981] E. E. Haller, W. L. Hansen, F. S. Goulding, "Physics of Ultra-Pure Germanium" Advances in Physics 30, 93 (1981).

[Haller 2006] E. E. Haller, "Germanium: From Its Discovery to SiGe Devices" Materials Science in Semiconductor Processing 9, 408 (2006).

[Hansen 1977] W. L. Hansen, E. E. Haller, "Amorphous Germanium as an Electron or Hole Blocking Contact on High-Purity Germanium Detectors," IEEE Transactions on Nuclear Science 24, 61 (1977).

[Hansen 1980] W. L. Hansen, E. E. Haller, G. S. Hubbard, "Protective Surface Coatings on Semiconductor Nuclear Radiation Detectors," IEEE Transactions on Nuclear Science 27, 247 (1980).

[Hansen 1983] W. L. Hansen, E. E. Haller, "High-Purity Germanium Crystal Growing," Materials Research Society Symposia Proceedings 16, 1 (1983).

[Hull 1995] E. L. Hull, R. H. Pehl, N. W. Madden, P. N. Luke, et al., "Temperature Sensitivity of Surface Channel Effects on High-Purity Germanium Detectors," Nuclear Instruments and Methods in Physics Research A 364, 488 (1995).

[Hull 2005] E. L. Hull, R. H. Pehl, "Amorphous Germanium Contacts on Germanium Detectors," Nuclear Instruments and Methods in Physics Research A 538, 651 (2005).

[IPC-4556 2013] IPC Plating Processes Subcommittee of the Fabrication Processes Committee, "Specification for Electroless Nickel/Electroless Palladium/Immersion Gold (ENEPIG) Plating for Printed Circuit Boards," IPC-4556 (2013).

[Jacoboni 1981] C. Jacoboni, F. Nava, C. Canali, G. Ottaviani, "Electron Drift Velocity and Diffusivity in Germanium," Physical Review B 24, 1014 (1981).

[Keithley 2020] Keithley, https://www.tek.com/keithley.

[Kierans 2016] C. A. Kierans, S. E. Boggs, J.-L. Chiu, A. Lowell, et al., "The 2016 Super Pressure Balloon Flight of the Compton Spectrometer and Imager," Proc. Sci. INTEGRAL2016, 075 (2016).

[Kingston 1956] R. H. Kingston, "Review of Germanium Surface Phenomena," Journal of Applied Physics 27, 101 (1956).

[Knoll 1989] G. F. Knoll, **Radiation Detection and Measurement**, 2nd Edition, John Wiley & Sons, New York, 1989.

[Knowles 2020] Knowles Precision Devices, https://www.knowlescapacitors.com/.

[Lapmaster 2020] Lapmaster Wolters, https://www.lapmaster-wolters.com/.

[Llacer 1964] J. Llacer, "Study of Surface Effects in Thick Lithium Drifted Silicon Radiation Detectors," IEEE Transactions on Nuclear Science 11, 221 (1964).

[Llacer 1966] J. Llacer, "Geometric Control of Surface Leakage Current and Noise in Lithium Drifted Silicon Radiation Detectors," IEEE Transactions on Nuclear Science 13, 93 (1966).

[Looker 2015a] Q. Looker, M. Amman, K. Vetter, "Leakage Current in High-Purity Germanium Detectors with Amorphous Semiconductor Contacts," Nuclear Instruments and Methods in Physics Research A 777, 138 (2015).

[Looker 2015b] Q. Looker, M. Amman, K. Vetter, "Inter-Electrode Charge Collection in High-Purity Germanium Detectors with Amorphous Semiconductor Contacts," Nuclear Instruments and Methods in Physics Research A 781, 20 (2015).

**M. Amman, 2020, "High Purity Germanium Based Radiation Detectors with Segmented Amorphous Semiconductor Electrical Contacts: Fabrication Procedures"**

[Luke 1992] P. N. Luke, C. P. Cork, N. W. Madden, C. S. Rossington, et al., "Amorphous Ge Bipolar Blocking Contacts on Ge Detectors," IEEE Transactions on Nuclear Science 39, 590 (1992).

[Luke 1994a] P. N. Luke, R. H. Pehl, F. A. Dilmanian, "A 140-Element Ge Detector Fabricated with Amorphous Ge Blocking Contacts," IEEE Transactions on Nuclear Science 41 976 (1994).

[Luke 1994b] P. N. Luke, C. S. Rossington, M. F. Wesela, "Low Energy X-Ray Response of Ge Detectors with Amorphous Ge Entrance Contacts," IEEE Transactions on Nuclear Science 41, 1074 (1994).

[Malm 1976] H. L. Malm, R. J. Dinger, "Charge Collection in Surface Channels on High-Purity Ge Detectors," IEEE Transactions on Nuclear Science 23, 76 (1976).

[McMaster-Carr 2020] McMaster-Carr Supply Company, https://www.mcmaster.com/.

[Micro Abrasives 2020] Micro Abrasives Corporation, https://microgrit.com/.

[Miner 1967] C. E. Miner, "A Semiconductor Detector Cryostat," Nuclear Instruments and Methods 55, 125 (1967).

[Mirion 2020] Mirion Technologies, Inc., https://www.mirion.com/.

[Momayezi 1999] M. Momayezi, W. K. Warburton, R. Kroeger, "Position Resolution in a Ge-Strip Detector" Proceedings of SPIE 3768, 530 (1999).

[Norton 2020] Norton Winter, https://www.nortonabrasives.com/.

[Ortec 2020] Ortec, Advanced Measurement Technology, http://www.ortec-online.com/.

[Ottaviani 1975] G. Ottaviani, C. Canali, A. Alberigi Quaranta, "Charge Carrier Transport Properties of Semiconductor Materials Suitable for Nuclear Radiation Detectors," IEEE Transactions on Nuclear Science 22, 192 (1975).

[PEI 2020] Photofabrication Engineering, Inc., https://www.photofabrication.com/.

[Photo Sciences 2020] Photo Sciences, Inc., https://www.photo-sciences.com/.

[Portable Technologies] Portable Technologies LLC, https://www.portable-technologies.com/.

[Qunitel 2020] Neutronix Quintel, http://neutronixinc.com/.

[RDMathis 2020] RD Mathis Company, https://rdmathis.com/.

[Rogers 2020] Rogers Corporation, http://www.rogerscorp.com/.

[SCS 2020] Specialty Coating Systems, https://scscoatings.com/equipment/.

[Shih 2012] A. Y. Shih, R. P. Lin, G. J. Hurford, N. A. Duncan, et al., "The Gamma-Ray Imager/Polarimeter for Solar Flares (GRIPS)," Proceedings of SPIE 8443, 84434H-1 (2012).

[Vaga 2020] Vaga Industries, https://www.vaga.com/.

[Vishay 2020] Vishay Intertechnology, http://www.vishay.com/.

[Walker 1991] P. Walker, W. H. Tarn, Editors, **Handbook of Metal Etchants**, CRC Press, New York, 1991.